\documentclass[final,onefignum,onetabnum]{journalarticle}

\usepackage{lipsum}
\usepackage{amsfonts}
\usepackage{graphicx}
\usepackage{epstopdf}
\usepackage{algorithmic}
\usepackage{xr}
\ifpdf
  \DeclareGraphicsExtensions{.eps,.pdf,.png,.jpg}
\else
  \DeclareGraphicsExtensions{.eps}
\fi

\newsiamremark{remark}{Remark}
\newsiamremark{hypothesis}{Hypothesis}
\crefname{hypothesis}{Hypothesis}{Hypotheses}
\newsiamthm{claim}{Claim}

\headers{Multilayer Modularity Belief Propagation}{Weir, Walker, Zdeborov\'{a} and Mucha}

\title{Multilayer Modularity Belief Propagation To Assess Detectability Of Community Structure\thanks{ Submitted to the editors DATE.
\funding{This work was supported by grant R01DK111930 from the National Institute of Diabetes and Digestive and Kidney Diseases and by the James S. McDonnell Foundation 21st Century Science Initiative - Complex Systems Scholar Award grant \#220020315, as well as by the NIH BD2K grant \#T32CA201159 and NIH BCB training grant \#T32GM67553.  Additional support was provided by the Army Research Office (MURI award W911NF-18-1-0244). We acknowledge the support and hospitality of Duke University, where this work was initiated. The content is solely the responsibility of the authors and does not necessarily represent the official views of any of the funding organizations. }}}

\author{William H. Weir\thanks{Bioinformatics and Computational Biology, University of North Carolina, Chapel Hill, NC 
  (\email{wweir@med.unc.edu}).}
\and Benjamin Walker\thanks{Carolina Center for Interdisciplinary Applied Mathematics, Department of Mathematics, University of North Carolina, Chapel Hill, NC}
\and Lenka Zdeborov\'{a}\thanks{Institut de Physique Th\'eorique, CEA, CNRS, Universit\'e Paris-Saclay, CNRS}
\and Peter~J.~Mucha$^{\dagger\ddagger}$}
\usepackage{amsopn}

\usepackage{amssymb}
\usepackage{epstopdf}
\usepackage{amsmath}
\usepackage{bbm}
\usepackage{nameref}
\usepackage{subcaption}
\usepackage{scalerel}
\usepackage[section]{placeins}
\usepackage[innerleftmargin = 4pt, innerrightmargin = 4pt, innertopmargin = 4pt, innerbottommargin = 4pt]{mdframed}

\makeatletter
\newcommand*{\rom}[1]{\expandafter\@slowromancap\romannumeral #1@}
\makeatother

\usepackage[color, rightbars]{changebar}
\cbcolor{blue}
\newcommand*{\fullref}[1]{\hyperref[{#1}]{ \textbf{\ref*{#1}.~\nameref*{#1}}}}

\DeclareMathOperator*{\argmax}{arg\,max}

\graphicspath{{./graphics/}}

\ifpdf
\hypersetup{
  pdftitle={Multilayer Modularity Belief Propagation To Assess Detectability Of Community Structure},
  pdfauthor={Weir, Walker, Zdeborov\`{a} and Peter J. Mucha}
}
\fi

\begin{document}
\maketitle

\begin{abstract}
Modularity based community detection encompasses a number of widely used, efficient heuristics for identification of structure in networks.  Recently, a belief propagation approach to modularity optimization provided a useful guide for identifying non-trivial structure in single-layer networks in a way that other optimization heuristics have not.  In this paper, we extend modularity belief propagation to multilayer networks. As part of this development, we also directly incorporate a resolution parameter.
We show that adjusting the resolution parameter affects the convergence properties of the algorithm and yields different community structures than the baseline.  We compare our approach with a widely used community detection tool, GenLouvain, across a range of synthetic, multilayer benchmark networks, demonstrating that our method performs comparably to the state of the art.  Finally, we demonstrate the practical advantages of the additional information provided by our tool by way of two real-world network examples.  We show how the convergence properties of the algorithm can be used in selecting the appropriate resolution and coupling parameters and how the node-level marginals provide an interpretation for the strength of attachment to the identified communities.  We have released our tool as a Python package for convenient use.

\end{abstract}

\begin{keywords}
community detection, modularity, belief propagation, networks, multilayer networks, message passing, resolution parameter
\end{keywords}

\begin{AMS}
  68Q25, 68R10, 68U05
\end{AMS}

\section{Introduction}
Networks provide useful models for understanding complex systems across a wide range of problems in different domains, including biology, engineering and the social sciences.  Recent attention has focused on developing tools to understand the expanded class of \textit{multilayer} networks.  The multilayer framework is quite flexible, allowing for the representation of multiplex pairwise interactions, dynamic networks, different classes of nodes, and ``networks of networks" \cite{Kivela:2014dm}.  One class of ongoing challenges in network science, in particular for multilayer networks, is the detection and representation of high-level structure and communities (for review, see e.g.\  \cite{Fortunato:2010vi,Fortunato:2016if,Porter:2009we,Schaub:2017gw,Shai:2017uc}).

There are many computational approaches to identifying structure within networks.  One family of approaches attempts to fit a statistical model to the observed network and uses hypothesis testing to assess the significance of proposed community structure.  Many of these models are derived from the stochastic block model (SBM) \cite{Holland:1983fu,Karrer:2011vd,Peixoto:2017fl}.  Another approach is to define an objective function that measures the quality of community structure in a particular sense, and then optimize that objective.
For example, the Infomap approach \cite{Rosvall1118} detects communities as groups of nodes that minimize the description length to encode random walks on a network.  Another popular quality function, the modularity score developed by Newman and Girvan \cite{Newman:2004ep}, compares the observed edge weight within groups to that expected under a null model of the network.  Modularity was extended to multilayer networks by incorporating interlayer edges as additional coupling between the layers~\cite{Mucha:2010vk}.  Generalizing a re-derivation of modularity from a Laplacian dynamics perspective \cite{Lambiotte:2008da,Lambiotte:2015da}, Mucha~\textit{et al.}~\cite{Mucha:2010vk} developed a formula for multilayer modularity that can be written generally in the supra-matrix form
\begin{equation}
Q(\gamma,\omega)=\sum_{i,j} \left( A_{ij} - \gamma P_{ij} + \omega C_{ij} \right)\delta(c_{i},c_{j})
\label{eq:multimod}
\end{equation}
where $i$ and $j$ each index distinct node-layer objects, possibly in different layers,
$\mathbf{A}$ is the supra-adjacency\footnote{In the supra-adjacency representation,~a single block diagonal matrix is used to represent all intralayer connections, each block representing a single-layer, with no connections between the blocks.  A different matrix, $\mathbf{C}$ encodes the interlayer connections.  Note that $\dim(\mathbf{A})=\dim(\mathbf{C})=\dim(\mathbf{P})$.  See Section~\ref{sec:notation} for further explanation of multilayer notation.} encoding the intralayer edges, $\mathbf{P}$ describes the expected number of intralayer edges based on the selected random model(s), and $\mathbf{C}$ encodes the interlayer connections.  The normalizing factor traditionally written in front of the summation above has been absorbed here into the constituent terms for notational convenience.  We here assume for simplicity the Newman-Girvan model for undirected edges within each layer, writing the null model contribution (prior to absorbing the normalizing factor) as

\begin{equation}
 P_{ij}=\begin{cases} \frac{d_i d_j}{2m_{l_{i}}} & l_i=l_j \\0 & l_i \ne l_j \end{cases}
\label{eq:nullmodel}
\end{equation}
where $l_i$ is the layer containing node-layer $i$, $d_i = \sum_j A_{ij}$, and $m_{l_i}=\sum_{i,j \in \mathcal{V}_{l_i}}A_{ij}$ is the total weight of edges in layer $l_i$ (\textit{i.e.}, between the set of node-layers denoted $\mathcal{V}_{l_i}$). We enforce on $\mathbf{A}$ that a given node-layer $i$ only participates in intralayer edges (\textit{i.e.}, edges within its own layer).  In the case where edge weights are binary ($A_{ij} \in \{0,1\}$), $d_i$ is the degree of node $i$.  For weighted networks, $A_{ij}$ may be discrete or continuous and $d_i=\sum_j A_{ij}$ is the `strength' of node $i$.  Similar null models are available for bipartite graphs, directed networks, and networks with signed edges (see, e.g., the supplement of \cite{Mucha:2010vk} for references to appropriate forms for $P_{ij}$ in different contexts).  

In general, maximizing Eq.~\ref{eq:multimod} over the combinatorially large space of possible partitions is {NP-hard} \cite{Brandes:2008ua}. Several fast and efficient algorithms exist for locally optimizing modularity, including Louvain \cite{Blondel:2008vn} and the \textit{GenLouvain} \cite{Jeub:2016uh} extension for optimizing multilayer modularity.  One of the main problems with optimizing modularity as a means of community detection is that partitions of high modularity often exist even in randomly generated networks without underlying structure (see for example \cite{Bagrow:2012hy,DeMontgolfier:2011jp}).
Zhang and Moore \cite{Zhang:2014gea} were able to surmount several of the issues with modularity-based methods on single-layer networks by treating modularity optimization in terms of the statistical physics of the  spin-glass system with Hamiltonian $ \mathcal{H}=-mQ(\{t_i\})$, where $m$ is the number of edges and $\{t_i\} = [t_1,\hdots,t_N]$ with $t_i \in \{1, ..., q\}$ indicates the assignment of node $i$ (of $N$) to one of $q$ communities. As such, the distribution of states of the system is given by the Boltzmann distribution
\begin{equation}
P(\{t_i\})  \propto e^{-\beta {\mathcal{H}} (\{t_i\} )}
\label{eq:gibbsprob2}
\end{equation}
where $\beta$ represents the nondimensional inverse temperature  that sets the sharpness of the energy landscape.  Maximizing the \textit{joint} distribution $P(\{t_i\})$ is then equivalent to globally optimizing modularity and identifying the ground state of the system.  Instead of searching for a global modularity maximum, Zhang and Moore attempt to solve for the marginals of each node, $P(t_i = q)$, in the finite temperature regime.  By looking for a ``consensus of good partitions" rather than seeking a single ``best" partition, the algorithm converges to non-trivial structures above a certain temperature only if there is broad underlying structure within the network.  If it exists, this parameter regime where belief propagation converges to non-trivial structure is called the retrieval phase by Zhang and Moore.  In particular, Zhang and Moore demonstrated that the algorithm's convergence properties distinguish between synthetic networks with and without known underlying structure, even when the nominal modularity values of the identified partitions are quite similar.  In this sense, a belief propagation approach is able to detect when a particular network has ``significant structure".  We note that throughout this paper we use the term `significant' to mean having an identifiable region in the $\beta$ domain where the algorithm converges to non-trivial community structure, i.e., the marginals are not all approximately equal to $\frac{1}{q}$. From a statistical physics perspective, this means that for a given network there exists a retrieval phase in the state space where the beliefs converge to a non-trivial solution.  The relationship between convergence of belief propagation and the detectability of communities in SBMs has been explored analytically \cite{Decelle:2011ek,Decelle:2011ik,Mossel:2014wm} and empirically  \cite{Zhang:2014gea,Ghasemian:2016hg} In this manuscript we show empirically that similar claims extend to the multilayer modularity approach employed here.  We emphasize that this differs from the standard notion of `statistical significance'
--- specifically, we do not assess the value of a statistic compared to any particular model. That is, we do not assume any explicit model of communities here in using the modularity objective function maximization approach.

 Beyond providing this notion of significance of structure, maximization of the marginals has the additional benefit of producing an interpretation of a soft partition wherein nodes are partially assigned across multiple communities. That is, the marginals reveal which node labels the algorithm is most uncertain about. Moreover, we can use the average entropy across all of the node labels as a measure of confidence in the predicted structure.  While several tools are available to compute the marginals, 
including Markov Chain Monte Carlo sampling and Gibbs sampling, the class of algorithms known as belief propagation (alternatively, the cavity method or sum-product algortihm) offers several unique advantages including computational efficiency and the tractability of asymptotic analysis \cite{mezard2009information}.  Belief propagation is a general algorithm for calculating the marginals of a joint distribution by a series of iterative updates.  Belief propagation was initially developed for trees \cite{Pearl:1982bw}, for which it is an exact algorithm, but has been shown to provide good approximations on graphs with loops (\textit{i.e.}\ ``loopy" belief propagation)~\cite{pearl1988probabilistic,mezard2009information} assuming loops are small and short range correlations decay exponentially \cite{Zdeborova:2016ff}.  Belief propagation was first successfully applied to community detection in solving the stochastic block model in \cite{Hastings:2006ce}, improved upon in \cite{Decelle:2011ik} using an expectation maximization process to update the model parameters, and rigorously analyzed in \cite{Decelle:2011ek}.  Zhang and Moore introduced the belief propagation updates for (single-layer) modularity maximization \cite{Zhang:2014gea}.

Our work introduces a belief propagation approach for the more general multilayer modularity framework, suitable for a variety of multilayer topologies.  Specifically, we extend Zhang and Moore's \textit{modbp} method in three ways.  We explicitly allow weighted edges, which can greatly influence the communities detected (see \cite{Newman:2004eb} and \cite{Shi:2018gc}).  We incorporate a resolution parameter $\gamma$ into the modularity quality function as done in~\cite{Reichardt:2006eh} and show that this can create a wider retrieval phase and achieve better performance in the case where the number of communities is not known \textit{a priori}.  Finally, we extend \textit{modbp} to the multilayer modularity framework developed by Mucha~\textit{et al.} \cite{Mucha:2010vk}. Our resulting method, \textit{multimodbp}, can be used on both multilayer networks and single-layer networks. We demonstrate the use of this tool on both synthetic and real-world data.  We have developed a \textit{multimodbp} Python package implementing our method in a fast, efficient manner that interfaces with other standard networks tools.  To our knowledge, the only other multilayer community detection method that incorporates belief propagation uses a specific dynamic stochastic block model (see Section~\ref{sec:dsbm}), which only captures temporal multilayer dynamics \cite{Ghasemian:2016hg}. 

The rest of this paper is organized as follows. In Section \ref{methods}, we introduce the original modularity belief propagation approach by Zhang and Moore.  Then we present the updates we have made including the incorporation of the resolution parameter and the extension to the multilayer framework.  Next, in Section \ref{sec:results}, we show how incorporation of the resolution parameter can improve performance in the context of synthetic data as well as a real-world network in the single-layer case. 
To demonstrate the ability of our approach to detect communities within the multilayer framework, we showcase \textit{multimodbp} on two types of synthetic multilayer models with differing interlayer topology, and compare the performance of our model with another popular multilayer modularity based approach, \textit{GenLouvain}~\cite{Jeub:2016uh}.  Finally, we demonstrate the utility of our model on two real-world data sets, showing how a belief propagation approach reveals additional information about a network's structure above and beyond other methods.  We conclude with a brief discussion and remark on other possible improvements.

\section{Methods}
\label{methods}
\subsection{Notation\label{sec:notation}}

Before introducing \textit{multimodbp}, we provide a brief introduction to the notation that we use to describe multilayer networks, much of which follows \cite{Kivela:2014dm}. A multilayer network is a general framework that can be used as a way to represent multiple types of relationships between nodes or to describe ``networks of networks."   In the multilayer formulation, all edges representing a certain kind of relationship are present within a ``layer."  A single ``node'' can exist in multiple layers, and we refer to a particular node as it exists in a single layer as a ``node-layer.''
In the single-layer case, the edges between pairs of nodes are represented by a matrix $A \in \mathbb{R}^{N\times N}$, with element $A_{ij}$ indicating whether an edge exists between nodes $i$ and $j$ (along with the weight of that edge, if applicable).  In the multilayer cases we consider here, we can represent each layer as $A^{l}$, with $l \in \{1,\ldots,L\}$ indexing the layer, and the whole of the multilayer network as a tensor $\mathbf{M} \in \mathbb{R}^{N \times N \times L}$, where L is the number of layers in the multilayer network. We note in particular here that in so doing we restrict our present attention to situations where we have a common set of nodes in each layer; where a node is not present in a given layer (\textit{i.e.}, does not have a node-layer in that layer), we add in a placeholder node-layer that remains unconnected to everything else in that layer. It is also common to flatten $\mathbf{M}$ into an $\mathbf{A} \in \mathbb{R}^{NL \times NL}$ ``supra-adjacency" matrix.  We use the notation $i\cong j$ to denote that two node-layers are identified with the same node.  Typically we index the node-layers such that $i \cong i+N \cong \ldots \cong i+N(L-1)$.  We use the ``supra-adjacency" notation in the rest of the paper and indices $i,j,k$ to refer to node-layers unless otherwise specified.  We also use the array $\vec{l}$ to keep track of layer assignments: $\vec{l} = [l_1, ... , l_{NL}] $ where $l_i\in \{1,\ldots,L\}$ specifies the layer that node-layer $i$ is in.  We use $\mathcal{V}_{l_i}$ to denote the set of node-layers in layer $l_i$ (\textit{i.e.}, nodes in the same layer as $i$ including $i$ itself).  In addition to the edges within each of the layers, we must also specify the interlayer topology, which we encode through the matrix $\mathbf{C} \in \mathbb{R}^{NL \times NL }$.   We only allow elements of $\mathbf{C}$ to be non-zero if the corresponding  node-layers are in different layers, that is, $C_{ij} = 0$ if $l_i=l_j$. The matrix $\mathbf{C}$ can be used to represent a wide variety of possible interlayer topologies. In the cases we consider here, each node-layer will be connected to a subset of the other node-layers that correspond to the same node in different layers.  For example in the unordered (categorical) multiplex case, we connect all pairs of nodes-layers that correspond to the same node. In the temporal (ordered) case, the simplest interlayer topology only connects pairs of node-layers between adjacent layers corresponding to the same node.  We discuss in more detail the structure of the coupling matrix used for each of the multilayer examples in Section~\ref{sec:multilayer_results}.

\subsection{Original modularity belief propagation}

Zhang and Moore \cite{Zhang:2014gea} apply belief propagation to modularity maximization, deriving update conditions for the node beliefs in terms of the message $\psi^{i \to k }_t $ from node $i$ to $k$ concerning community $t$ that helps determine what node $k$ ``believes" its own community to be:\footnote{See supplement section~\ref{bpeq_deriv} for derivation of these equations.} 
\begin{equation}
\psi^{i \to k }_t = \frac{1}{Z_{i\to k}} \exp {\left [ \frac{\beta d_i }{2m}\theta_t  + \sum_{j\in \partial_i \setminus k} \log \left( 1+\psi^{j \to i }_t(e^{\beta}-1) \right) \right ]}
\label{eq:modbp}
\end{equation}
where $m$ is the total number of edges; $d_i = \sum_j { A_{ij}}$ is the degree of node~$i$; $\partial_i\setminus k$ is the neighborhood of node $i$ except node $k$; $Z_{i \to k}$ is a normalization constant such that $\sum_s{\psi^{i \to k}_s}=1$; and $\theta_t=\sum_{i}{d_i \psi^i_t}$ serves as a field-like term, approximating the null model contribution in modularity in terms of each node's belief about its own community, $\psi^i_t$, given by its marginal
\begin{equation}
\psi^{i }_t =\frac{1}{Z_i} \exp {\left [ \frac{\beta d_i }{2m}\theta_t  + \sum_{j\in \partial_i } \log \left( 1+\psi^{j \to i }_t(e^{\beta}-1) \right) \right ]}
\label{eq:modbpmarginal}
\end{equation}
where the belief includes contributions from all neighbors of node~$i$, with normalization $Z_i$ such that $\sum_s{\psi^i_s}=1$. That is, $\psi^i_t$ can be thought of as the belief that node $i$ sends to itself, insofar as $i$ is already not a member of its own neighborhood, $\partial_i$, so there is no excluded element in the sum over $j$.

Fixed points of the ``loopy" belief propagation algorithm are minimizers of the Bethe free energy
\begin{equation}
f_{\rm Bethe}=-\frac{1}{N\beta}\left(\sum_{i\in\mathcal{V}}{\log{Z_i}} - \sum_{(i,j)\in \mathcal{E}}{\log{Z_{ij}}}+\frac{\beta}{4m}\sum_t{\theta^2_t} \right)\,,
\label{eq:bethe}
\end{equation}
where $\mathcal{V}$ is the set of $N$ nodes, $\mathcal{E}$ is the set of edges, and $Z_{ij} = \sum_{st}e^{\beta\delta_{st}}\psi^i_s\psi^j_t $ is the normalization constant for the pairwise joint marginals.

Computing marginals for each node, Zhang and Moore defined a ``retrieval partition'' by assigning each node to a community according to its greatest marginal $t_i = \argmax_t \psi^i_t$, with randomly broken ties.  Retrieval modularity can be computed from the retrieval partition using Eq.~\ref{eq:multimod}.  We note that while this approach uses the modularity score to establish the energy landscape over which optimization is performed, ultimately the belief propagation minimizes the free energy; while lower free energy often corresponds to higher modularity for the retrieved partition, this relationship is in no way required and indeed is sometimes violated.

\subsection{Multilayer modularity belief propagation update equations}

We describe here the modifications to the \textit{modbp} equations used in \textit{multimodbp}. Formal justification for these modifications is provided in Section~\ref{bpeq_deriv} of the supplement.  
First, by incorporating a resolution parameter~\cite{Reichardt:2006eh}, $\gamma$, we effectively treat the field term and the edge term in the update equations as though they are at different temperatures.  Second, to appropriately handle the null model in the multilayer modularity framework (see Eq.~\ref{eq:nullmodel}), we have adapted the field term, $\theta_t^l$ to be layer specific and to only contribute to the beliefs originating from nodes within a given layer,~$l$.  Having a separate null model for each layer is one of the differentiating features between the original modularity and multilayer modularity defined in Eq.~\ref{eq:multimod}.\footnote{If there is no interlayer coupling ($\omega=0$), then multilayer modularity is equivalent to running original modularity on each layer separately, treated as independent networks.}   Finally, we introduce an additional interlayer contribution, scaled by interlayer coupling parameter $\omega$, to account for interlayer edges in a manner similar to the interlayer contributions to multilayer modularity, leading to the new update equation:

\begin{equation}
\label{multimodbp_eq}
\psi^{i \to k }_t \propto \exp {\left [ \gamma \frac{\beta d_i }{2m_{l_i}}\theta^{l_i}_{t}  + \sum_{j\in \partial_i \setminus k}  \log{(1+\psi^{j \to i }_t(e^{\tilde{A}_{ij}\beta} -1))} \right]} 
\end{equation}

where $l_i$ is the layer containing node $i$ (\textit{i.e.}\ $i\in\mathcal{V}_{l_i}$);
the field term $\theta^{l_i}_{t}=\sum_{j\in \mathcal{V}_{l_i}}{d_j\psi^j_t}$ and node strength\footnote{Modularity belief propagation for weighted networks was developed in \cite{Shi:2018gc}.  The update equations are similar to the Zhang and Moore version, however now $A_{ij} \in [0,\infty)$ and we use weighted degree (strength) as defined above.}\ 
$d_i=\sum_{j\in\mathcal{V}_{l_{i}} A_{ij}}$ only include contributions from layer $l_i$;
$\tilde{A}_{ij}=A_{ij}\delta(l_i,l_j)+\omega C_{ij}(1-\delta(l_i,l_j))$, with $l_j$ the layer containing $j$, combines the intralayer and interlayer edges according to whether $i$ and $j$ are present in the same layer; 
and we again define $\psi^i_t$ to be the normalized ($Z_i=\sum_s{\psi^i_s}$) version of $\psi^{i\to i}_t$ with $\partial_i\setminus i = \partial_i$ for the sum in Eq.~\ref{multimodbp_eq}. 
We note that the block description of $A_{ij}$ and $C_{ij}$ considered here makes the $\delta(\cdot,\cdot)$ indicators in $\tilde{A}_{ij}$ unnecessary; but we include them to help clarify the notation in terms of the layers containing $i$ and $j$.

The solution to the above iterative equations is the minimizer of the following Bethe free energy equation, as derived in  Section~\ref{bfe_deriv} of the supplement:
\begin{align}
\label{eq:bfe_multi}
f_{\text{Bethe}}= -\frac{1}{N\beta}  \left( \sum_i{\log Z_i} -  \sum_{i,j\in \mathcal{E}} { \log Z_{ij}}  + \sum_l \frac{\beta}{4m_l}\sum_t  (\theta^l_t)^2 \right)
\end{align}
where $Z_{ij}= \sum_{st}e^{\beta\delta_{st}}\psi^i_s\psi^j_t $ is the normalization factor for the pairwise joint marginals.

While we demonstrate \textit{multimodbp} below in the context of specific multilayer topologies, our formulation is flexible enough to handle any type of multilayer network consisting of two classes of edges (i.e., intralayer and interlayer edges). In particular, we remark that, similar to the weights in $A_{ij}$, the contribution from $C_{ij}$ is explicitly included here, allowing for different interlayer weights.  In principle, the method could also be extended to networks with multiple types of edges, such as encountered in representing network data that is both longitudinal and multiplex, with each new edge type introducing its own coupling parameter, $\omega_x$ (appropriately indexed).

\subsection{Choice of $\boldsymbol{\beta}$}
\label{sec:beta_star}

By analyzing the linearized stability of the fixed point to small, uncorrelated perturbations, Zhang and Moore provided a heuristic for selecting an appropriate value of $\beta=\beta^*$ at which point the trivial, factorized solution ($\psi_t^{j\rightarrow i}= 1/q$ for all beliefs) is no longer stable, assuming a random distribution of edges conditioned on the degree distribution.  If significant structure is \emph{not} present within the network, for values of $\beta>\beta^{*}$, the algorithm enters the `spin-glass' phase in which convergence never occurs.  In contrast, if the network has detectable community structure, then there is a range of values, $\beta^{R}<\beta<\beta^{SG}$ where a retrieval state has lower free energy than the trivial solution and is stable.  Typically, $\beta^*$ is greater than $\beta^{R}$ and is within the retrieval phase.  We demonstrate empirically that $\beta^*$ is indeed within the retrieval phase in supplement Figures~\ref{sbm2com_betascan}, \ref{multilayer_betascan}, and \ref{fig:beta_scan_senate}.  However, in principle for real-world networks, $\beta^*$ could exist outside of the retrieval phase, in which case it would be necessary to scan a wider range of $\beta$ values.

In practice, this can be used to eliminate or at least reduce one of the free parameters involved in running the algorithm.   Shi et al. \cite{Shi:2018gc} recently expanded the stability analysis around the fixed point for the case where random weights are added on the edges. We have adopted their heuristic for selecting $\beta^*$ in the multilayer context, as the intralayer field term does not contribute to the linearized form of the update equations in the limit of small perturbations.    The linear stability of the factorized solution  is characterized by the derivatives of the messages with respect to each other at the fixed point ($1/q$).  To identify $\beta^{*}$, the critical value for instability with respect to random, uncorrelated perturbations, we linearize the \textit{multimodbp} update equations (Eq.~\ref{multimodbp_eq}) and then analyze the stability of the equations under repeated iteration.  We use the notation from Zhang and Moore and Shi \textit{et al}.   Suppose that each belief is perturbed by a small random amount, $\psi_{t_j}^{i\rightarrow j} = \frac{1}{q} + \epsilon^{i \rightarrow j}_{t_j}$. To first order, these perturbations will propagate by
\begin{equation}
\epsilon_{t_i}^{i \rightarrow j} = \sum_{k\in \partial i \setminus j}\sum_{t_k} T^{i \rightarrow j,k \rightarrow i}_{t_i,t_k} \epsilon^{k\rightarrow i}_{t_k} \,,
\end{equation}
where
\begin{equation}
T^{i \rightarrow j,k \rightarrow i}_{t_i,t_k}= \frac{\partial \psi_{t_i} ^{i \rightarrow j}}{\partial \psi_{t_k} ^{k \rightarrow i}} \biggr \rvert_{1/q} \,.
\end{equation}
We provide a derivation for the form of $T^{i \rightarrow j,k \rightarrow i}_{t_i,t_k}$ in the supplement, section~\ref{beta_star_deriv}, and show that its largest eigenvalue is
\begin{equation}
\eta_{ij}=\frac{e^{\beta \tilde{A}_{ij}} -1 }{e^{\beta \tilde{A}_{ij}} + q -1} \,,
\end{equation}
where again $\tilde{A}_{ij}=A_{ij}\delta(l_i,l_j)+\omega C_{ij}(1-\delta(l_i,l_j))$ defines the appropriate weight and connectivity between nodes $i$ and $j$.   Shi \textit{et al.}\ show that the message will only remain stable if the variance of the perturbations remains less than one over an arbitrary-length path in the graph, providing the following equation: 
\begin{equation}
\label{eq:bstar_cond}
\left< \left(\frac{e^{\beta^* \tilde{A}_{ij}} -1 }{e^{\beta^* \tilde{A}_{ij}} + q -1}\right)^2 \right>_{ij}\hat{c} =1 \,,
\end{equation}
where $\hat{c} = \frac{<d^2>}{c}-1$ is the average excess degree of the network, and the expectation is taken over all non-zero edge weights. We can solve this equation to identify the $\beta^*$ that appropriately incorporates both the weights on the edges of the networks as well as the interlayer coupling $\omega$.  We use Newton's method to solve Equation~\ref{eq:bstar_cond} for $\beta^*(\tilde{A}_{ij} | q,\omega)$.

We have found that this heuristic works well in identifying values of $\beta$ for which our method converges. We note that $\beta^{*}$ represents the boundary for stability of the solution for uncorrelated perturbations in the beliefs.  In the case when detectable community structure exists, the messages become correlated with each other and the transition from the trivial paramagnetic phase to the retrieval phase is generally lower than $\beta^{*}$ \cite{Zhang:2014gea}.  Thus, choosing values of $\beta$ near $\beta^*$ works well in practice.  Additionally, Sch\"ulke \textit{et al.}\ showed that in many (single-layer) networks there can be multiple zones of the retrieval phase corresponding to detecting communities at different scales~\cite{Schulke:2015hq}.  Therefore, in our experiments, we run the algorithm for a range of $\{\beta^*_q\} = \{ \beta^* (q=2) , \hdots, \beta^*(q=q_\mathrm{max}) \}$, where $q_\mathrm{max}$ is some reasonable upper limit for the number of communities expected in a particular network.  We have found that this approach identifies a reasonable retrieval phase for the networks tried in this paper.  For example in Figure~\ref{fig:beta_scan_senate}, we show how several of the $\{\beta^*_q\}$ consistently lie within the retrieval phase for the US Senate voting network discussed in Section~\ref{sec:realworld_results}.  

We emphasize that, like the original Zhang and Moore approach as well that that by Shi \textit{et al.}, our heuristic assumes a sparse, tree-like network as well as randomly distributed edges and edge weights and provides no guarantees that $\beta^{*}$ will be found within the retrieval phase.  For certain networks, scanning a larger range of $\beta$ will be necessary, though in practice we have found that the approach above is fairly robust.   We note that while it is possible that a fairly small retrieval phase could be missed by such an approach, in our experiments this approach for selecting $\beta^*$ has identified values of $\beta$ for which the algorithm converges close to the known detection limit (see Figure~\ref{fig:multilayer_detection}).  In running the algorithm, we also set an upper limit to the number of message passing iterations allowed, after which we assume that the algorithm has not converged.  We generally select this to be several hundred times the number of iterations at which the algorithm converges to the trivial fixed point ($\psi^i_t=1/q$) for smaller values of $\beta$.  

\subsection{Selection of number of communities, $\boldsymbol{q}$\label{numcom}}

One critical issue with many community detection algorithms is in selecting the appropriate number of communities.  In the context of modularity, adjusting the resolution parameter $\gamma$ can reveal communities of different scale and size, overcoming the ``resolution limit" first raised in \cite{Fortunato:2007js}.  Since then there have been several approaches showing how the scale of the community structures identified varies with the resolution parameter (see, e.g., the discussion and references in \cite{Weir:2017dha}).

Zhang and Moore do not include a resolution parameter in deriving their \textit{modbp} algorithm (thereby implicitly setting $\gamma=1$ in Eq.~\ref{eq:modbp}), instead suggesting an alternative approach for selecting the appropriate number of communities.  They show in several examples that the maximum modularity achieved in the retrieval phase of the algorithm peaks at certain numbers of communities.  They suggest that this peak identifies the correct value for $q$, the number of communities, where there is no additional increase in the retrieval modularity $Q(\{t_i\})$. However, this approach requires running \textit{modbp} for many possible values of $q$, and then choosing an arbitrary threshold when modularity is no longer sufficiently increasing to establish the correct value of $q$.  In many cases, selecting an exact value of $q$ is made difficult because of fluctuations in the retrieval modularity near the $\beta^*$ value derived by Zhang and Moore.  Figure~\ref{fig:chooseqfootball} in the supplement illustrates how choosing $q$ is challenging in practice for these reasons.  Meanwhile, selecting the number of communities in this manner implicitly uses the value $\gamma=1$, which has been shown to return non-ideal partitions in synthetic and real-world networks (see, e.g.,  \cite{Arenas:2008hq,Fortunato:2007js,Newman:2016il,Traag:2013hf}).  We show in Section~\ref{sec:results} the positive impact of using different values for $\gamma$ on several different networks.

There have been two other approaches to selecting the appropriate number of communities using \textit{modbp} without having to run the algorithm at many values of $q$.  Both approaches involve selecting a $q_\mathrm{max}$, the largest possible number of communities, and then using similarities in the marginal probabilities of assignments to evaluate the true number of communities.  Lai \textit{et al.}~\cite{Lai:2016el} noted that in the event that $q$ is too large, many of the marginal community assignments will be highly correlated, and highly correlated states (community assignments) can be condensed into a single group.  Similarly, Ref.~\cite{Schulke:2015hq} condenses the community assignments on the basis of the average distance between the marginals across all nodes in the network.  In practice, we have found that for the default resolution ($\gamma=1$), choosing the number of communities this way all but obliterates the retrieval phase if $q_\mathrm{max}$ is chosen to be too much larger than the actual number of communities. We have implemented the method in Ref.~\cite{Schulke:2015hq}, letting the number of communities float up to a pre-specified $q_{\mathrm{max}}$ (see Section~\ref{sec:beta_star}), and condensing together communities that have closely aligned marginals.  We show that incorporation of a resolution parameter $\gamma$ restores the width of the retrieval phase and returns values closer to the correct number of communities.  As previously mentioned, because we do not specify a single value of $q$, we run the algorithm across a range of $\beta=\{\beta^*(c,q=2),\hdots,\beta^*(c,q=q_\mathrm{max})\}$ where the value of $\beta^*(c,q)$ is obtained using Eq.~\ref{eq:bstar_cond}. We have found that this provides a reasonable range of $\beta$ values to search within and that performance of the algorithm does not depend on the precise value of $\beta$, as long as it is within the retrieval phase.

\subsection{Assessing partition alignment with AMI}

We use the information theoretic measure Adjusted Mutual Information (AMI) \cite{Vinh:2010uc} throughout our analysis to assess the agreement between the predicted partition and either (1) the known underlying ground truth in the case of the generative models tested, or (2) relevant metadata for real-world networks.  Mutual information measures how much entropy or uncertainty is removed from one variable by observation of another. 
The adjusted mutual information measures the overlap between two partitions with a value of $1$ representing perfect agreement and a value of $0$ representing overlap no better than expected under random chance.   We have chosen AMI here as a more conservative measure because it is less biased than normalized mutual information or the Rand Index towards partitions with a larger number of communities \cite{Vinh:2009hs}.  With our multilayer examples, we have applied AMI in two ways to assess different aspects of the alignment of the discovered partitions.  We calculate the AMI across all node-layers in the partitions, each taken as a single array.  We refer to this metric simply as AMI when we use it throughout this paper.  We also use a layer-averaged version of AMI where we compute the AMI of partitions induced within each layer separately, weighting the contribution by the size of the layer:
\begin{equation}
\left<\text{AMI}\right> = \sum_{l=1}^{L} \text{AMI} ( c_l , c^*_l ) \frac{| \mathcal{V}_{l} |}{N} \,
\end{equation}
where $c_l$ and $c^*_l$ are the partitions being compared on the set $\mathcal{V}_{l}$ of node-layers restricted to layer $l$, $|\mathcal{V}_{l}|$ is the cardinality of that set, and $N$ is the total number of node-layers. Layer-averaged $\left<\text{AMI}\right>$ is useful in assessing how well multilayer community detection methods are leveraging information across layers to detect communities within each layer (see discussion in \cite{Bazzi:vn} for advantages of a layer-averaged metric).  

\subsection{Cross-layer community alignment}

When running \textit{multimodbp} at low levels of interlayer coupling ($\omega$) on multilayer networks with  temporal or multiplex coupling topology (\text{e.g.}\ the dynamic stochastic block model described in Section~\ref{sec:dsbm} and the multiplex networks in Section~\ref{sec:mulitlayer_benchmarking}), we frequently observed that the intralayer marginals would rapidly converge to communities that remained misaligned between layers.  Such misalignment would then typically lead to ``fragmented" partitions as shown in Figure~\ref{fig:mlsplit} as well as a lower AMI.   For these partitions, within any single-layer the AMI of the partition with the ground truth with that layer would be very high, but the total AMI over the entire multilayer data would become much lower.  To correct for this issue, we implemented a greedy heuristic to explicitly permute the community assignments within certain layers in order to maximize local alignment between neighboring layers.  Specifically, we identify the layer $x$ that has the greatest number of nodes (of those present in both layers) that change community identity from the previous layer, $y$. We then find the matching of community labels in $x$ that best matches those observed in $y$; that is, we minimize the total number of mismatches across layers $x$ and $y$:
\begin{equation}
\label{eq:comswitchobj}
C(x,y) = \sum_{i \in \mathcal{V}_{x}}\sum_{j \in \mathcal{V}_{y}} \left[ {(c_i\ne c_j) \wedge \mathbb{I}( (i,j) \in \mathcal{E}_{inter}}) \right]\,.
\end{equation}
Once the optimal bipartite matching has been identified \cite{Kuhn:1955hh}, the community labels in layer $x$ and every subsequent layer are rearranged according to that matching (with community labels in subsequent layers that are not present in either layer $x$ or $y$ remaining unchanged). We then repeat this procedure until no further labels are changed (\textit{i.e.}\ the optimal matching is the identity at the layer where the greatest change occurs).  We note that this procedure does not alter the community structure identified within any particular layer, maintaining nodes that have been grouped together.  Rather, this procedure aligns the community labels between layers in a way that always increases the retrieval modularity, thereby improving the computed results.  This method is the same as the \textit{interlayer merging} approach, developed by Bazzi \textit{et al.} to overcome a similar problem encountered when optimizing multilayer modularity with the \textit{GenLouvain} algorithm \cite{Bazzi:2016era}.  It assumes a notion of persistent community across inherently \emph{ordered} layers which is appropriate in the temporal multilayer setting.  However, the approach needs modification in the multiplex case.  In networks that are multiplex, for each layer we permute communities in order to minimize that layers differences with all other layers $\sum_{y\ne x}C(x,y)$, cycling through the layers in random order until no permutations are found.  This procedure as applied to the multiplex case is the same as implemented in \textit{GenLouvain} \cite{Jeub:2016uh}.

\section{Results}
\label{sec:results}
\subsection{Single-layer networks}
We begin by examining how our modifications affect the ability of \textit{modbp} to detect communities within synthetically generated data in the single-layer case.  For single-layer networks, our method is equivalent to Zhang and Moore's apart from two main differences (see also \fullref{methods}).
First, we have included a resolution parameter  $\gamma$ that adjusts the relative balance of the terms in the update equation.  Like other implementations of modularity, this effectively controls the size of the identified partitions.
Second, we have set an upper limit $q_\mathrm{max}$ on the number of communities and incorporated the approach from~\cite{Schulke:2015hq} to select an effective number of communities based on the overlap of the marginals (see Section \ref{numcom}).

\subsubsection{Single-layer stochastic block model}
\begin{figure}[!htb]
\centering
\begin{subfigure}{\textwidth}
	\begin{mdframed}
		\includegraphics[width = .73\textwidth]{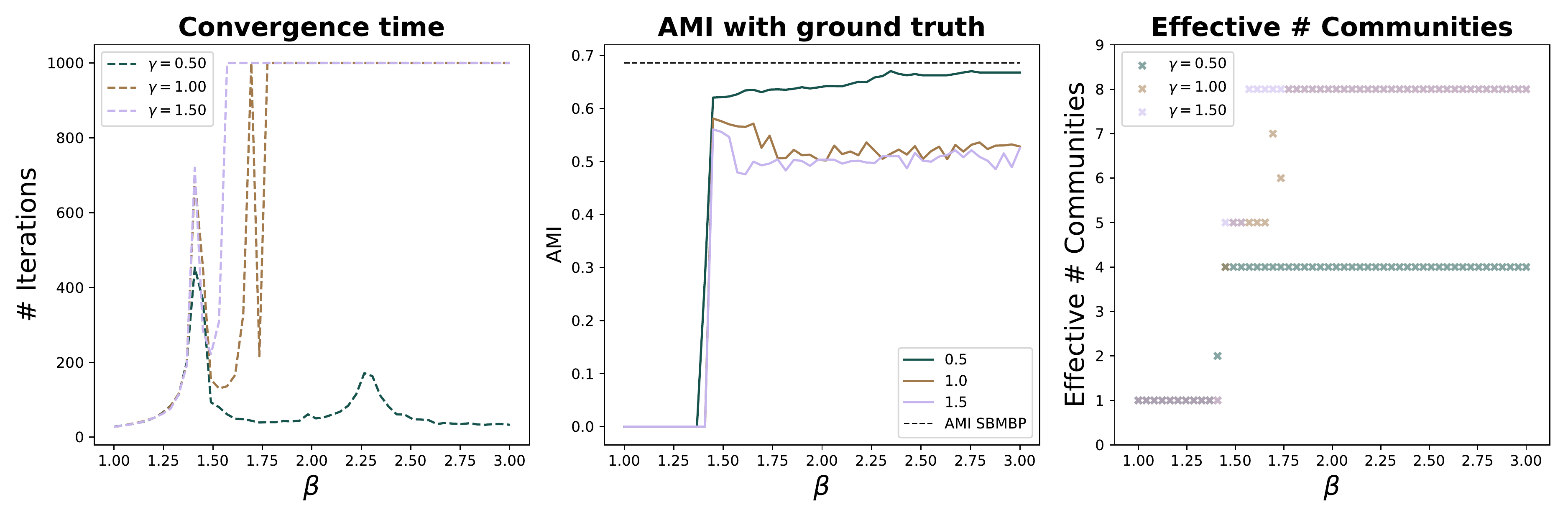}
		\includegraphics[width = .23\textwidth]{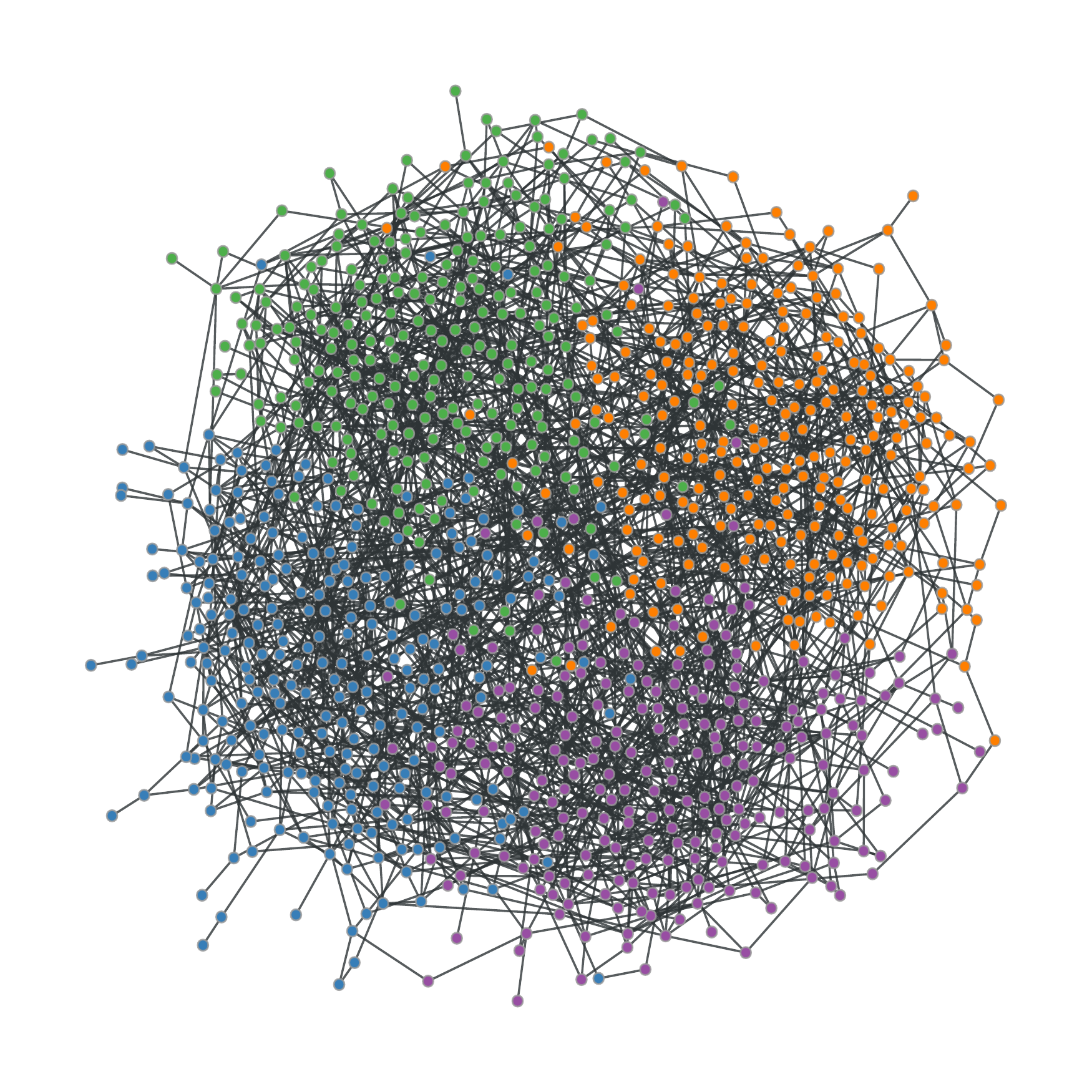}
		\hspace{-.2in}
		\begin{minipage}[l]{0.05\textwidth}
		\caption{\label{fig:igeven}}
		\end{minipage}%
	\end{mdframed}
\end{subfigure}%

\begin{subfigure}{\textwidth}
	\begin{mdframed}
		\includegraphics[width = .73\textwidth]{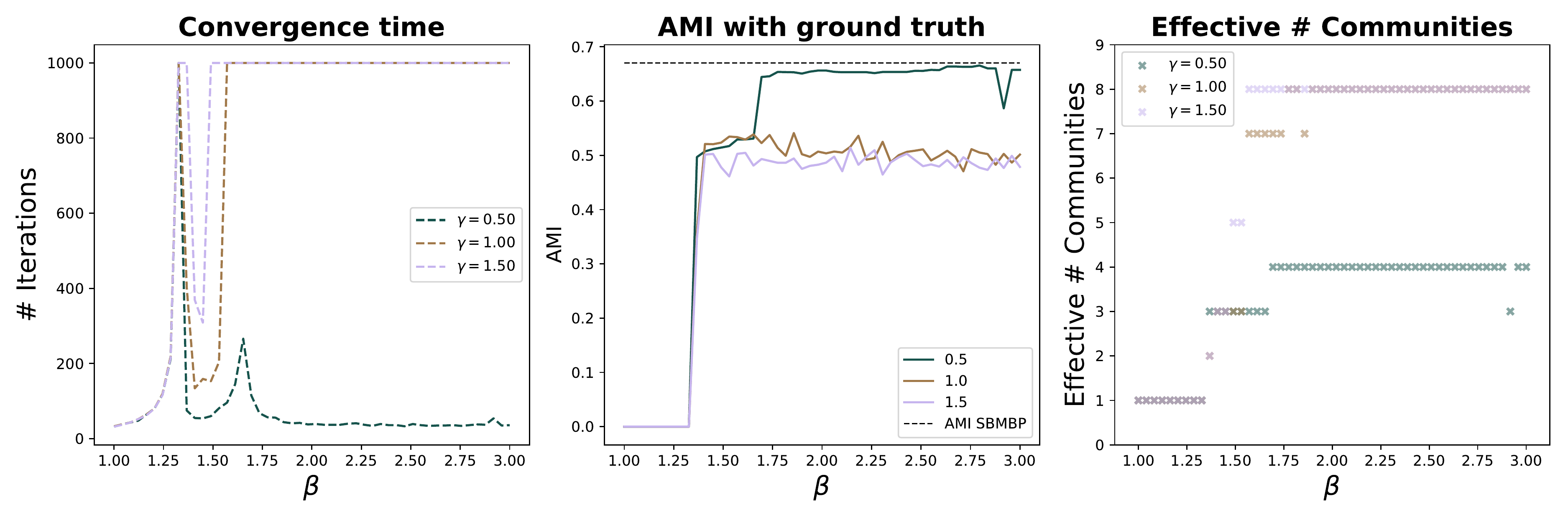}
		\includegraphics[width = .23\textwidth]{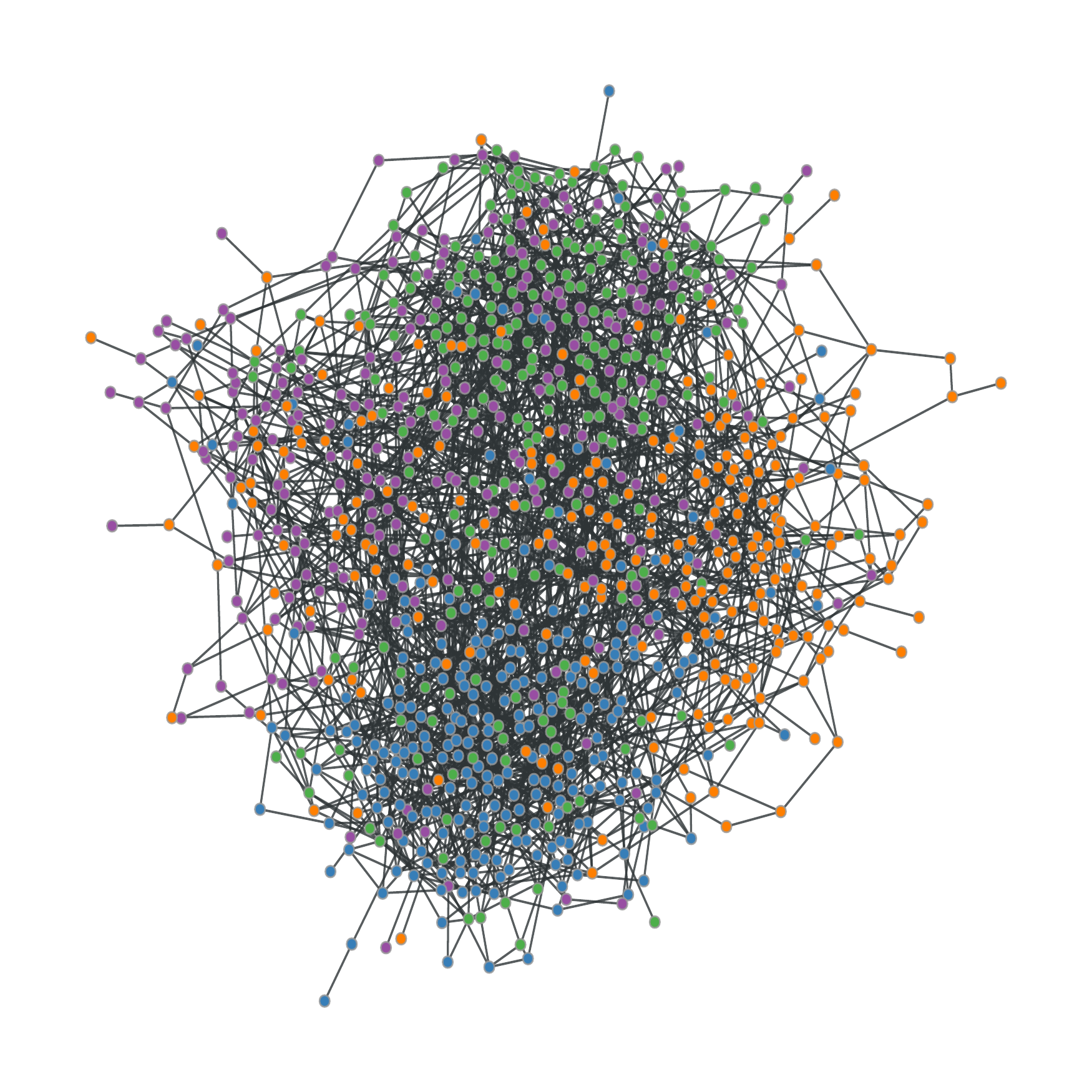}
		\hspace{-.2in}
		\begin{minipage}[l]{0.05\textwidth}
		\caption{\label{fig:iguneven}}
		\end{minipage}\hfill%
		\end{mdframed}
\end{subfigure}
\vspace{-.1in}
 \caption {  Demonstration of \textit{multimodbp} on two realizations of a (non-degree corrected) stochastic block model (SBM).  From left to right, the plots show the retrieval modularity, number of iterations to convergence, and the AMI of the retrieval partition with known community assignments and the effective number of communities.  
  \textbf{(a)} A 4-community planted partition SBM with $n=1000$,  $\epsilon=\frac{p_{\rm out}}{p_{\rm in}}=.1$, mean degree $c_{\rm avg}=4$, and equal sized communities. \textbf{(b)} A 4-community SBM with $n=1000$,  $\epsilon=.1$, $c_{\rm avg}=4$, with uneven community sizes ($[300,200,300,200]$).  For each network we also show the performance of \textit{sbmbp} with parameters for the SBM supplied (middle plot, dotted black line, see Section~\ref{sec:LFRsbmbp} for details of the \textit{sbmbp} method.)  }
 \label{fig:singlelayer}
\vspace{-.2in}
\end{figure}

We examine the behavior of \textit{multimodbp} on different instances of a 4-community stochastic block model (SBM) (using the original, non-degree corrected SBM) for different values of the resolution parameter $\gamma$.  First, we show that in the setting with multiple smaller communities, a lower value of $\gamma$ produces a much wider retrieval phase and thus makes detection of communities more robust to selection of $\beta$.  To investigate this robustness, we generated a single realization of each SBM and scanned a range of $\beta$ values to characterize the behavior of the algorithm seen in {Figure~\ref{fig:singlelayer}}. For an SBM network with four even-sized communities, {Figure~\ref{fig:igeven}} shows that the retrieval phase for both $\gamma=1.0$ and $\gamma=1.5$ are very narrow (leftmost panel) with a small corresponding peak in the AMI of detected communities (middle panel).  In contrast, for $\gamma=0.5$ the retrieval phase widens out with a broader and higher set of AMI values for the detected communities.  Furthermore, the number of communities identified for $\gamma=0.5$ plateaus at the correct number, 4, as shown in the far right panel of \textbf{Figure~\ref{fig:igeven}}.

We also tested the performance of the algorithm in the case where the sizes of the planted communities were uneven, shown in \textbf{Figure~\ref{fig:iguneven}}.  The relative performance for varying $\gamma$ is even more disparate in this case.  There is a small retrieval phase for $\gamma=1$, but it is much smaller than that for $\gamma=0.5$ and the AMI is again consistently lower.  For $\gamma=0.5$ we actually detect two retrieval phases. In the first retrieval phase (approximately $\beta \in [1.4,1.7]$), only nodes within the two larger communities are labeled correctly.  Then, as $\beta$ increases ($\beta \in [1.7,3.0]$), the smaller two communities also become identifiable.  This is consistent with the multiphase behavior observed in \cite{Schulke:2015hq}, though we note that in their example, the phase transition is observed for the default value of $\gamma=1$.  In both of these examples the AMI of the identified partition by \textit{multimodbp} is close to the result achieved by a belief propagation implementation of the SBM model, which has been shown to achieve the optimal bounds for this model \cite{Decelle:2011ek,Decelle:2011ik}.

In both of these experiments the value of $\beta^*$ marking the transition from the paramagnetic phase to either the retrieval phase or the spin-glass phase appears to be independent of the value chosen for the resolution parameter, $\gamma$.  However, the width of the retrieval phase is dependent on the particular value of the resolution parameter $\gamma$ (see upper left panel in Figure~\ref{fig:igeven}).  Thus the detection of significant communities in this case relies on the appropriate selection of the value of $\gamma$.

\subsubsection{Comparison of \textit{multimodbp} with SBMBP on LFR benchmark networks} \label{sec:LFRsbmbp}

We compare the performance of our algorithm \textit{multimodbp}, with a belief propagation approach to fit the Stochastic Block Model (SBM) developed and implemented in Ref.~\cite{Decelle:2011ek}, which we refer to as \textit{sbmbp}.
This Expectation-Maximization (EM) implementation of \textit{sbmbp} alternates between iteratively updating the marginals using belief propagation with fixed SBM parameters, and updating the SBM parameters using likelihood maximization for the fixed marginals.  Their implementation requires setting a fixed $q$ however, so for testing we ran \textit{sbmbp} across a range of $q$ values ($q \in \{ 2,3,...,8\}$) and selected the partition with the lowest free energy.

\begin{figure}[!htbp]

\begin{center}
	\begin{mdframed}
		\includegraphics[width =
\textwidth]{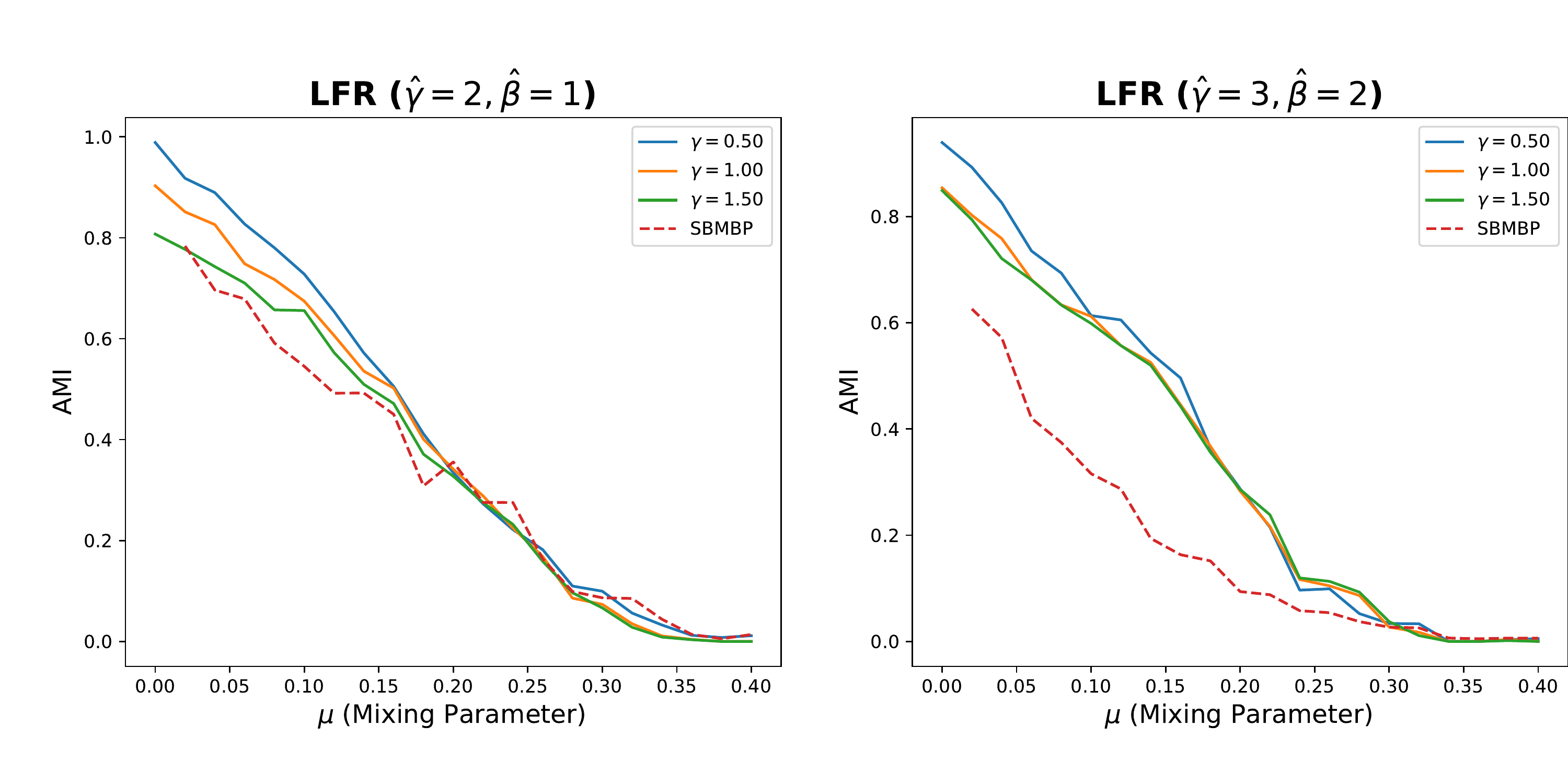}
	\vspace{-.3in}
		\caption{Performance of \textit{multimodbp} and \textit{sbmbp} over many LFR benchmark realizations with a range of values for the mixing parameter $\mu$.  Each point represents an average over 100 realizations of LFR with 1000 nodes, an average degree of 3 (with a max of 10), and other parameters set to default values.
	\label{fig:lfr_bench}}
	\vspace{-.1in}
	\end{mdframed}
\end{center}
\vspace{-.1in}
\end{figure}

Our test data set is the Lancichinetti-Fortunato-Radicchi (LFR) benchmark generator \cite{Lancichinetti:2008ge}, an algorithm developed to generate networks with more diverse community structures.
We tested our \textit{multimodbp} with several values of the resolution parameter $\gamma$ against \textit{sbmbp} across a range of parameters of the LFR model.  We vary the LFR mixing parameter $\mu$, which sets the detectability of the underlying communities.  The LFR algorithm also has a parameter $\hat{\gamma}$ to set the exponent of the power law for the degree distribution and a parameter $\hat{\beta}$ to set the exponent of the community size distribution.  We tested both algorithms for two sets of $(\hat{\gamma},\hat{\beta})$ in {Figure~\ref{fig:lfr_bench}}.

{Figure~\ref{fig:lfr_bench}} shows that the modularity based approach outperforms the stochastic block model across a range of $\mu$, the mixing parameter, all the way down to the detectability limit.  The flexibility of the modularity approach allows for better identification of communities with real-world degree distribution (since the classic SBM assumes homogenous degree distribution within a community). 
The comparison was done using \textit{sbmbp}'s EM approach which is not well suited to determine the number of communities.  In contrast, using our approach as described in Section~\ref{numcom}, the \textit{multimodbp} algorithm was able to identify the correct number of communities and get more accurate community assignments using a resolution parameter value of $\gamma=0.5$ (though other values of $\gamma$ also performed well).

\subsubsection{NCAA Division I-A College Football network}

We now demonstrate that inclusion of the resolution parameter $\gamma$ in the modularity objective function can significantly improve performance on real-world networks.  As an example of a real-world network with stable community structure we selected the 2000-2001 NCAA Division I-A College football network, which has 115 nodes representing teams (schools) and 613 unweighted edges connecting teams that played at least one game~\cite{Evans:2010ga,Girvan:2002ez}.  Our previous work suggests that modularity optimization produces the best community partition in a range $\gamma \in [1.4,4]$~\cite{Weir:2017dha,WeirCHAMP}.

\begin{figure}[!htbp]
\begin{center}
	\begin{mdframed}
		\includegraphics[width = .66\textwidth]{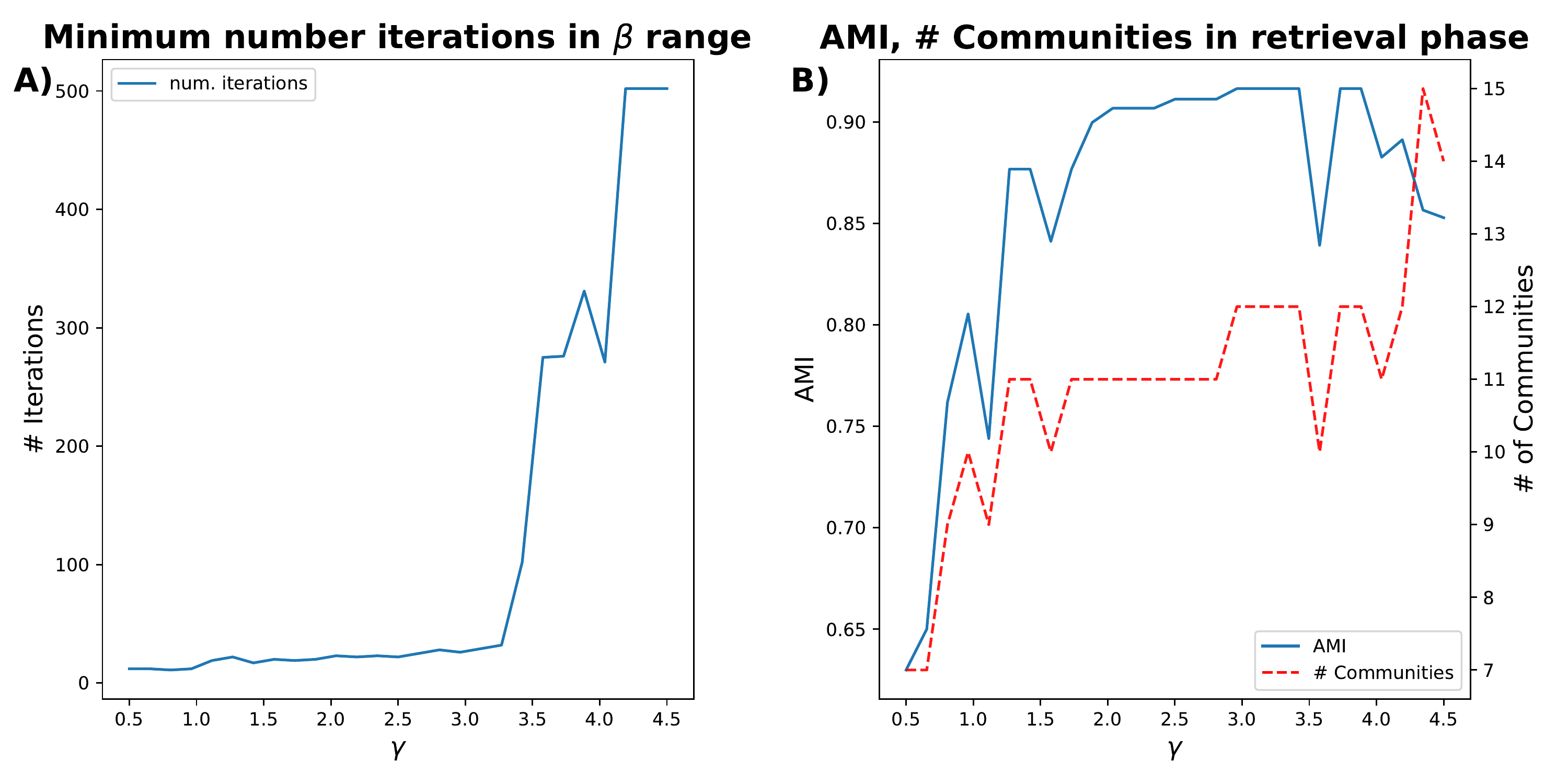}
		\includegraphics[width = .33\textwidth]{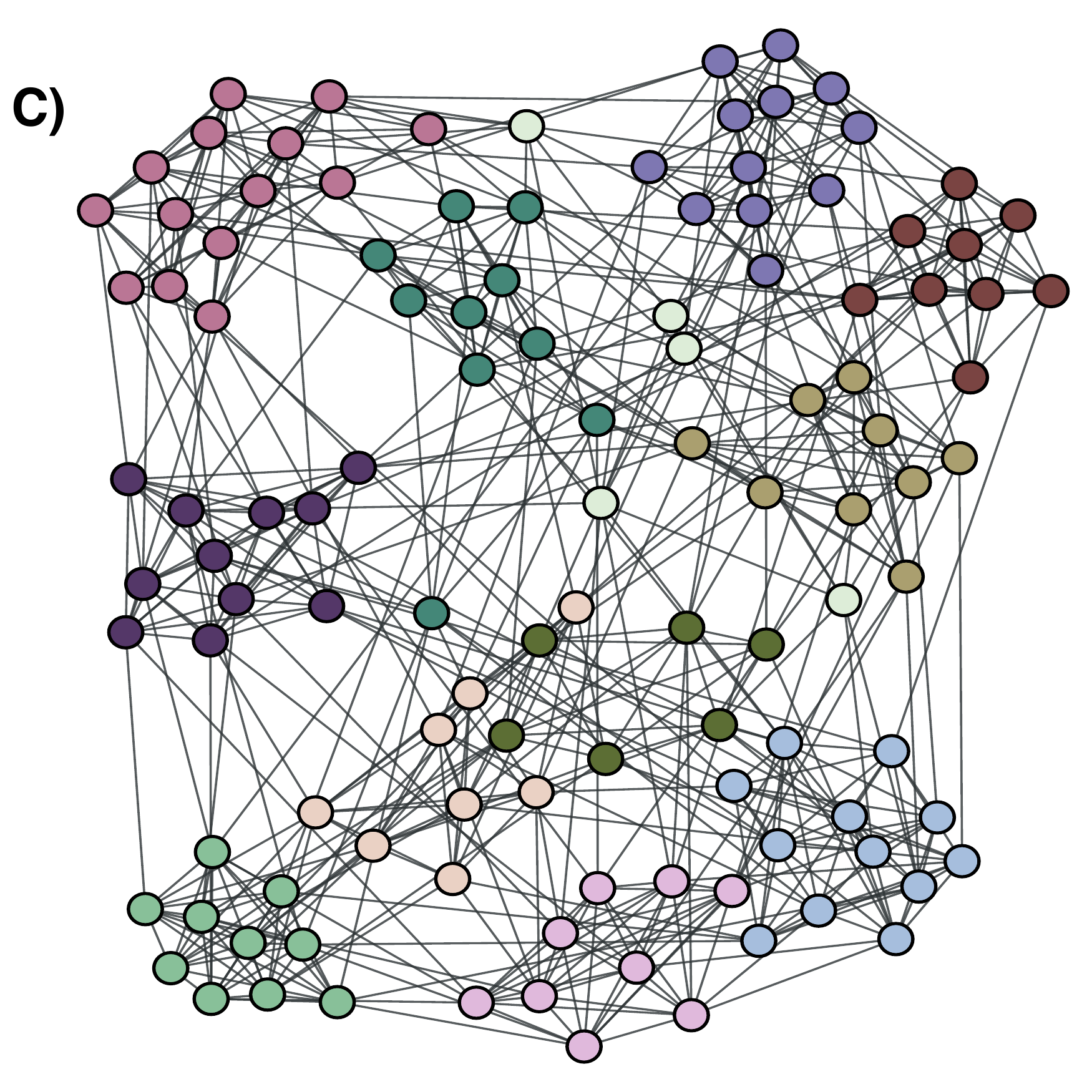}
\vspace{-.3in}
\caption{ Testing \textit{multimodbp} on the 2000-2001 Division I-A College Football network \cite{Evans:2010ga,Girvan:2002ez}.  \textbf{A)}~The average number of iterations until convergence in the retrieval phase across a range of $\gamma$ values. \textbf{B)}~The average number of communities detected in the retrieval phase as $\gamma$ increases and the corresponding adjusted mutual information (AMI) of those partitions.  \textbf{C)}~ForceAtlas2 \cite{Jacomy:2014iw} layout of the football network with each node colored according to a partition identified using $\gamma=3.0$, demonstrating excellent alignment to the conference structure. \label{fig:footballgamma}}
\vspace{-.1in}
\end{mdframed}
\end{center}
\end{figure}

\begin{figure}[htbp]
\begin{center}
\begin{mdframed}
\includegraphics[width=\linewidth]{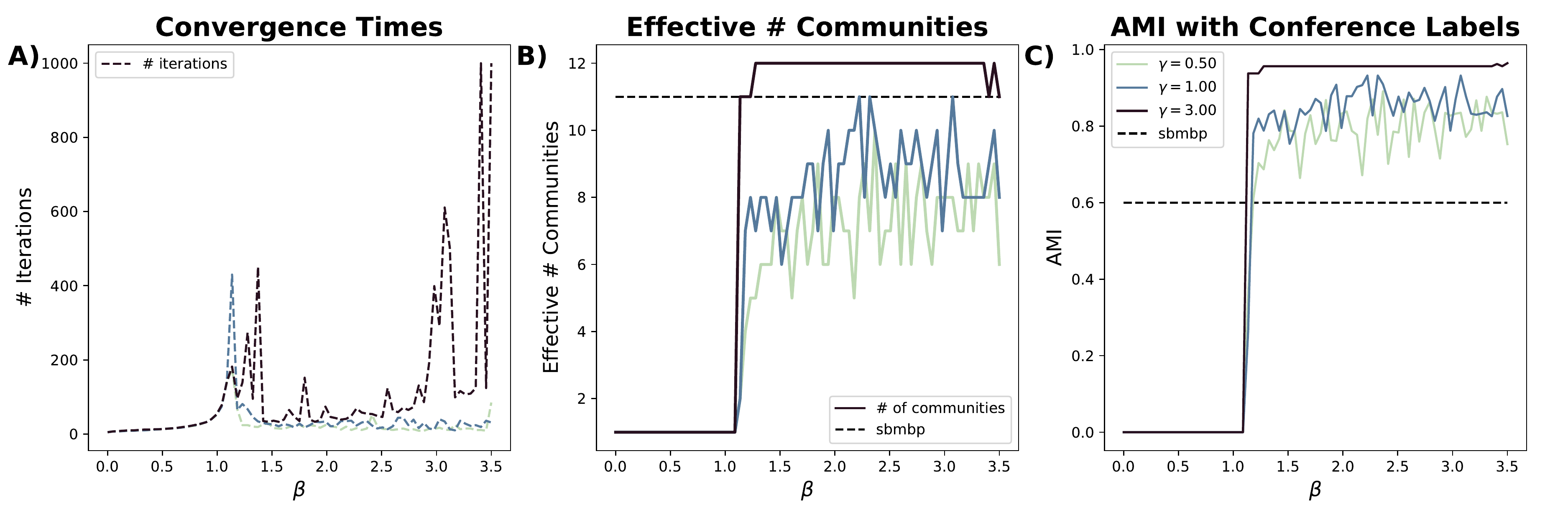}
\vspace{-.3in}
\caption{ \label{fig:col_scan} We show the performance characteristics of the algorithm for 3 different values of $\gamma$ on the 2000-2001 NCAA Division I-A College football network   \textbf{A)} Although all three values of $\gamma$ produce a wide retrieval phase, the communities identified within each retrieval phase are different.  \textbf{B)} The number of non-redundant communities is higher as $\gamma$ increases with $\gamma=3$ producing the number of communities that lines up well with the ground truth (the conferences) in this example, with \textbf{C)} showing corresponding higher values of AMI for $\gamma=3$.  Horizontal black dashed line shows that \textit{sbmbp} identifies correct number of communities in \textbf{B)} but has less agreement with the known conference structure of the network. }
\label{fig:scanq}
\vspace{-.2in}
\end{mdframed}
\end{center}
\vspace{-.1in}
\end{figure}

To investigate how the value of $\gamma$ affects the retrieval phase, we ran \textit{multimodbp} for a range of values of the parameter $\gamma$ and examined the minimum number of iterations for which non-trivial structure was identified, shown in {Figure~\ref{fig:footballgamma}}.  For each value of $\gamma$, \textit{multimodbp} was run over 30 evenly-spaced values of $\beta \in [0.5,4.5]$.  For each value of $\gamma$ we show the minimum number of iterations over all values of $\beta$ for which non-trivial structure was identified and the AMI of the partition of the corresponding partition (the partition identified with the minimum number of iterations).  Runs that did not converge after 500 iterations suggest that for that value of $\gamma$ the retrieval phase was either very small or nonexistent.  It is possible that a retrieval phase exists outside the chosen range for $\beta$, though we verified for a few arbitrary values of $\gamma$ that the algorithm did not have a retrieval phase.  Furthermore, {Figure~\ref{fig:footballgamma}} demonstrates that the AMI of the retrieval partition increases as a function of $\gamma$  from $\gamma=1$ up until it plateaus from $\gamma=[1.7,3.4]$ at a stable 11/12 community partition\footnote{There are 11 conferences, with a few schools also labeled as independent.} (shown in the far right panel).  In {Figure~\ref{fig:col_scan}}, we show the algorithm convergence properties as well as performance for a few values of $\gamma$ on this network.  We also compare the performance of the \textit{multimodbp} algorithm with the \textit{sbmbp} approach, showing that even when the SBM approach identifies the correct number of communities (middle panel dashed line), \textit{multimodbp} still achieves more accurate identification of the underlying community structure (right panel).  

\FloatBarrier
\subsection{Multilayer modularity belief propagation}
\label{sec:multilayer_results}
\subsubsection{Dynamic stochastic block model}
\label{sec:dsbm}
We test the multilayer functionality of \textit{multimodbp} by application to a multilayer SBM called the dynamic stochastic block model (DSBM) as described in \cite{Ghasemian:2016hg}.  The DSBM represents a temporal multilayer network where each node in the network is represented by a single node-layer within each layer.  The correspondence between identified node-layers is represented by a single interlayer edge between adjacent layers.  In the DSBM, each layer is drawn from a regular stochastic block model with $q_\mathrm{true}$ communities and edge probabilities described by probabilities $p_{\rm in}$ within communities and $p_{\rm out}$ between communities.  Each node-layer's community assignment has a fixed probability $\eta$ of remaining the same between subsequent layers (and $1-\eta$ probability of choosing a new community).  Conditioned on the node community assignments, each layer's edges are independent of all other layers.  For a fixed average degree $c$, the strength of community structure within each layer is given by the parameter $\epsilon = { p_{\rm out}}/{p_{\rm in}}$.  

\begin{figure}
\includegraphics[width=\textwidth]{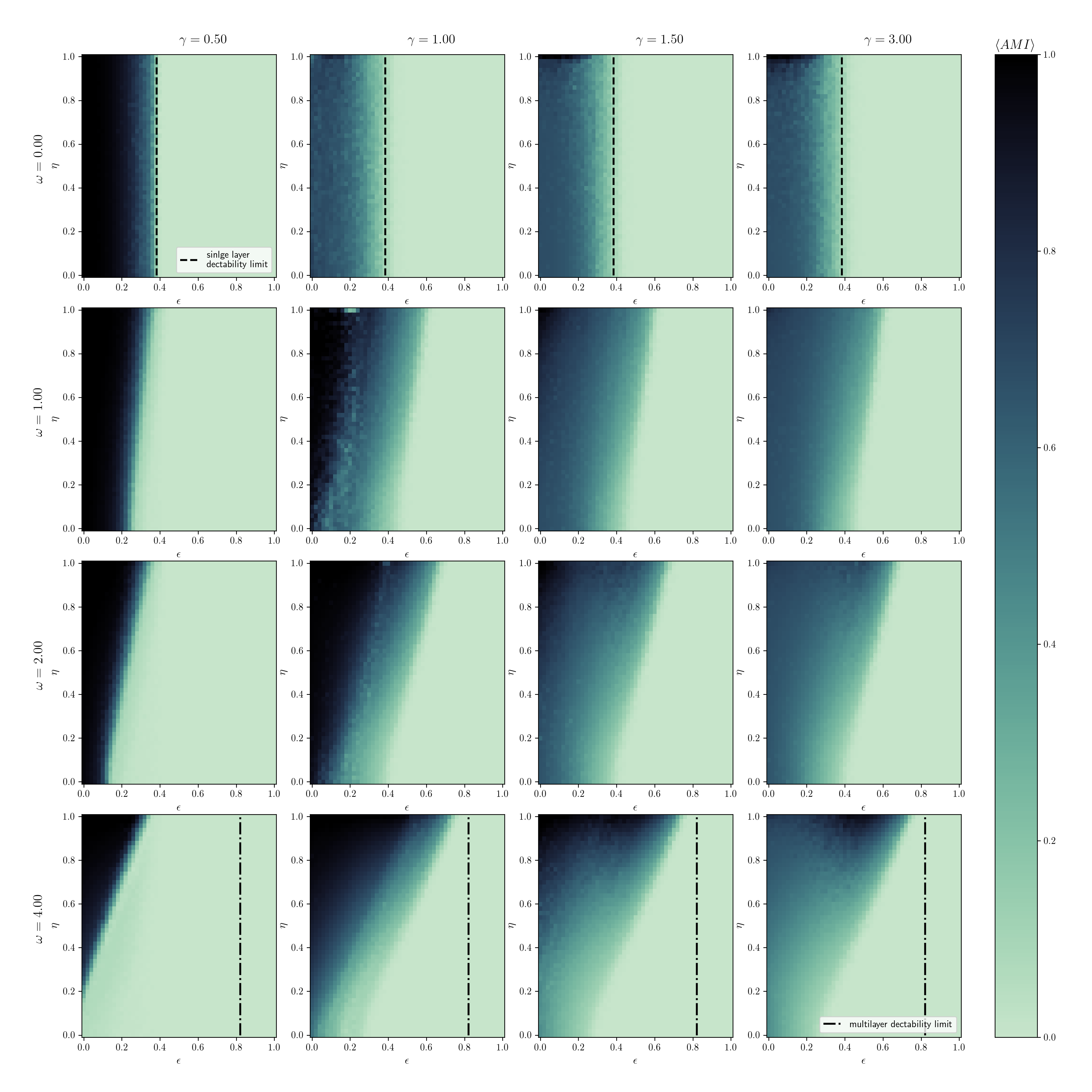}
\caption{ Accuracy of \textit{multimodbp} on a multilayer SBM  (assessed by $\left<\text{AMI}\right>$ averaged over 50 independently sampled networks at each point) for different values of model parameters $\epsilon$, and $\eta$ (horizontal and vertical axes respectively) and parameters $\gamma$ and $\omega$ (moving horizontally and vertically across panels). The networks generated here are with $N=250$ nodes, $L=20$ layers, average degree $c=10$, and $q_\mathrm{true}=2$ communities.}
\label{fig:multilayer_detection}
\vspace{-.4in}
\end{figure}

In Figure~\ref{fig:multilayer_detection} we show the $\left<\text{AMI}\right>$ score of the \textit{multimodbp} algorithm averaged over many realizations of the dynamic stochastic block model for a range of parameters. We consider DSBM networks created using values of $\epsilon$ and $\eta$ ranging from 0 to 1. For each choice of $\epsilon$ and $\eta$ we created 50 networks and computed the $\left<\text{AMI}\right>$ between partitions identified using \textit{multimodbp} and the ground truth.  Because the value of $q$ is usually not known beforehand, for each $(\gamma,\omega)$ point we scan a range of possible values of $\beta^*$ corresponding respectively to possible values of $q$ as given by Eq.~\ref{eq:bstar_cond} with $q_\mathrm{max}=4$ set to twice the true number of communities ($q_\mathrm{true}=2$).  For each trial, we select the partition with the highest retrieval modularity among all that converged.

We apply the \textit{multimodbp} algorithm in this analysis with several choices of the resolution parameter, $\gamma$ (columns in Figure~\ref{fig:multilayer_detection}) and coupling parameter, $\omega$ (rows of Figure~\ref{fig:multilayer_detection}). Figure~\ref{fig:multilayer_detection} shows that incorporation of a resolution parameter makes a large difference for detectability of community structure based on the DSBM parameters used to generated the network.  For lower values of $\epsilon$ (i.e., increased intralayer community signal) with frequent community switching (decreased $\eta$), $\gamma=0.5$ clearly outperforms the higher values of $\gamma$.   However, for $\gamma=0.5$ the algorithm does not appear to utilize information across the layers and performance drops off as $\omega$ is increased.  For a lower $\gamma$ with higher $\omega$, the algorithm would converge to a single, large community across all layers at values of $\epsilon$ that would ordinarily be detectable.    For $(\gamma=0.5,\omega=0)$, \textit{multimodbp} performs quite well all the way down to the limit of detection for single layer networks, given by the condition $ N(p_{\rm in}-p_{\rm out}) > q\sqrt{c}$ in \cite{Decelle:2011ik} and \cite{Nadakuditi:2012kx} (depicted by the vertical dashed line at $\epsilon=.38$ in upper row of Figure~\ref{fig:multilayer_detection}).

For higher values of $\gamma$ we appear to get better aggregation of information across the layers with increasing $\omega$.  At $\gamma=1.0$ and $\omega=4.0$ the range of detection is increased to a maximum of $\epsilon=.75$  for $\eta=1.0$ (no community switching).  This behavior is consistent with the limits of detectability that are achieved through aggregation of layers as discussed in \cite{Taylor:2016kj}.  They derive a modified limit of detectability in the case where each layer is drawn from the same 2-block SBM with the community labels fixed throughout the layers, unlike our model where each nodes' community assignment is allowed to vary. 
They compute a detectability threshold of $N L ( p_{\rm in}-p_{\rm out}) = \sqrt{4NL\rho(1-\rho)}$ where $\rho=\frac{1}{2}(p_{\rm in}+p_{\rm out})$.  For parameters used in this experiment the theoretical detectability limit is $\epsilon\approx.82$ (shown by the dashed lines in Figure~\ref{fig:multilayer_detection}).  These results demonstrate how the additional flexibility provided by tuning $\gamma$ and $\omega$ allows for achieving near optimal performance depending on the parameters of the underlying model.

\subsubsection{Comparison with \textit{GenLouvain} on Heterogeneous Benchmarks}
\label{sec:mulitlayer_benchmarking}

To assess the performance of \textit{multimodbp} on more realistic synthetic data, as well as on different multilayer topologies, we have applied it to the generative models described in \cite{Bazzi:vn} and implemented in MATLAB \cite{jeubmatlab} and Python \cite{jeubpython}.  In \cite{Bazzi:vn}, Bazzi \textit{et al.}\ present a multilayer generative model that allows for the coupling of mesoscale structures across a variety of interlayer topologies.  In their approach, a multilayer partition is sampled from a distribution defined by a given null model as well as the specified interlayer dependencies.  For the multilayer networks shown here, community assignments are drawn from a Dirichlet distribution in an arbitrary starting layer, and then either copied or resampled based on the interlayer coupling probability ($p$) to the other layers. The intralayer edges are then drawn independently for each layer conditioned on the assigned communities: specifically, the edges within each layer are sampled according to a degree corrected stochastic block model (DCSBM) conditioned only on the community assignments within that layer.\footnote{Other models could be used for sampling the intralayer connections.  The network generation process described by Bazzi \textit{et al.}\ is modular in nature, allowing for a large combination of inter- and intralayer structures.  We have chosen the DCSBM for direct comparison with the results in \cite{Bazzi:vn}.}  That is the edges within each layer are independent of each other given the community assignments.  Interlayer dependencies are introduced only through the probability that a given node keeps the same community assignment from layer to layer.  Within each layer, the strength of community structure is given by the mixing parameter, $\mu\in [0,1]$.  If $\mu=0$, communities are perfectly separated (no edges between) while if $\mu=1$ edges are placed without regard to the communities.  We specify the interlayer coupling topology and parameters for each experiment below and the parameters for sampling the intralayer edges from the DCSBM.  For each experiment in this section, we have used the same parameters detailed in Section~\rom{5}.A of \cite{Bazzi:vn} for the corresponding multilayer topologies.  

We have compared \textit{multimodbp} to \textit{GenLouvain} \cite{Jeub:2016uh} across a range of values for the interlayer coupling parameter, $\omega$ (keeping $\gamma=1.0$).  \textit{GenLouvain} is a fast, stochastic, greedy heuristic to optimize multilayer modularity (Eq.~\ref{eq:multimod}) and is one of the most widely used tools for multilayer community detection.  We have used the iterated version of \textit{GenLouvain} where solutions from a run are used to initialize the next run, until no improvements in modularity are obtained.  We also used the random move setting which allows the algorithm opportunities to break out of local optima.   We have chosen not to include here another commonly used multilayer detection tool, \emph{Infomap}, as its performance metrics on these benchmark tests were typically below those of \textit{GenLouvain} (see results in~\cite{Bazzi:vn}).  

Within the multiplex experiments detailed below, we found that as the number of layers became larger, the all-to-all connections between node-layers representing a single node quickly became stuck in a local minimum where node-layers with interlayer connections were all strongly forced into the same community, washing out the weak community information from the intralayer neighbors.  To counteract this, we used a spectral clustering approach to initialize the beliefs as was suggested in \cite{Zhang:2012dv}.  We compute the top $q_{max} -1$ eigenvectors of $\textbf{B} = \textbf{A}-\gamma\textbf{P} + \omega\textbf{C}$, the supra-modularity matrix, and then use K-means to find $q_{max}$ different clusters.  This does not add significant additional runtime as computing the leading eigenvectors of sparse matrices can be done efficiently.  All incoming beliefs to a given node-layer are then initialized to a soft version of the identified spectral partition where the belief representing the node-layer's associated community is set to be some factor (5 in this case) times larger than the other beliefs.  We found in our experiments that starting \textit{multimodbp} with even relatively weak alignment provided by the spectral partitioning greatly improved the results for the higher values of $\omega$.  We show in supplement Figure~\ref{fig:multiplex_noinit} that \textit{multimodbp} improves on the baseline provided by the spectral initialization (even when the spectral clustering by itself performs quite poorly).  Although the improvements were most notable for the two multiplex experiments, for consistency, we have employed spectral initialization in all three of the multilayer layer models used below.\footnote{We found that the discrepancies between \textit{multimodbp} and \textit{GenLouvain} were largest for higher values of $\omega$, which is where the spectral initialization adds the most benefit.  See supplement Figure~\ref{fig:multiplex_noinit} for performance of \textit{multimodbp} without spectral initialization.}  

For each combination of $\mu$, $p$, and $\omega$ in the experiments below, we run each approach once on each of 100 independently sampled networks from the benchmark model.  Our results below for \textit{GenLouvain} closely mirror the findings in \cite{Bazzi:vn}.     
 
\begin{figure}[htbp]
\begin{center}
\includegraphics[width=1.0\linewidth]{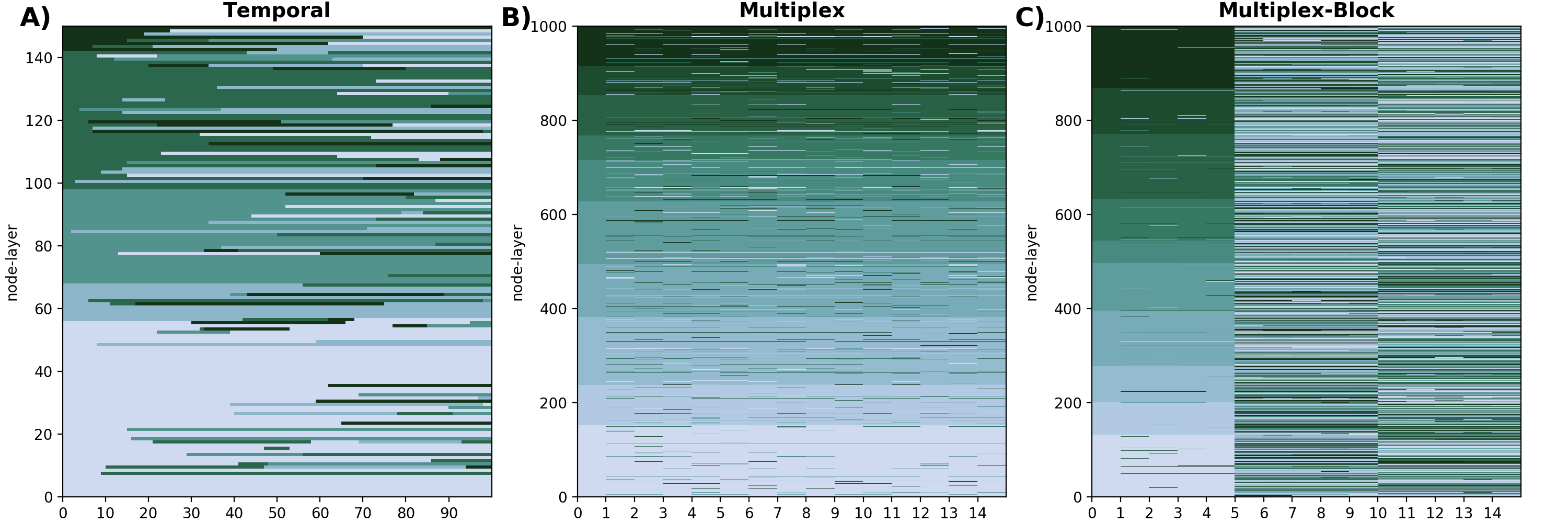}
\caption{\label{fig:multi_struct} Graphical representation of the community structures for networks sampled from different interlayer topologies available with the multilayer-generative model in \cite{Bazzi:vn}.  In each subfigure, each row represents a particular node, with each column representing a layer of the network.  Each node-layer is colored according to its multilayer partition.  Thus we can see how the different communities persist across the layers of the network.}
\end{center}
\end{figure}

\paragraph{Temporal Multilayer Network} ~This network has a similar interlayer topology as the DSBM detailed in Section~\ref{sec:dsbm} with an ordering on the layers and each node-layer connected to only the node-layers in the layers adjacent to it.  That is,
\begin{align}
C^{temporal}_{ij} =
\begin{cases}
1 & \text{ if } l_j = l_i\pm 1 \text{ AND } i\cong j  \\
0 & \text{otherwise}
\end{cases} \ .
\end{align}

\begin{figure}[htbp]
\begin{center}
\includegraphics[width=\linewidth]{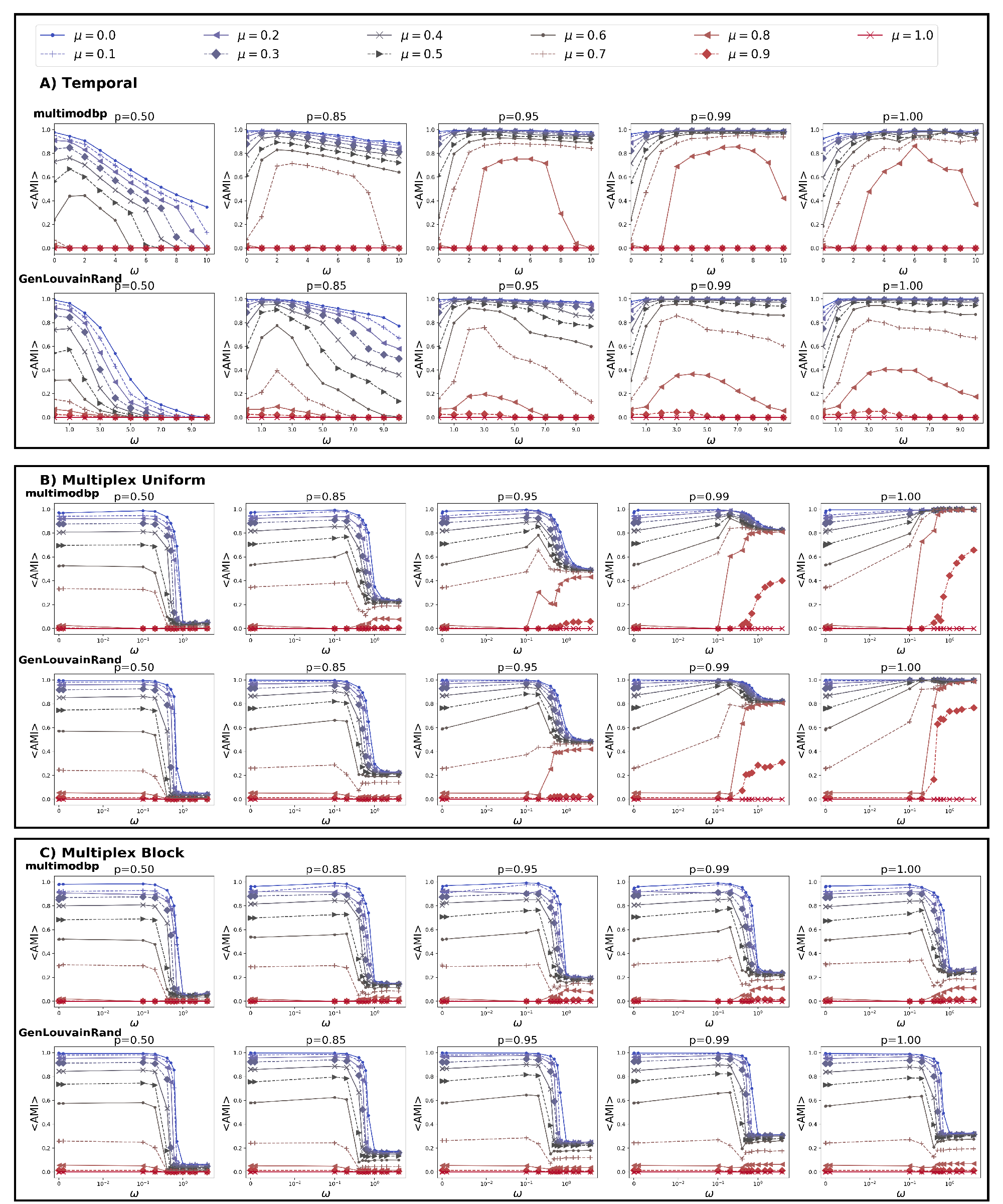}
\caption{\label{fig:multilayer_benchmarks} (\textit{Caption on next page.}) }
\end{center}
\end{figure}

\noindent Similar to the DSBM, each node has a probability $p$ of copying its identified node-layer's community in the preceding layer.  The major difference with this experiment is that community assignments are drawn from a more realistic Dirichlet null distribution with $\theta=1$ ($\theta$ is the concentration parameter of the Dirichlet distribution) and $q=5$.\footnote{We use $q$ to denote the number of communities drawn from the null distribution (referred to as $n_{set}$ in \cite{Bazzi:vn}).  See Appendix A of Ref~\cite{Bazzi:vn} for a more detailed description of the Dirichlet null distribution for the multilayer partition.}  The intralayer connections are drawn from an SBM with degree correction (DCSBM) with $\eta_k=-2$, $k_{max}=30$, and $k_{min}=3$.  Each sampled network has 150 node-layers in each layer with 100 layers for a total of 15000 node-layers.  Visualization of an example temporal network\footnote{In Ref~\cite{Bazzi:vn} this interlayer topology is referred to as \textit{uniform} temporal because there are no temporal breaks.  One could also consider \textit{block} temporal networks analogous to the block multiplex model considered below, with or without corresponding different weights of the $C^{temporal}_{ij}$ interlayer coupling (which \textit{multimodbp} would then use directly).  We demonstrate the performance of \textit{multimodbp} on the temporal block model in supplement Figure~\ref{fig:temp_block}.} is shown in Figure~\ref{fig:multi_struct}.A.  We have run \textit{multimodbp} across a range of $p$ and $\mu$ and compared how increasing the interlayer coupling parameter $\omega$ affects the performance of the algorithm.  In Figure~\ref{fig:multilayer_benchmarks}.A (the top two rows) we see that \textit{multimodbp} with the spectral initialization tends to outperform \textit{GenLouvain} for a wide variety of model parameters.  We see that the peak AMI obtained for $\mu=.8$ is higher for \textit{multimodbp} across most values of $p$, in some cases notably so ($\left<\text{AMI}\right>  \approx .8 $ vs $\left<\text{AMI}\right>  \approx .4$ at $p=.99$).  That is, it appears in these cases that \textit{multimodbp} is better able to utilize the information across the adjacent layers to inform community prediction.    

\addtocounter{figure}{-1}
\begin{figure} [t!]
  \caption{(Previous page.) We compare the performance of \textit{multimodbp} (top rows of each panel) and \textit{GenLouvain}  \cite{Jeub:2016uh} (bottom row of each panel) across a range of multilayer benchmark networks developed by Bazzi \textit{et al.} \cite{Bazzi:vn}.  For each model, we vary both $\mu$, the intralayer mixing parameter (strength of communities) denoted by the different markers and colors, as well as $p$, the persistence of communities across layers going from left to right, across the subfigures.  Each point represents the $\left<\text{AMI}\right>$ averaged over 100 independent realizations of the model.  \textbf{A)} Temporal network topology with ordered layers and interlayer connections only present between adjacent layers.  Multilayer community partitions are drawn from Dirichlet distribution with $\theta=1$ and $q=5$.  Intralayer edges are samples from a DCSBM with $\eta_k=-2$, $k_{max}=30$, $k_{min}=3$.  Each network has 100 node-layers in each layer with 150 layers for a total of 15000 node-layers. \textbf{B)} Uniform multiplex multilayer network with unordered layers and all to all interlayer connections among identified node-layers across all layers.  Multilayer partitions are sampled from Dirichlet distribution with $\theta=1$ and $q=10$. Intralayer connections are drawn from DCSBM with $\eta_k=-2$, $k_{min}=3$, and $k_{max}=150$.  Each network has 1000 node-layers in each layer with 15 layers for a total of 15000 node-layers.  \textbf{C)} Block multiplex model with the same parameters as the uniform multiplex models but with a discontinuity between each block of 5 layers where community labels are completely independent. }
\end{figure}

\paragraph{Uniform Multiplex and Block Multiplex}  
\label{sec:mulitlayer_benchmarking:multiplex}
We also sampled graphs from two different multiplex interlayer topologies.  Unlike in the temporal multilayer networks, in the multiplex topology, there is no inherent ordering to the layers. Each node-layer is connected with interlayer edges to all other node-layers in the identified set:
\begin{align}
C^{multiplex}_{ij} =
\begin{cases}
1 & \text{ if } i\cong j  \\
0 & \text{otherwise}
\end{cases} \ .
\end{align}

\noindent In the uniform multiplex coupling, a node-layer's community assignment is copied with a given probability $p$ to all of its identified node-layers.  A visualization of an example network with this structure is shown in Figure~\ref{fig:multi_struct}.B.  In contrast, for the block multiplex, we divide the layers into a specified number of blocks, and only copy the layer assignments with probability $p$ for layers within a given block.  Within each block the structure is the same as the uniform multiplex, however there is a complete discontinuity in node-layer community assignments from block to block.  Note that while the interlayer coupling probabilities are set to $0$ between blocks in the model, the interlayer edges between blocks are still present in the network.  Figure~\ref{fig:multilayer_benchmarks}.C shows an example of a multiplex block network.  For both of these examples we use a Dirichlet null model with $\theta=1$ and $q=10$ and generate the intralayer edges from the DCSBM with parameters $\eta_k=-2$, $k_{min}=3$, and $k_{max}=150$. 

We compare \textit{multimodbp} with \textit{GenLouvain} on the uniform multiplex benchmark networks in Figure~\ref{fig:multilayer_benchmarks}.B (middle two rows).  We find that in most of the parameter regimes, performance is relatively comparable between the two methods, with \textit{GenLouvain} having a slight edge overall, especially for higher values of $\mu$ and lower values of $\omega$.  However, for some parameters on the block multiplex networks in Figure~\ref{fig:multilayer_benchmarks}.C, \textit{multimodbp} outperforms GenLouvain, especially at lower values of $\mu$ (see $\mu=.8$, $p=.95$).  Overall, we see that \textit{multimodbp} is able to utilize information across layers to detect community structure where it would be undetectable if each layer was considered independently.  These benchmarks demonstrate that \textit{multimodbp} performs comparably with and in some cases better than \emph{GenLouvain}.   The belief propagation approach is also quite fast with computational complexity $\mathcal{O}(qm)$.  Supplement Figure~\ref{fig:time_trial} shows that the runtime of \textit{multimodbp} is actually lower than \textit{GenLouvain} if the number of possible communities ($q$) is not too large. 

\begin{figure}[htbp]
\begin{center}
\includegraphics[width=.7\textwidth]{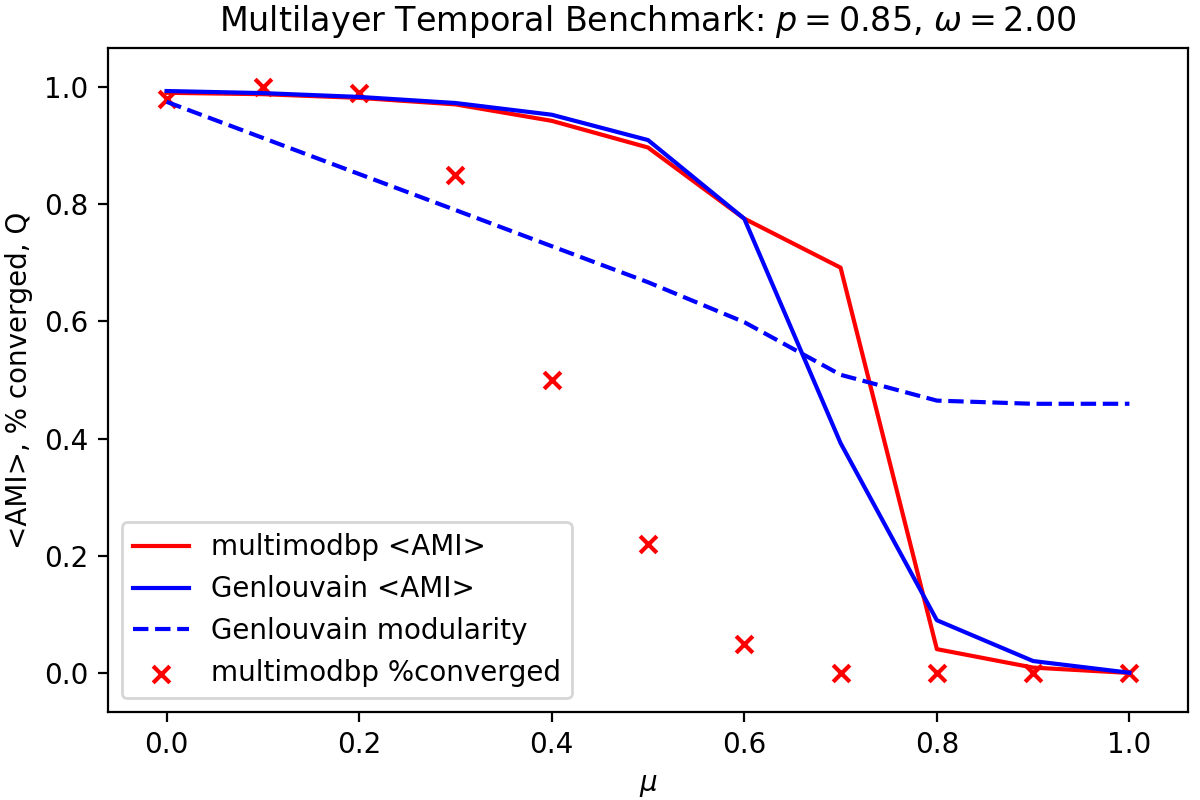}
\caption{Detectability of communities in the multilayer temporal benchmarking network (with $p=.85$) as $\mu$ is varied.  We plot the average $\left<\text{AMI}\right>$ of the detected communities for both \textit{multimodbp} (solid red line) and for \textit{GenLouvain} (solid blue line).  We note that a partition can be computed from the final marginals of the belief propagation even in the case where the algorithm has not converged (\textit{i.e.} the communities are not detectable).  We also show the average modularity of the partitions identified by \textit{GenLouvain} (dashed blue line) as well as percentage of trials that converged to a non-trivial solution for \textit{multimodbp}. }
\label{fig:ami_v_conv}
\end{center}
\end{figure}

In addition to the above results, it is important to emphasize that the convergence properties of \textit{multimodbp} provide additional information about whether there is significant community structure in the first place.  In Figure~\ref{fig:ami_v_conv} we show that \textit{multimodbp} stops converging when the planted structure is undetectable.  In contrast, the modularity of the communities detected by \textit{GenLouvain} remains relatively stable, even as the communities become increasing undetectable for higher values of $\mu$.  As is well understood, modularity maximization by itself may overfit the noise within a network and cannot reliably assess its own performance.  In networks without known communities (\textit{e.g.}\ most real-world networks), \textit{multimodbp} might be used as a complement to other methods (including \emph{GenLouvain}) to better determine whether community structure is actually present.   In supplement Figure~\ref{fig:temp_hist_conv} we compare the distribution of AMI for partitions that do converge versus those that do not for a range of parameter values in the temporal multilayer benchmark network.  This demonstrates that for a large range in the parameter space, convergence of individual runs corresponds with detection of the underlying community structure, though we do see cases where the returned marginals are aligned with the community despite the algorithm not having converged. 

Furthermore, while we have measured the performance of our algorithm here using a hard partition of the network, one of the advantages of belief propagation is the ability to generate a soft partitioning by using the computed marginals for each node-layer.  In the next section we showcase how this information can be used to interpret the structure of two real-world multilayer networks.

\subsubsection{Real-world multilayer networks}
\label{sec:realworld_results}

We conclude our results by demonstrating the inferences that can be made on real-world networks using the additional information provided by \textit{multimodbp}.  We begin with the US Senate voting similarity network as introduced by \cite{Waugh:2009vz} and analyzed in \cite{Mucha:2010bja}.  This data set represents the voting similarity patterns of 1,884 U.S.~Senators over 110 Congresses starting in 1789.  Each 2-year Congress beginning in the January following an election is represented as a layer within this network. A node within a layer represents a Senator serving in that Congress with Senators serving in consecutive Congresses linked through interlayer edges. In the analysis performed here, the network was modified to sparsify the intralayer connections by taking the K-nearest neighbors (KNN) of each Senator based on the strength of voting correlations (using K=10) while keeping the edges with the original weights based on voting similarities (that is, each node retains the 10 highest-weight edges around it).

\begin{figure}
\begin{center}
\begin{mdframed}
\hspace{-.07in}\includegraphics[width=.3\textwidth]{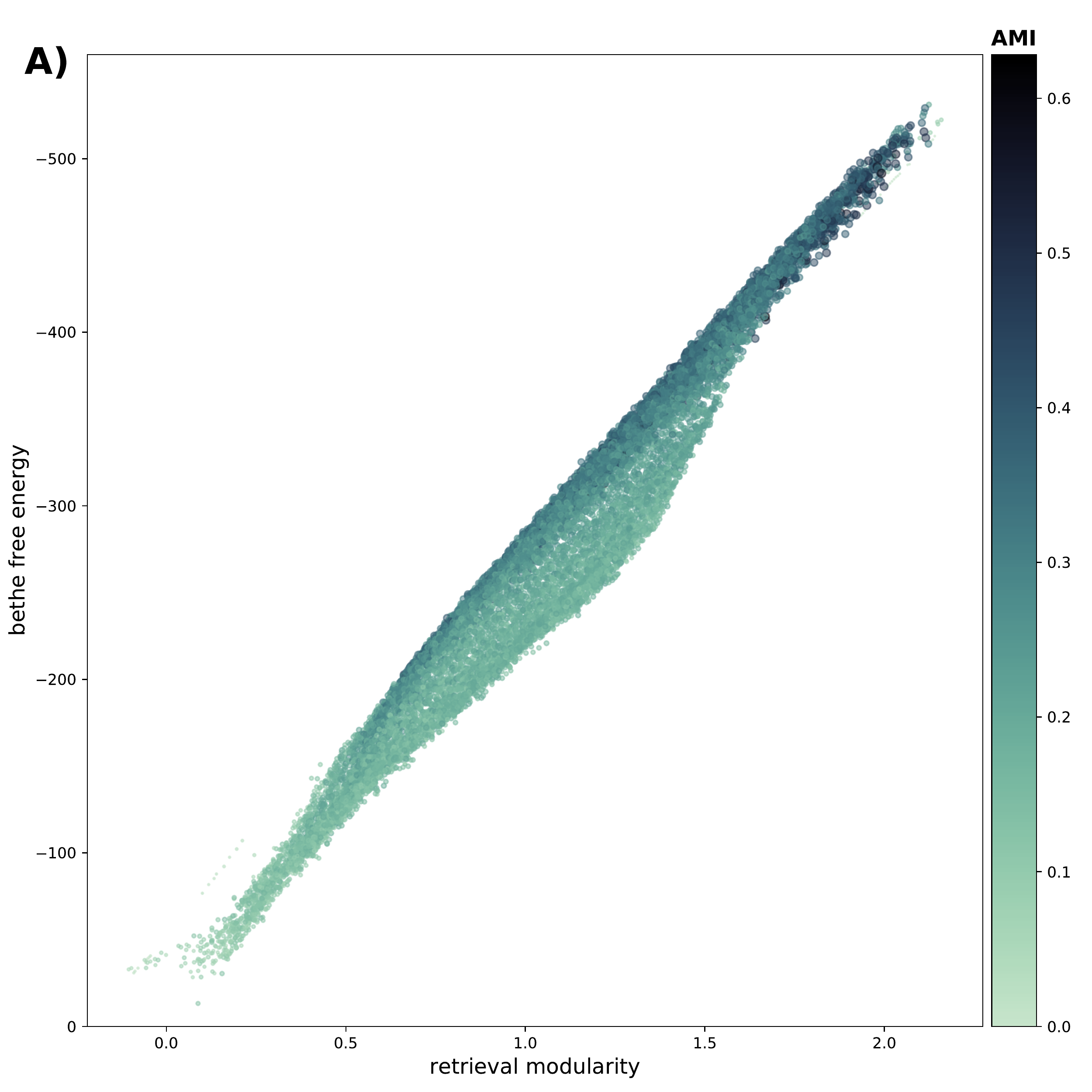}
\hspace{-.05in}\includegraphics[width=.7\textwidth]{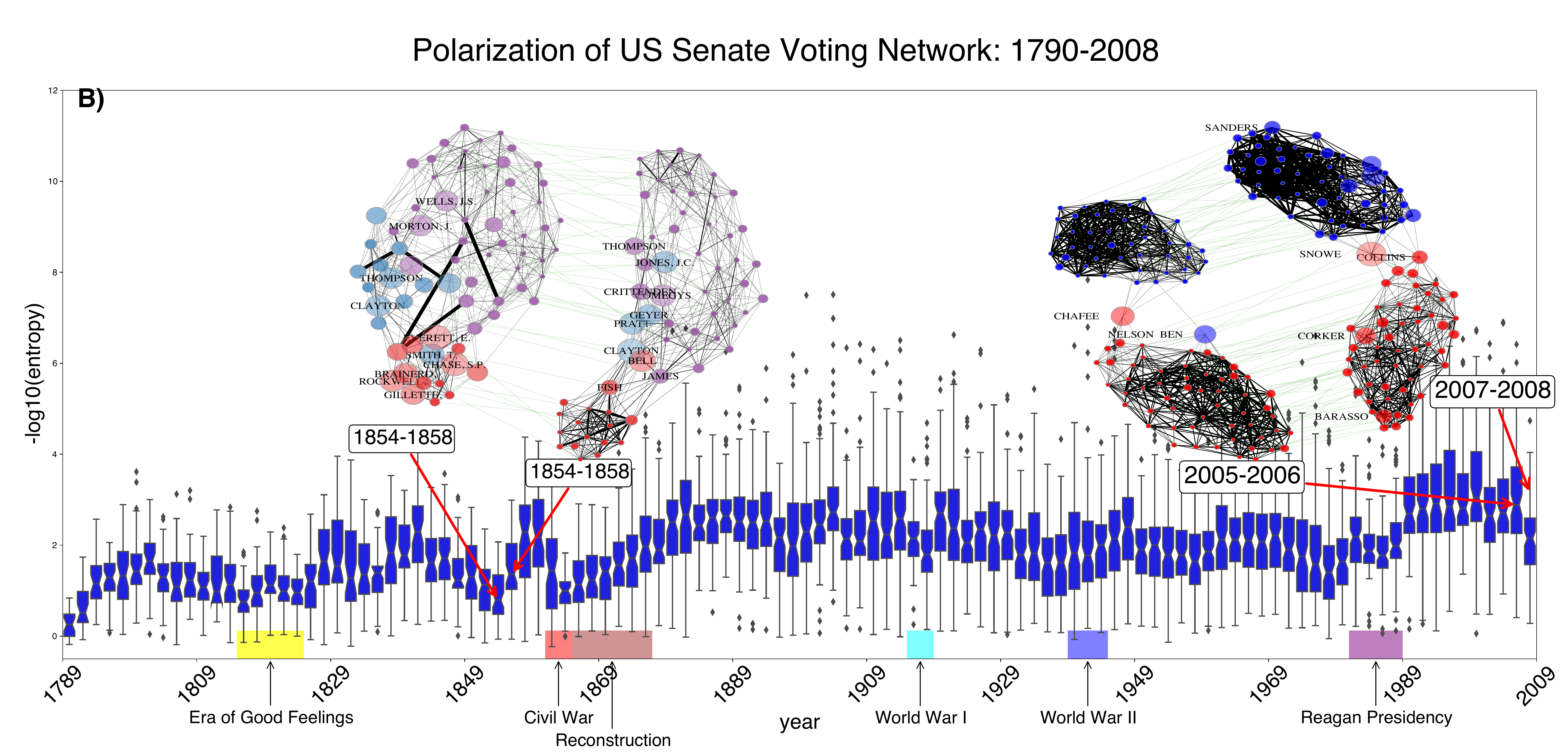}
\end{mdframed}
\caption{We ran \textit{multimodbp} on the US Senate voting similarity network comprised of 1884 Senators across the first 110 Congresses \cite{Mucha:2010bja,Waugh:2009vz}.  \textbf{A)}  The relationship between Bethe free energy and retrieval modularity is given by equation Eq.~\ref{eq:bfe_multi}.  We see that the Bethe free energy correlates strongly with retrieval modularity, and the partitions with the lowest free energy tend to correspond best with the underlying party structure.  \textbf{B)} We examined the distribution of the average Senator entropy for each Congress (layer) in the network.  Inset graphs depict how changes in average entropy correspond with network structure and the overall level of polarization within the network.  Node size depicts the average entropy level of Senators with ``high entropy" Senators labeled.}
\label{fig:senate_retmod}
\end{center}
\vspace{-.3in}
\end{figure}

In {Figure~\ref{fig:senate_retmod}}.A we show the correspondence between the retrieval modularity, the Bethe free energy (Eq.~\ref{eq:bfe_multi}), and the AMI with the political party labels of partitions identified across a range of the $(\gamma,\omega)$ parameter space.  Each point represents a partition identified using \textit{multimodbp}. The belief propagation algorithm fixed points are actually minimizers of the Bethe free energy (rather than optimizers of the retrieval modularity).  We see in general that partitions that minimize the Bethe free energy produce high retrieval modularities.  Optimizing the Bethe free energy also produces partitions that accurately reflect the underlying known structure in the data set (\textit{i.e.}, the political party affiliations of the Senators), shown by the color of the scatter points in Figure~\ref{fig:senate_retmod}.    We show a comparison of these partitions with the real party layouts in Figure~\ref{fig:senate_state}.  It appears that the most appropriate choice (in this sense) of the \textit{multimodbp} parameters are around $(\omega=6,\gamma=0.5)$.

One of the main benefits of using the belief propagation approach for community detection is that we can obtain a measure of how confident we are in the predicted community for each node-layer.  In Figure~\ref{fig:senate_retmod}.B, we show the distribution of Senator entropies for each Congress, averaged over the top 200 partitions identified (by AMI with parties).  On the y-axis we plot the distribution of $-\log_{10}(\mathrm{entropy})$ across all Senators as a measure of how strongly identified the communities are and thus how polarized Congress is along party lines.  We have highlighted several periods of American history such as the \textit{Era of Good Feelings} with corresponding low polarization/high entropy, or the high level of polarization immediately preceding the Civil War. The insets show how the corresponding changes in entropy from Congress to Congress are reflected in the community structure of the graph.  This is consistent with the increasing level of polarization identified in previous study of this data set \cite{Waugh:2009vz,Moody:2013wv}.  Our method gives the further benefit of providing a node level metric to identify how strongly each node-layer is connected with its community.  In  Figure~\ref{fig:senate_retmod}.B we have labeled the ``high entropy" Senators, those whose voting patterns indicate a measure of bipartisanship (or independence from the party as in the case of Bernie Sanders in the 2007--08 Congress).   Thus node-level information contained in the marginals allows for additional assessment and interpretation of the obtained community structure. In particular, this additional assessment is distinct from comparing detected communities with the node-level metadata.  This is especially notable given the now well-established difficulties with assessing community detection approaches solely on the basis of alignment with metadata attributes \cite{Peel:2017kj}.   
\begin{figure}
\includegraphics[width=\textwidth]{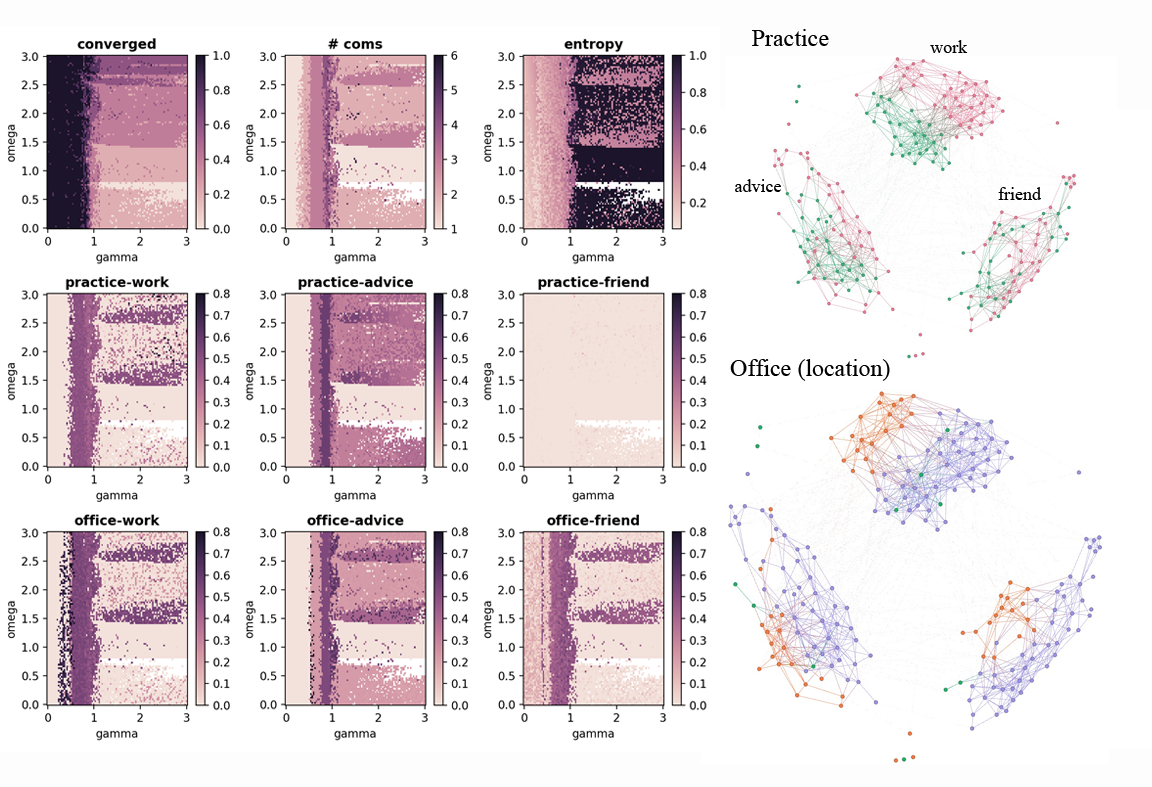}
\vspace{-.4in}
\caption{Several visualizations of the Lazega Lawyer network~\cite{lazega2001collegial}.  On the left we show several characteristics of partitions identified with \textit{multimodbp} at various values of $\gamma$ (x-axes) and $\omega$ (y-axes).  In the top row, from left to right, we show how many times the algorithm converged over 10 runs across the range of $\beta$ values, the number of communities identified by the best run for each set of parameters (based on lowest Bethe free energy), and the average entropy of the marginals across all of the nodes for each of these partitions.  In the next two rows we show the AMI of the identified partition within a single-layer and a specified metadata attribute.  For example in the left most panel of the second row, we show how the ``practice" (which type of law practiced by each node) attribute lines up with the partitioning of work layer. Showing the partitions in this manner demonstrates how different metadata attributes affect the community structure in the different layers and how this is best captured by \textit{multimodbp} for different values of $\gamma$ and $\omega$.  White spots are where \textit{multimodbp} did not converge for any value of $\beta$.  On the right we show the three layers of the network (advice, work, friends) colored by two of the metadata attributes: Practice (which specialty of law each person is involved in) and Office (which location the person works in).}
\label{fig:lawyer_ml}
\vspace{-.4in}
\end{figure}

The second real-world network that we have analyzed is the Lazega Lawyer network introduced by \cite{lazega2001collegial}.  We scan the $(\gamma,\omega)$ parameter range  $ [0 , 3 ] \times [ 0 , 3 ]$
and select the partition with the highest retrieval modularity, $Q(\{\hat{t}\})$ at each point.  In Figure~\ref{fig:lawyer_ml}, we show the number of iterations taken by the converged partitions for different parameter choices of $(\gamma,\omega)$. Within the lower right quadrant (high $\gamma$, low $\omega$) the algorithm only converged for a small range of $\beta$ values.  In the top row, middle panel, we see that for this network three communities were chosen for a large portion of the parameter space, although the structure of the identified partitions varied widely.  In the top right panel of Figure~\ref{fig:lawyer_ml}.A, we look at the average entropy per node across the parameter space to identify regions where node ambiguity is minimized.   These results suggest another possible way to identify regions of the $(\gamma,\omega)$ with corresponding strong community structure.  We see that for places where the entropy is quite low, the detected communities are aligning with the attributes in at least one of the layers and that regions where the algorithm does not converge also have a high average entropy.  This provides a useful way to compare the strength of the communities detected in different regions of $(\gamma,\omega)$, because the modularity score is itself a function of these variables (and thus is unhelpful in determining the appropriate parameter values by itself).    

In the bottom two rows we have explored how each partition overlays with a particular metadata attribute within a given layer.  For instance the panel titled ``office-friends" shows the AMI of all partitions with the office attribute only within the friends layer.  We see that within different parts of the parameter space, different features of the metadata align more closely with the partitions identified.  For instance there is a narrow band from $\gamma \in [0.4,0.6]$ for which the office attribute strongly aligns with the community structure of the work layer, with a corresponding low band of average entropy.  In contrast, the structure in the advice layer lines up most strongly with the practice attribute at a higher resolution range: $\gamma \in [0.8,1.0]$.   We see also that the friendship layer really only aligns with the office attribute (the location where people work).  Our results for this network complement those derived in \cite{Peel:2017kj} suggesting that no single metadata attribute explains the structure of this network.  These results highlight the need to explore and summarize partitions across different parameter ranges.

\section{Discussion}

We have presented \textit{multimodbp}, an extension of the modularity-based belief propagation framework to multilayer modularity.  Like the original belief propagation framework for modularity~\cite{Zhang:2014gea}, there are a number of features of \textit{multimodbp} that make it a useful tool for identifying community structure in real-world multilayer networks.   At its core, modularity and its multilayer extension are objective functions for assessing community structure and do not allow for true statistical inference\footnote{That said, we highlight that Newman has shown that optimizing modularity is equivalent to the MLE for a planted partition of the degree corrected stochastic block model for a certain value of $\gamma$ \cite{Newman:2016il}.  Likewise Pamfil \textit{et al.} showed an equivalence for multilayer modularity for a multilayer SBM with both temporal and multiplex topologies for specific values of $(\gamma,\omega)$ \cite{Pamfil:2019gg}.} (cf.\ generative approaches like the stochastic block model, e.g., \cite{Han:2015wz,Peixoto:2017fl,Stanley:2016bja,Taylor:2016kj}).  However, by formulating multilayer modularity optimization from the perspective of a Boltzmann ensemble, we can obtain an estimate of the uncertainty of assignment at each node from its marginal.  The marginals reflect how much shifting a node from one community to another changes the modularity and thus is a measure of how strongly a node prefers a certain community.  In this sense we can find a ``soft" partitioning of the nodes, in which one node may belong to multiple communities, along with confidence levels corresponding to each community.  We have shown in two real-world examples how knowledge at the node level about the confidence in the community prediction can inform interpretation of the community structure. 
Most modularity-based algorithms do not allow for overlapping communities with a few notable exceptions including OverMod~\cite{Bennett:2014eh} and the fuzzy c-means \cite{Zhang:2007hm}, both of which require an initial disjoint partitioning of the network in order to identify overlaps.  Other versions of overlapping modularity-like approaches include \cite{Havens:fz,Liu2010}.  Our approach is useful in that it can be used for either a hard or soft partitioning of the network depending on the desired context.  

Meanwhile, although the method of Zhang and Moore allows for the selection of the number of communities by identifying the value of $q$ for which the retrieval modularity plateaus~\cite{Zhang:2014gea}, we have shown that this approach fails to perform optimally in a number of cases.  This underscores the need for greater flexibility as provided by incorporation of the resolution parameter $\gamma$.  Rather than searching along the domain of $q$, we allow $q$ to float (up to a certain point $q_\mathrm{max}$) and search along the $\gamma$ domain to characterize network structure.  The flexibility added by the resolution parameter becomes even more important in the multilayer context.  We have shown that performance of \textit{multimodbp} is optimized by different combinations of $(\gamma,\omega)$ in different parameter regimes of the dynamic stochastic block model.  This is consistent with the work of Newman who demonstrated a link between the resolution parameter $\gamma$ of modularity and the $p_{\rm in}$ and $p_{\rm out}$ parameters of a degree-corrected stochastic block model \cite{Newman:2016il}.  Recently, Pamfil \textit{et al.}\ extended this approach to multilayer modularity, deriving a similar mapping between the coupling parameter, $\omega$, and the parameters of a model very similar to the DSBM studied here \cite{Pamfil:2019gg}.

One of the greatest benefits of the \textit{multimodbp} approach is that the convergence of the algorithm to non-trivial solutions reveals the existence of significant community structure beyond what would be expected at random.  Several prior works have shown that even in randomly-generated networks without underlying structure, modularity optimization heuristics are capable of finding high-modularity partitions \cite{Bagrow:2012hy,DeMontgolfier:2011jp,Zhang:2014gea}. For this reason alone we believe an extension of modularity belief propagation for multilayer networks provides a valuable new tool for network analysis.  We have shown that our algorithm performs comparably to \textit{GenLouvain} across a range of multilayer topologies and that its convergence can in some cases reveal when no detectable structure is present, as compared to \emph{GenLouvain} continuing to report high-modularity but possibly meaningless community partitions.

There remain a number of technical challenges for implementing \textit{multimodbp} at scale.  The runtime of the algorithm depends greatly on the number of iterations of belief propagation that are required to run before convergence.  As described in Zhang and Moore, this tends to spike as you approach the retrieval phase, and the formula for $\beta^{*}$ we have used tends to yield values slightly above where this spike occurs.  Ideally, one could have an adaptive solution, identifying a value of $\beta$ for which the algorithm appears to be converging quickly early on and adjusting $\beta$ once the algorithm is closer to converging.  Eventually, we would like to devise an automatic method of selecting an appropriate value for $\beta$ based on a preliminary scan of convergence rates across the $\beta$ domain, similarly to how we iteratively select the appropriate number of communities as the algorithm runs.   Another issue is the dependency of the runtime and memory of the algorithm on the number of marginals being optimized.  We try to reduce the dimension of the marginals after the algorithm has run, by combining redundant dimensions (those that are highly correlated).  One could imagine attempting such a reduction earlier on after a few course-grained runs of the algorithm to produce additional performance gains.  

To facilitate use of (and possible improvements on) our method, we have written and distributed a Python package available on PyPI \cite{multimodbp_software}.


\newpage

\renewcommand{\thesection}{S}
\renewcommand\thefigure{\thesection.\arabic{figure}}

\section{Supplement}

\makeatletter
\def\Let@{\def\\{\notag\math@cr}}
\makeatother

\subsection{Derivation of Multilayer Belief Propagation Update Equations}
\label{bpeq_deriv}

To derive the update equations for the multilayer belief propagation, we have relied heavily on the approach employed by Zhang and Moore~\cite{Zhang:2014gea}.  In the belief propagation algorithm, also known as the sum-product algorithm or cavity method, each node sends a ``message" ($\psi^{i\rightarrow k}_{t}$ ) to its neighboring nodes encoding the marginal probability of it occupying a given state, $t$ (or, in our case, belonging to a given community).  These updates are iterated over all nodes until the messages converge to a fixed point of the update equations (if it exists). 

\ In general, belief propagation can be written for any factor graph with interactions up to an arbitrary order.  However, in the case of only pairwise interactions, the belief propagation equations simplify to the following the form:
\begin{eqnarray} 
\psi^{i\rightarrow k}_{t} &=\frac{1}{Z_{i\rightarrow k }} \prod_{ j \in \partial i \setminus k } \sum_{s} f_{ij}(s,t) \psi^{j\rightarrow i}_{s} \,
\end{eqnarray}
where  $f_(ij)$ is a function representing the interaction between nodes $i$ and $j$.  We note that the product is over all neighbors of $i$ excluding $k$, which is to say every node that $i$ interacts with.  We will see that interactions are not strictly limited to nodes that are connected within the graph we are optimizing over.  The original formula for modularity developed by Newman and Girvan is given by
\begin{eqnarray} 
\label{eq:mod}
Q(c) &= \frac{1}{2m} \sum_{i,j} ({ A_{ij}  - \frac{d_i d_j}{2m})\delta_{c_i,c_j} }
\end{eqnarray} 
where $A_{ij}$ is the (possibly weighted) adjacency matrix, $c_i \in \{1,...,q\}$ denotes the community assignment of node $i$, $d_i = \sum_{j} A_{ij}$ gives the (weighted) degree/strength of node $i$, and $m=\sum_{i<j}A_{ij}$ the number (total weight) of edges in the graph. We distinguish here between an interaction between two nodes in our model and an actual edge in our graph.  In the case of modularity, the sum is over all pairs of nodes (rather than only the edges) because pairs of nodes within the same community still contribute to the score through the null model term, $\frac{d_i d_j}{2m}$. 

\begin{figure}[htbp]
\begin{center}
\includegraphics[width=.4\linewidth]{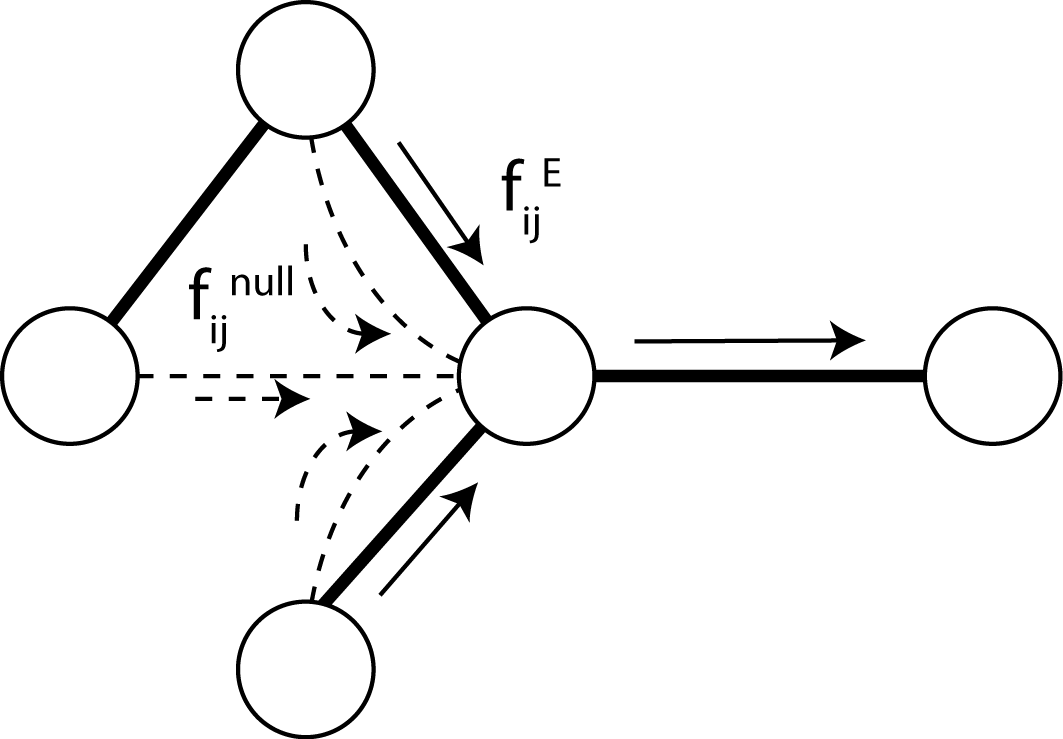}
\caption{ \label{fig:bp_example} Schematic of modularity belief propagation.  We have split the contributions to the modularity into two kinds of interactions: strong interactions represented by edges in the original graph (shown as dark, solid lines in figure) and weaker, all-to-all connections given by the null model term (shown as dashed lines).  Beliefs (shown as arrows) are summed from all interacting nodes except the one who is receiving the message (far right node).}
\end{center}
\end{figure}

We could imagine Equation~\ref{eq:mod} as defining the contribution to the energy for a single type of interaction between nodes with corresponding interaction term: $f_{ij} = e^{\beta(A_{ij} - \frac{d_i d_j}{2m})\delta_{st}}$, where we have dropped the factor of $\frac{1}{2m}$ because only the relative energies matter for the probability of occupying a given state, and switched notations from $c_i , c_j \leftrightarrow s, t$ to match the belief propagation notation.   However, this leads to a much more computationally intensive form of belief propagation where updates for each node depend weakly on every other node in the graph requiring $\mathcal{O}(qn^2)$ computations for each round of updates.  Instead we represent two different kinds of interactions between nodes shown in Figure~\ref{fig:bp_example}: those corresponding to the edges in the original graph with contribution $f^{\mathcal{E}}_{ij} = e^{\beta A_{ij}\delta_{st}}$ and the weak but dense interactions between all pairs of nodes in the graph that arise from the null-model term $f^{\text{null}}_{ij} = e^{-\beta \frac{d_i d_j}{2m}\delta_{st} }$.  By splitting the interaction into contributions from the separate terms in Equation~\ref{eq:mod}, we factor out the contribution from these, giving the belief update equations introduced by Zhang and Moore \cite{Zhang:2014gea}: 
\begin{eqnarray}
\psi_t^{i\rightarrow k} &=\frac{1}{Z_{i\rightarrow k }} \prod_{j \in \partial i \setminus k } \sum_{s=1}^{q} e^{\beta \delta_{st}}\psi^{j\rightarrow i}_{s} \prod_{j \ne i \setminus k } \sum_{s=1}^{q} e^{-\beta (d_i d_j / 2m)\delta_{st}}  \psi^{j\rightarrow i}_{s} \,.
\end{eqnarray}
where the first product is only over the neighbors of node $i$ in the network excluding $k$ (\textit{i.e.}\ $ j \mid (i,j) \in \mathcal{E}$).  

These two terms can be simplified as follows: 
\begin{align}
\sum_{s=1}^{q} e^{\beta \delta_{st}}\psi^{j\rightarrow i}_{s} &= e^{\beta}\psi^{j\rightarrow i}_{t} + \sum_{s\ne t}^{q} \psi^{j\rightarrow i}_{s} = (e^{\beta} - 1 ) \psi_t^{j \rightarrow i} + 1\,, \\
\sum_{s=1}^{q} e^{-\beta \frac{d_i d_j}{2m} \delta_{st}}\psi^{j\rightarrow i}_{s} &= e^{- \beta \frac{d_i d_j}{2m}}\psi^{j\rightarrow i}_{t} + \sum_{s\ne t}^{q} \psi^{j\rightarrow i}_{s} = (e^{-\beta \frac{d_i d_j}{2m}} - 1 ) \psi_t^{j \rightarrow i} + 1\,.
\end{align}

Further simplification of the second term occurs by replacing all of the weak interactions with a single field term that is updated after each round of belief propagation updates.  In the event that the network is sparse,  the degree of any given node will be small compared to the square root of the total number of edges ($d_i,d_j \ll \sqrt{2m}$).  In this case we can approximate the message a node sends to non-neighboring nodes (along a dashed line in Figure~\ref{fig:bp_example}) with its own marginal: 
\begin{align}
\psi_t^i &= \psi_t^{i\rightarrow k} \sum_{s} e^{-\beta \frac{d_id_j}{2m}\delta_{st}}  \psi_s^{k\rightarrow i} \\
&= \psi_t^{i\rightarrow k} (\sum_{s\ne t} \psi_s^{k\rightarrow i}  + e^{-\beta \frac{d_id_j}{2m}}  \psi_t^{k\rightarrow i})\\
&\approx \psi_t^{i\rightarrow k} (\sum_{s\ne t} \psi_s^{k\rightarrow i}  + (1- \beta\frac{d_i d_j}{2m} ) \psi_t^{k\rightarrow i})\\
&= \psi_t^{i\rightarrow k} (\sum_{s} \psi_s^{k\rightarrow i}  - \beta\frac{d_i d_j}{2m}  \psi_t^{k\rightarrow i})\\
&= \psi_t^{i\rightarrow k} ( 1- \beta\frac{d_i d_j}{2m} \psi_t^{k\rightarrow i})\\
&\approx \psi_t^{i\rightarrow k}\,.
\end{align} 

We can thus write our contribution from the null-model as follows:
\begin{align} \label{eq:field_trick}
\prod_{j \ne k } \sum_{s} e^{-\beta \frac{d_i d_j } {2m}}\delta_{st} \psi^{j\rightarrow i}_{s} &= \prod_{j \ne k } ((e^{-\beta \frac{d_i d_j}{2m}} - 1 ) \psi_t^{j \rightarrow i} + 1 )\\
& \approx \prod_{j \ne k } ((e^{-\beta \frac{d_i d_j}{2m}} - 1 ) \psi_t^{j} + 1 )\\
& \approx \prod_{j \ne k } ( \beta \frac{d_i d_j}{2m}  \psi_t^{j} + 1 )\\
& \approx \prod_{j \ne k } ( e^{ \beta \frac{d_i d_j}{2m}  \psi_t^{j}}  )\\
& \approx \exp \left (-\beta \frac{d_i}{2m} \sum_j {d_j \psi^{j}_t} \right)\\
&=\exp \left (-\beta \frac{d_i}{2m} \theta_t \right) \,
\end{align}
where $\theta_t = \sum_{j} \psi_t^j d_j$ and is treated as constant for each round of belief propagation and then updated accordingly with each node's marginal.  This ``field trick" originally applied in \cite{Decelle:2011ek} and \cite{Decelle:2011ik} is made possible by the way we split the contributions from the edges of the network into a separate term from the interactions that come from the null-model term in the modularity formula.  This reduces the computational complexity to a much more manageable $\mathcal{O}(qm)$.  Combining these simplifications gives Zhang and Moore's original update equation: 
\begin{align}
\psi_t^{i\rightarrow k } \propto \exp \left( \frac{-\beta d_i }{2m } \theta_t + \sum_{j \in \partial i \setminus k}\log \left( 1+ (e^{\beta} - 1 ) \psi_t^{j\rightarrow i} \right) \right)\,.
\end{align} 

We have employed a very similar approach as Zhang and Moore, however now we use the formula for multilayer modularity in Equation~\ref{eq:multimod} as the Hamiltonian to represent interactions in our model.  We now use $i,j, \text{ and } k$ to index the node-layers in our multilayer network. For more details about the multilayer notation used here, please see Section~\ref{sec:notation} in the main text.  First, we account for the additional contribution of the interlayer edges in a similar manner to the intralayer edges: $f_{ij}^{\text{multi}} = e^{\beta \tilde{A}_{ij}}$, where $\tilde{A}_{ij}=A_{ij}\delta(l_i,l_j)+\omega C_{ij}(1-\delta(l_i,l_j))$ is the appropriate weight for the inter/intralayer edge the message is traveling along.  We note that we have allowed for weights along the intralayer edges by the same method as Shi \textit{et al.}\  \cite{Shi:2018gc} and then similarly include interlayer contributions scaled by $\omega$ in $\tilde{A}_{ij}$. Thus the contributions from the edges in the network are now given by
\begin{align}
\prod_{j \in \partial i \setminus k } \sum_{s=1}^{q} e^{\beta \tilde{A}_{ij}} \psi^{j \rightarrow i}_s  = \prod_{j \in \partial i \setminus k } \left( (e^{\beta\tilde{A}_{ij}}-1)\psi^{j\rightarrow i}_t) +1\right) \,.
\end{align}  
This product is over all node-layers in the neighborhood of $i$ including its interlayer neighbors (but excluding node-layer $k$). 

We also modify the belief update equations in accordance with the different contributions from the null model in multilayer modularity.  In the null model for multilayer modularity given by Equation~\ref{eq:multimod}, only pairs of node-layers that are within the same layer contribute to $P_{ij}$ with the denominator being the total number of edges within $i$'s layer, $m_{l_i}$ (see Equation~\ref{eq:nullmodel}).  Thus the product in the null-model component only goes over the indices of node-layers within the same layer as node-layer, $i$: 
\begin{align}
\prod_{j \ne k } \sum_{s} e^{-\beta \frac{d_i d_j } {2m}}\delta_{st} \Rightarrow 
\prod_{j \in  \mathcal{V}_{l_i} \setminus k,i } \sum_{s} e^{-\beta \frac{d_i d_j } {2m_{l_i}}}\delta_{st}
\end{align}

We have also incorporated a resolution parameter, $\gamma$, to the contribution from the null model.  The resolution parameter effectively balances the contributions of edges that are internal to communities and the strength of the field from the null model. This gives the null-model interaction: $f_{ij}^{\text{null}} = e^{-\beta \gamma \frac{d_i d_j}{2m_{l_i}}}$.  We can still incorporate the field trick detailed above as long as $d_i,d_j \ll \sqrt{2m_{l_i}/\gamma}$.  In our experience we have found that the algorithm generally doesn't converge if $\gamma$ is too large and have generally used $\gamma \le 3$ for experiments in this manuscript.  We can apply the same line of reasoning as Equation~\ref{eq:field_trick} above, substituting $\psi_t^j$ for $\psi_t^{j\rightarrow i}$ and Taylor's theorem to arrive at
\begin{align}
\prod_{j \in \mathcal{V}_ {l_i} \setminus k } \sum_{s} e^{-\beta \gamma \frac{d_i d_j } {2m_{l_i}}}\delta_{st} \approx \exp \left(-\beta \frac{d_i}{2m_{l_i}} \theta_{l_i}^t\right) \,,
\end{align} 
where $\theta^t_{l_i} = \sum_{j \in \mathcal{V_{l_i}} } \psi_t^j d_j$, is the layer specific field term that is treated as constant for each round of message passing, then updated according to the new marginals.  These modifications combined give us the update equations~\ref{multimodbp_eq} for \textit{multimodbp}.  We note that the message passed from node-layer $i$ to node-layer $k$, $\psi_t^{i\rightarrow k}$ does \emph{not} depend on the type of edge $(i,k)$.  Node-layer $i$ integrates information from its neighboring node-layers (except node-layer $k$), handling both edge weights and types appropriately, and passes this information to node-layer $k$.  The edge type (and weight) between node-layer $i$ and node-layer $k$ only comes into play when node-layer $k$ integrates all the information coming in from its neighboring node-layers.

\

\subsection {Derivation of Bethe Free Energy}
\label{bfe_deriv}
We derive here the formula for the free energy of the single layer model given in Zhang and Moore \cite{Zhang:2014gea}.  In the next section we will show how this naturally extends to the multilayer case with interlayer edges.  For any model which has only pairwise interactions, the formula for the Bethe free energy approximation is given by 
\begin{align}
f_{\text{Bethe}}&=-\frac{1}{N\beta} \left( \sum_i{\log Z_i} - \sum_{i,j\in \mathcal{E}} { \log Z_{ij}} \right)\,.
\end{align}

In the modularity model, there are really two types of edge interactions:  those that are given explicitly by the underlying graph (\textit{i.e.}\ the $A_{i,j}\delta_{c_i,c_j}$ term), and the pairwise interaction term that comes from the null model (\textit{i.e.}\ $P_{i,j}=\frac{d_id_j}{2m}\delta_{c_i,c_j}$).  We can split these two apart:
\begin{align}
f_{\text{Bethe}}=-\frac{1}{N\beta} \left( \sum_i{\log Z_i} - \sum_{i,j\in \mathcal{E}} { \log Z_{ij}} - \sum_{ i\ne j} { \log \hat{Z}_{ij}} \right)
\end{align}
where we refer to the edges in the underlying graph as $\mathcal{E}$ and split out the non-edge interactions into another term with normalization $\hat { Z}^{ij}$.
We write out the joint distribution for these ``non-edges":
\begin{align}
\psi_{st}^{i  j } = \frac{1}{\hat{Z}_{ij}} e^{-\beta(d_id_j/2m)\delta_{st}}\psi_s^{i}\psi_t^{j}
\end{align}
We use this to compute $\hat{Z}_{ij}$:
\begin{align}
\hat{Z}_{ij} &= \sum_t \sum_s e^{-\beta(d_id_j/2m)\delta_{st}}\psi_s^{i}\psi_t^{j}\,,
\end{align}
\begin{align}
 \sum_{i < j } \log{\hat{Z}_{ij}} &=\sum_{i < j} \log{ \sum_t \sum_s e^{-\beta(d_id_j/2m)\delta_{st}}\psi_s^{i}\psi_t^{j}} \\
 &\approx \sum_{ i < j } \log{ \left( \sum_t \sum_s  1-\beta(d_id_j/2m)\delta_{st}\psi_s^{i}\psi_t^{j} \right)} \\
 &\approx -\sum_{i < j} \left( \sum_t \sum_s  \beta(d_id_j/2m)\delta_{st}\psi_s^{i}\psi_t^{j} \right) \\
&= -\sum_t \sum_{i < j}  \beta(d_id_j/2m)\psi_t^{i}\psi_t^{j} \\
&= -\frac{\beta}{4m}\sum_t  \sum_{i \ne j} d_id_j\psi_t^{i}\psi_t^{j} \\
&= -\frac{\beta}{4m}\sum_t  \theta_t^2 \,.
\end{align}

This gives us the expected full formula,
\begin{align}
f_{\text{Bethe}}=-\frac{1}{N\beta} \left( \sum_i{\log Z_i} - \sum_{i,j\in \mathcal{E}} { \log Z_{ij}} + \frac{\beta}{4m}\sum_t  \theta_t^2 \right)\,.
\end{align}

\subsection{Multilayer Bethe Free Energy}

We now extend the Bethe Free Energy equation to multilayer networks. The formula for multilayer modularity for undirected networks is given by  Equation~\ref{eq:multimod} in the main text:
\begin{equation}
Q(\gamma,\omega)=\sum_{i,j} \left( A_{ij} - \gamma P_{ij} + \omega C_{ij} \right)\delta(c_{i},c_{j})\,.
\end{equation}
As before we only have pairwise interactions within the model.   However, note that in the multilayer formulation there are now both intra- and interlayer edges.  We can split the edge term of  $f_{\text{Bethe}}$ into the contributions from interlayer and intralayer edges:
\begin{align}
 \sum_{i,j\in \mathcal{E}} { \log Z_{ij}} =  \sum_{i,j\in \mathcal{E}_{inter}} { \log Z^{inter}_{ij}}+ \sum_{i,j \in \mathcal{E}_{intra}} { \log Z^{intra}_{ij}}
\end{align}
where $\mathcal{E}_{intra}$ and $\mathcal{E}_{inter}$ are given by the non-zero elements of $A_{ij}$ and $C_{ij}$ respectively.   For the non-edge term in the multilayer case, we note that the non-edge interaction terms are all restricted to within a given layer.  This means that nodes within different layers of the model only interact through the interlayer edge term and not through the null model term: 
\begin{align}
\frac{1}{N\beta} \sum_{ i\ne j} { \log \hat{Z}_{ij}}= \frac{1}{N\beta} \sum_{l} \sum_{ i\ne j , i,j \in l} { \log \hat{Z}^l_{ij}}\,.
\end{align}
We can therefore split this term into a sum over the contributions from each of the layers with a similar form as from before:
\begin{align}
\sum_{l} \sum_{ i\ne j , i,j \in l} { \log \hat{Z}^l_{ij}} =  - \sum_l \frac{\beta}{4m_l}\sum_t  (\theta^l_t)^2
\end{align}
and we can write the full Bethe free energy as
\begin{align}
f_{\text{Bethe}} = -\frac{1}{N\beta} & \left(  \sum_i{\log Z_i} -  \sum_{i,j\in \mathcal{E}_{inter}} { \log Z^{inter}_{ij}} \right. \cr & - \left. \sum_{i,j \in \mathcal{E}_{intra}} { \log Z^{intra}_{ij}} + \sum_l \frac{\beta}{4m_l}\sum_t  (\theta^l_t)^2 \right)
\end{align}
where the $Z_{ij}^{inter}$ can be computed from the pairwise marginals of the interlayer interactions:
\begin{align}
\psi^{i,j}_{s,t} = \frac{1}{Z_{ij}^{inter}} e^{\beta \omega C_{ij} \delta_{s,t}} \psi^{i \rightarrow j}_s \psi^{ j \rightarrow i}_t \,.
\end{align}

\subsection{Formula for selection of  $\beta^*$ on weighted multilayer networks}
\label{beta_star_deriv}
\ 
In this paper, we have used the approach by Shi \textit{et al.} to identify the value of $\beta^*$ where the trivial solution is no longer stable \cite{Shi:2018gc} which is an extension of the reasoning to the original approach presented in Zhang and Moore \cite{Zhang:2014gea} (see Section~\ref{sec:beta_star} in main text).  Here we present the form of the linearized approximation of the messages as well as its eigenvalue.         

Consider the update equation for \textit{multimodbp}:
\begin{equation}
\psi^{i \to k }_t \propto \exp {\left [ \gamma \frac{\beta d_i }{2m_{l_i}}\theta^{l_i}_{t}  + \sum_{j\in \partial_i \setminus k}  \log{(1+\psi^{j \to i }_t(e^{\tilde{A}_{ij}\beta} -1))} \right]} \,.
\end{equation}
We compute the derivative $\frac{\partial \psi_t^{i \to k}}{\partial \psi_s^{j \to i}}$ assuming both $(i,j)$ and $(i,k)$ are edges:
\begin{align}\label{eq:beta_derive}
    \frac{\partial \psi_t^{i \to k}}{\partial \psi_s^{j \to i}} &= \frac{\partial}{\partial \psi_s^{j \to i}} \left[ \frac{1}{Z} \exp {\left [ \gamma \frac{\beta d_i }{2m_{l_i}}\theta^{l_i}_{t}  + \sum_{j\in \partial_i \setminus k}  \log{(1+\psi^{j \to i }_t(e^{\tilde{A}_{ij}\beta} -1))} \right]} \right] \\
    &= \frac{1}{Z} \frac{\partial}{\partial \psi_s^{j \to i}} \exp {\left [ \gamma \frac{\beta d_i }{2m_{l_i}}\theta^{l_i}_{t}  + \sum_{j\in \partial_i \setminus k}  \log{(1+\psi^{j \to i }_t(e^{\tilde{A}_{ij}\beta} -1))} \right]} \\
    &\quad+ \exp {\left [ \gamma \frac{\beta d_i }{2m_{l_i}}\theta^{l_i}_{t}  + \sum_{j\in \partial_i \setminus k}  \log{(1+\psi^{j \to i }_t(e^{\tilde{A}_{ij}\beta} -1))} \right]} \frac{\partial}{\partial \psi_s^{j \to i}} \left[ \frac{1}{Z} \right]\,.
\end{align}
We will consider each of these two derivatives separately.  To help condense notation we define $F_{\text{arg}}=\gamma \frac{\beta d_i }{2m_{l_i}}\theta^{l_i}_{t}  + \sum_{j\in \partial_i \setminus k}  \log{(1+\psi^{j \to i }_t(e^{\tilde{A}_{ij}\beta} -1))}$.  First,
\begin{align}
    &\frac{\partial}{\partial \psi_s^{j \to i}} \exp {\left [ \gamma \frac{\beta d_i }{2m_{l_i}}\theta^{l_i}_{t}  + \sum_{j\in \partial_i \setminus k}  \log{(1+\psi^{j \to i }_t(e^{\tilde{A}_{ij}\beta} -1))} \right]} \\
    &= \exp\left[ F_{\text{arg}} \right]\frac{\partial}{\partial \psi_s^{j \to i}}\left [ \gamma \frac{\beta d_i }{2m_{l_i}}\theta^{l_i}_{t}  + \sum_{j\in \partial_i \setminus k}  \log{(1+\psi^{j \to i }_t(e^{\tilde{A}_{ij}\beta} -1))} \right]\,.
\end{align}
The derivative of the first term here, 
\begin{equation*}
    \frac{\partial}{\partial \psi_s^{j \to i}} \gamma\frac{\beta d_i}{2m} \theta^{l_i}_t 
\end{equation*}
is $O\left(\frac{d_i d_j}{2m}\right)$, which we can ignore given our assumption that the network is sparse ($d_i\ll \sqrt{m}$ for all $i$). 

We are then left with
\begin{equation*}
    \frac{\partial}{\partial \psi_s^{j \to i}}\left[ \sum_{j\in \partial_i \setminus k}  \log{(1+\psi^{j \to i }_t(e^{\tilde{A}_{ij}\beta} -1))}\right]\,.
\end{equation*}
The only term in this sum that will lead to a non-zero derivative is if $s=t$, leading to 
\begin{equation}
    \frac{\partial}{\partial \psi_s^{j \to i}} \log{(1+\psi^{j \to i }_t(e^{\tilde{A}_{ij}\beta} -1))}\delta_{st}
    = \delta_{st} \frac{e^{\tilde{A}_{ij}\beta} - 1}{1 + \psi_s^{j \to i}\left(e^{\tilde{A}_{ij}\beta} - 1\right)}\,.
\end{equation}
Evaluating at the fixed point, and combining with the previous $\frac{1}{Z}\left. \exp(F_{\text{arg}})\right\rvert_{\frac{1}{q}}=\frac{1}{q}$, this term becomes
\begin{equation}
    \delta_{st} \frac{e^{\tilde{A}_{ij}\beta} - 1}{q+e^{\tilde{A}_{ij}\beta} - 1}\,.
\end{equation}

Next we move on to the second term from the previous product rule expansion (Eq~\ref{eq:beta_derive}):
\begin{align*}
    \frac{\partial}{\partial \psi_s^{j \to i}} \frac{1}{Z} &= -\frac{1}{Z^2} \frac{\partial Z}{\partial \psi_s^{j \to i}} \\ &= -\frac{1}{Z^2}\frac{\partial}{\partial \psi_s^{j \to i}} \left[ \sum_t \exp {\left [ \gamma \frac{\beta d_i }{2m_{l_i}}\theta^{l_i}_{t}  + \sum_{j\in \partial_i \setminus k}  \log{(1+\psi^{j \to i }_t(e^{\tilde{A}_{ij}\beta} -1))} \right]} \right]\,.
\end{align*}
We follow the same line of reasoning as before, dropping the $\theta^{l_i}$ term to arrive at
\begin{equation}
    -\frac{\exp(F_{\text{arg}})}{Z^2} \frac{e^{\tilde{A}_{ij}\beta} - 1}{1 + \psi_s^{j\to i}(e^{\tilde{A}_{ij}\beta} - 1)}\,.
\end{equation}
If we bring the extra $\exp(F_{\text{arg}})$ from before back in, and evaluate at the fixed point, this leads to
\begin{equation}
\label{eq:tst}
    -\frac{1}{q}  \frac{e^{\tilde{A}_{ij}\beta} - 1}{q+e^{\tilde{A}_{ij}\beta} - 1}\,.
\end{equation}
In total, we find the linear approximation of the messages is given by the $q \times q$ matrix:
\begin{equation}
    T^{i\rightarrow k, j \rightarrow i }_{st} =  \frac{e^{\tilde{A}_{ij}\beta} - 1}{q+e^{\tilde{A}_{ij}\beta} - 1} \left(\delta_{st} - \frac{1}{q}\right) \,.
\end{equation}
The leading eigenvalue of this matrix is given by
\begin{equation}
\eta_{ij}=\frac{e^{\beta \tilde{A}_{ij}} -1 }{e^{\beta \tilde{A}_{ij}} + q -1} \,.
\end{equation}

\vspace{.1in}
\
\subsubsection{Testing selection of $\boldsymbol{\beta^*}$}

\begin{figure}[htbp]
\begin{center}
\includegraphics[width=\linewidth]{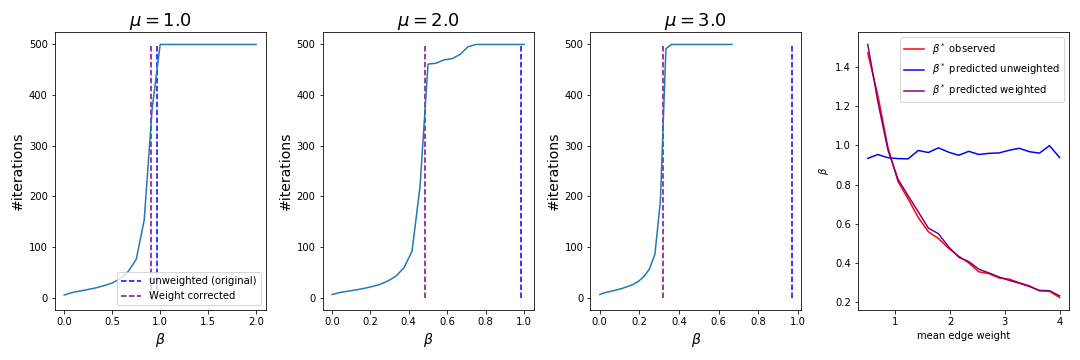}
\caption{Stability boundary for Erd\H{o}s-R\'enyi graph with weights assigned randomly from a $\mathcal{N}(\mu,\sigma=.5)$ normal distribution.  Left three plots depict convergence curves of the algorithm for three different means of the normally distributed edge weights ($\mu=$1,2, and 3 respectively).  Each curve represents the average over 10 realizations of the ER random graph.   The unweighted prediction for $\beta^*$ is given by the black dashed line, while the weight adjusted prediction is given by the dashed green line.  On far right plot $\beta^*$ was empirically determined for several different mean weights (red line) and compared with the predicted values (blue line) showing good agreement.}
\label{er}
\end{center}
\end{figure}

As part of testing the formula for $\beta^*(q,w)$, we look at the effect of adding normally distributed edges weights on an Erd\H{o}s-R\'enyi graph shown in Figure~\ref{er}. For the Erd\H{o}s-R\'enyi graph with normally distributed weights, Equation~\ref{eq:bstar_cond} gives a very good estimate of where the divergence occurs, while the unmodified equation becomes less accurates as the weights differ from 1.

In the far right panel of Figure~\ref{er}, we show that formula derived by Shi \textit{et al.} well predicts the point where the trivial solution is no longer stable (shown by the red line).  Below in Figure~\ref{sbm2com_betascan}, we also demonstrate that for a 2-community SBM the modified formula for $\beta^*$ occurs within the retrieval phase, detecting the communities with high accuracy.        

\begin{figure}[htbp]
\begin{center}
\includegraphics[width=\linewidth]{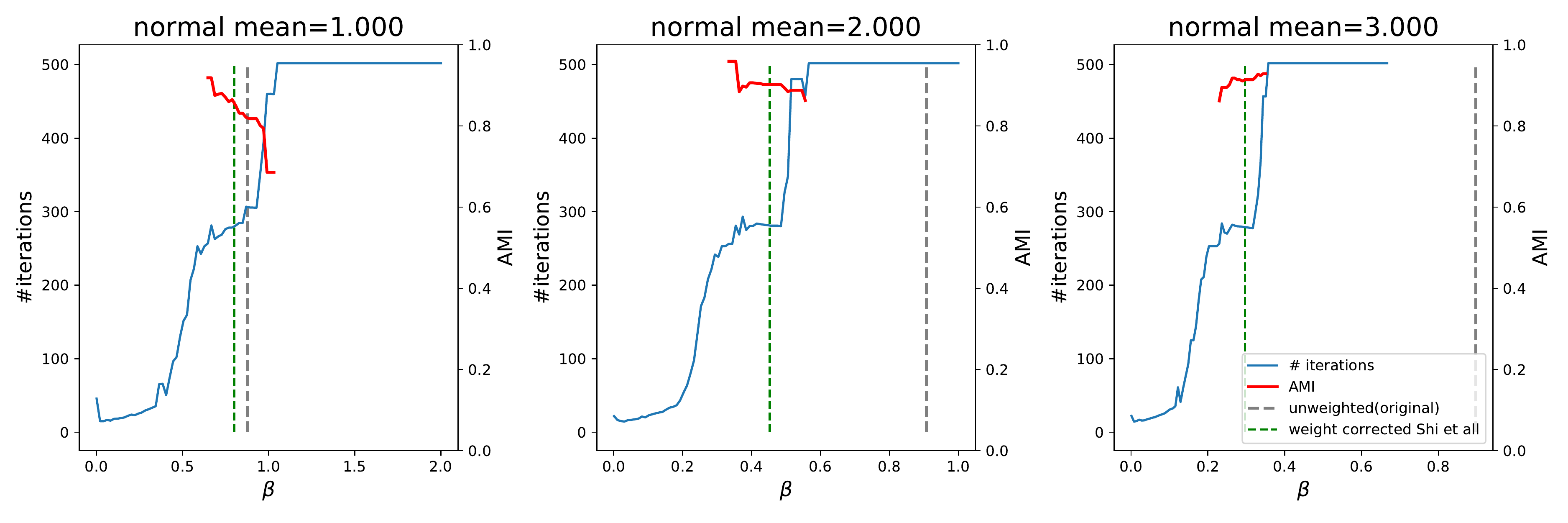}
\caption{Stability boundary for 2-community planted partition stochastic block model graphs with weights assigned randomly from a $\mathcal{N}(\mu,\sigma=.5)$ normal distribution.  These SBMs had $N=200$ nodes with mean degree $c=6$, and $\epsilon=\frac{p_{out}}{p_{in}}=.1$.  Each convergence curve was averaged over 10 realizations of the SBM with different means of the normally distributed edge weights ($\mu=1$, $2$, and $3$ respectively).  The unweighted prediction for $\beta^*$ is given by the black dashed line, while the weight adjusted prediction is given by the dashed green line.  The red curves show the adjusted mutual information with the underlying ground truth.}
\label{sbm2com_betascan}
\end{center}
\end{figure}

We have used Equation~\ref{eq:bstar_cond} to identify the value of $\beta$ to run the algorithm at in all of the experiments within this manuscript.  Since \textit{a priori} the number of communities, $q$, isn't known in advance, we run the algorithm at several values $\beta = [ \beta^*(q=2,c,\langle w\rangle), ...  \beta^*(q=q_{max},c,\langle w\rangle)]$ for a range of expected numbers of communities, $[2,q_{max}]$.  
We reiterate that the heuristic derived works well in most cases, but makes no guarantees that $\beta^*$ will be inside the retrieval phase for all degree distributions and distribution of edge weights.  For some networks scanning a range of $\beta$ values might be required.  

\begin{figure}[htbp]
\begin{center}
\includegraphics[width=\linewidth]{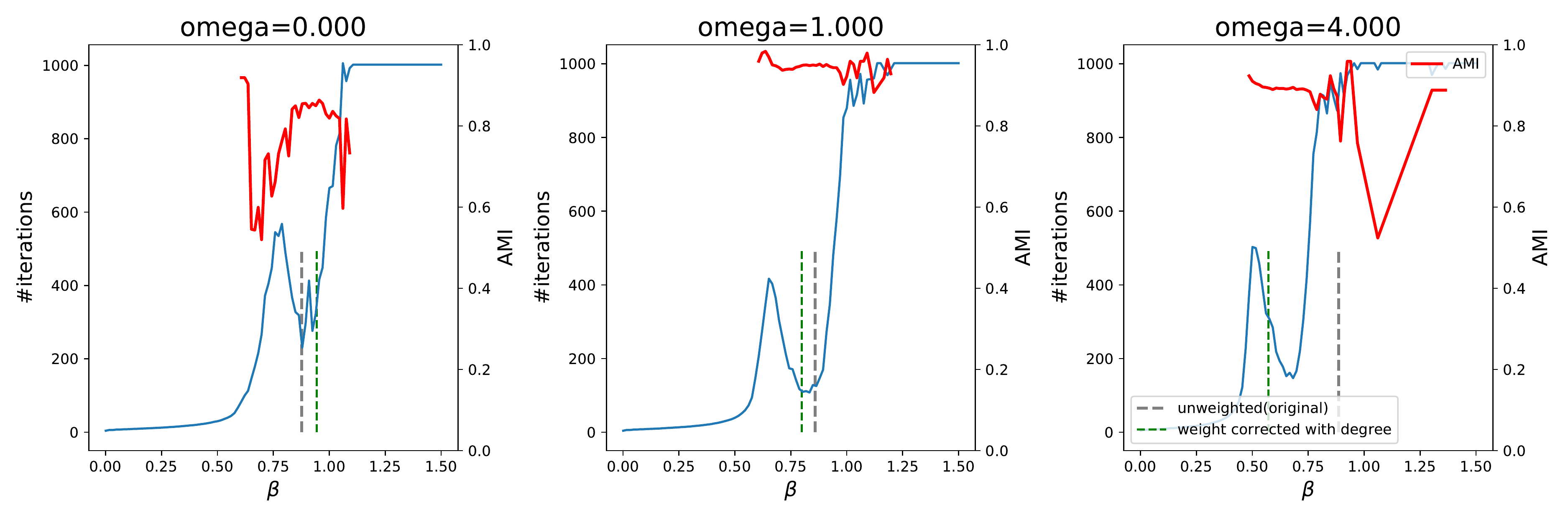}
\caption{Stability boundary for 2-community unweighted multilayer dynamic stochastic block model graph.  Network had $N=100$ nodes within each layer with mean degree $c=6$ and $\epsilon=\frac{p_{out}}{p_{in}}=.1$.  Each convergence curve was averaged over 10 realizations of the SBM model with the algorithm run with different interlayer edge couplings ($\omega=0$, $1$, and $4$ respectively).  The unweighted prediction for $\beta^*$ is given by the black dashed line, while the weight adjusted prediction is given by the dashed green line.  Red curve shows the adjusted mutual information with the underlying ground truth.}
\label{multilayer_betascan}
\end{center}
\end{figure}

In Figure~\ref{multilayer_betascan}, we also show that the retrieval phase of multilayer networks also varies with the strength of the coupling parameter, $\omega$.  The $\beta^*$ predicted by Equation~\ref{eq:bstar_cond} consistently lies within the retrieval phase even as $\omega$ increases (in contrast to the value of $\beta$ given from the unmodified equation).

\subsection{Supplement Figures}

\FloatBarrier
\begin{figure}[!htbp]
\includegraphics[width=\textwidth]{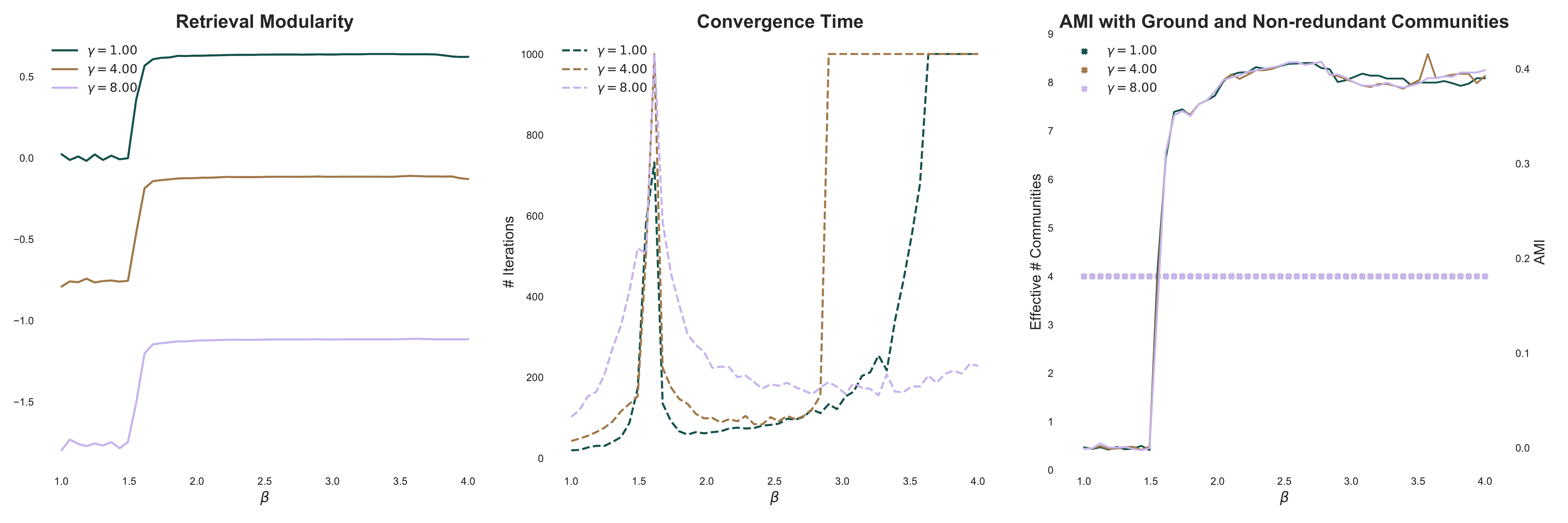}
\caption{We compare the performance of the algorithm for a wide range of $\gamma$ values in the event that the number of communities is fixed at the correct number ($q=4$).  Here we do not allow $q$ to float as described in Section~\ref{numcom} \label{fig:fixqvarygamma}}
\end{figure}

 \begin{figure}[!htbp]
\begin{center}
\includegraphics[width = .5\textwidth]{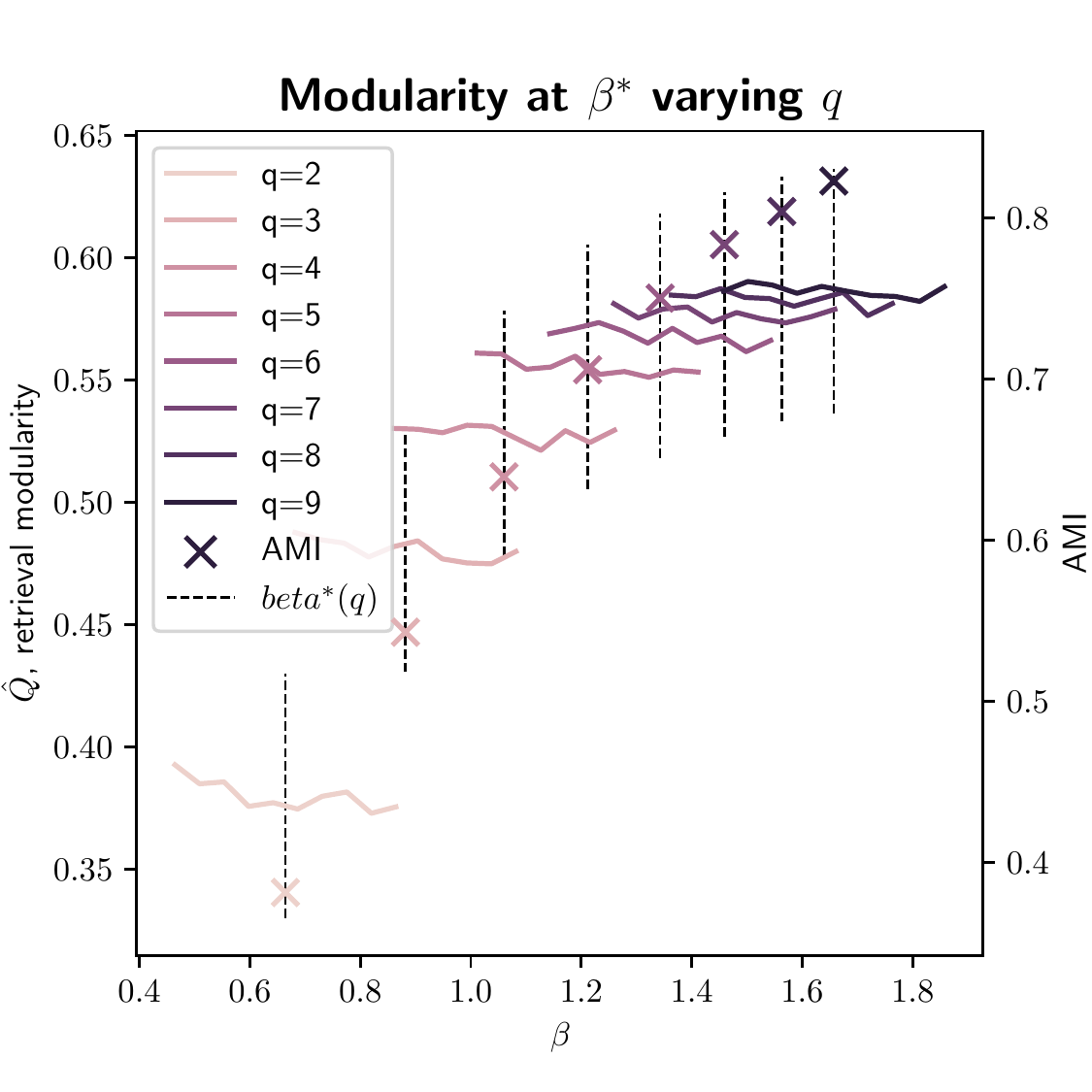}
\caption{ Using the method recommended by Zhang and Moore to select the appropriate value of $q$ for the American NCAA Div-IA College Football Network \cite{Girvan:2002ez,Evans:2010ga}.  Each colored line corresponds to running \textit{modbp} for a given value of $q$ across a window of $\beta$ around $\beta^*(q)$ (shown by black dashed line). Using this method would suggest an appropriate $q \in [6,8]$ depending on the threshold selected.  We note that here, we do not collapse community labels as described in Section~\ref{numcom}; for each run a single fixed value of $q$ is used as well as the default resolution ($\gamma=1$).  AMI with the school conferences is denoted for each $q$ by the colored $\times$ symbols.  \label{fig:chooseqfootball}}
\end{center}
\end{figure}

\
\begin{figure}[!htbp]
\begin{center}
\includegraphics[width =1.0\textwidth]{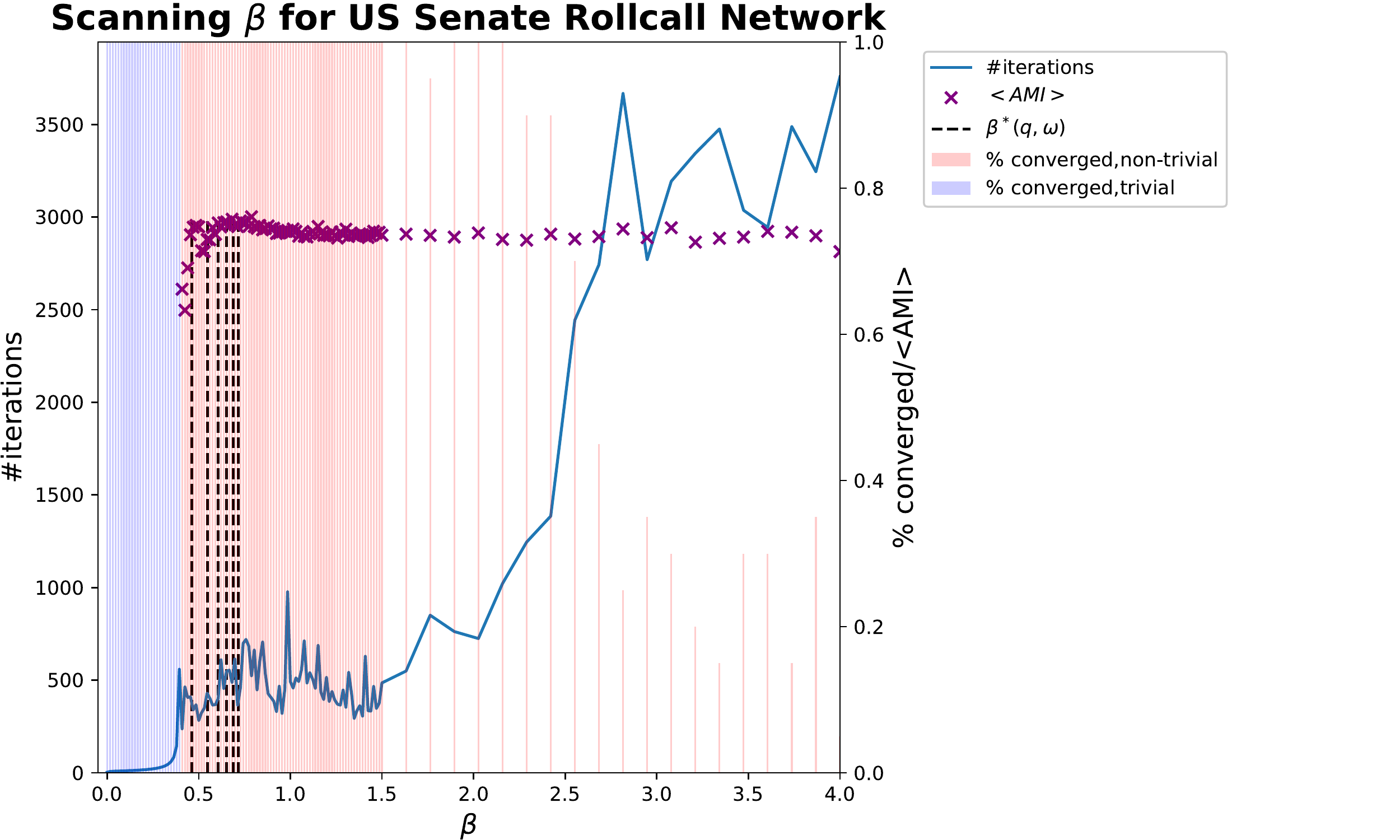}
\caption{ We run \textit{multimodbp} on the US Senate Voting similarity network \cite{Waugh:2009vz}, using the KNN (k=10) as described in Section~\ref{sec:realworld_results} of the main text.  We ran \textit{multimodbp} for a maximum of 4000 iterations across 100 evenly spaced values of $\beta \in [0,1]$.  For each value of $\beta$ we ran \textit{multimodbp} 5 different times.  We show that Shi \textit{et al}'s approach to selecting $\beta^*$ \cite{Shi:2018gc} identifies regions where the algorithm is in the retrieval phase (\textit{i.e} converges to non-trivial partitions).  Vertical dashed black lines show calculated value for $\beta^*(q)$ for $q \in \{4, 6, 8 ,10 ,12, 14\}$.  Vertical blue and red bars denote the percentage of runs for that value of $\beta$ that ultimately converged (percentage is shown by the proportion of the space under the number of iterations curve occupied by the bar).  Bar color denotes whether the identified partitions were trivial ($\psi_t^i=\frac{1}{q}$).   We see that several of these lie within the observed retrieval phase ($q \in \{8,10,12,14\}$)  \label{fig:beta_scan_senate}. }
\end{center}
\end{figure}
\

 \begin{figure}[!htbp]
\begin{center}
\hspace*{-0.125\textwidth}\includegraphics[width = 1.25\textwidth]{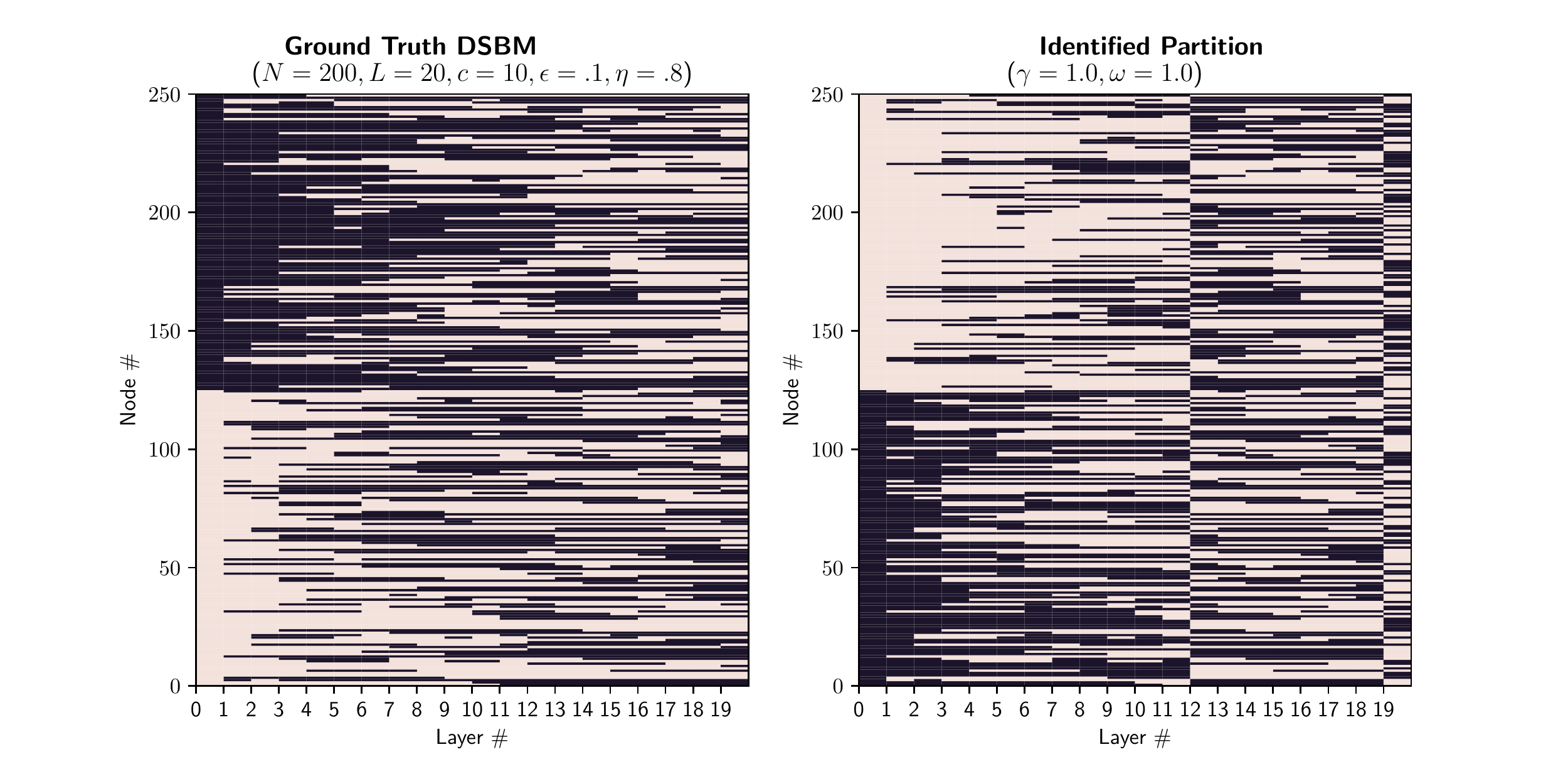}
\caption{ Demonstration of layer ``splitting" on the multilayer dynamic stochastic block model (DSBM).  Left shows the ground truth planted community assignments while the right shows the communities identified by \textit{multimodbp} without the cross layer assignment procedure.  We reiterate that this cross layer label permuting preserves all identified structure within a layer and always results in higher modularity. \label{fig:mlsplit}}
\end{center}
\end{figure}

\begin{figure}[!htbp]
\begin{center}
\hspace*{-0.01\textwidth}\includegraphics[width = 1.3\textwidth]{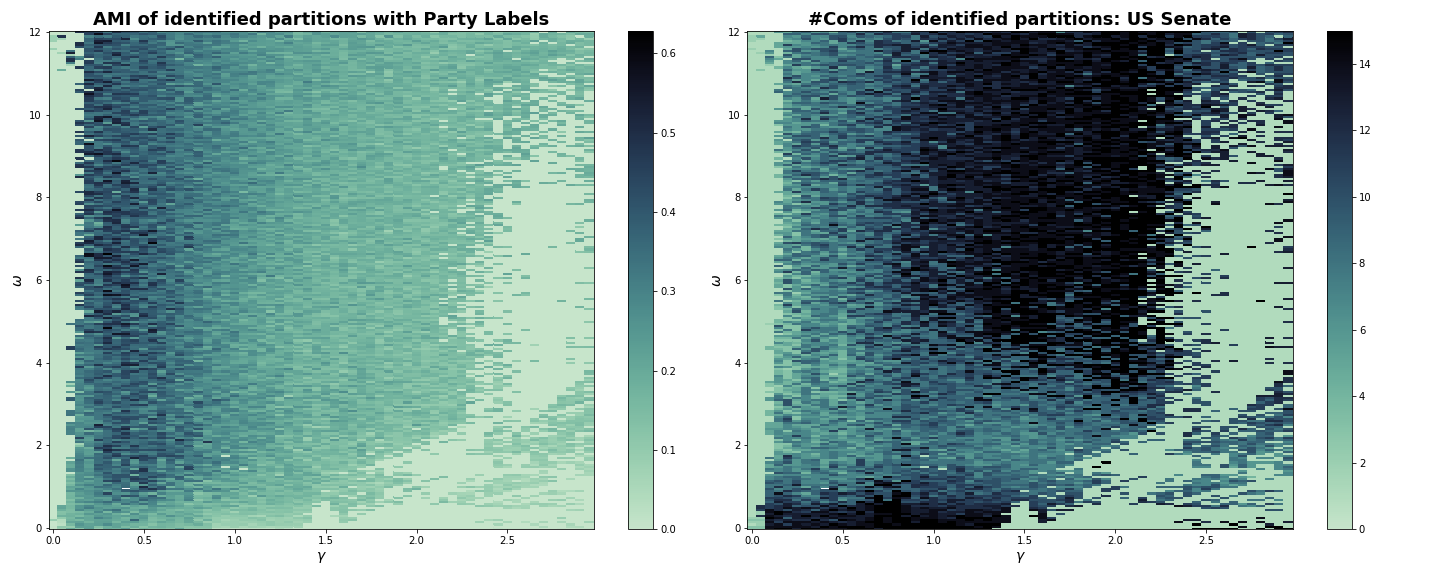}
\caption{ \textit{multimodbp} applied to the US Senate Voting similarity network \cite{Waugh:2009vz}. Left: AMI of identified partitions with the political party labels using \textit{multimodbp} across a range of $\gamma$ and $\omega$ values.  Right: the number of communities identified by the algorithm as a function of the parameters $(\gamma,\omega)$.   \label{fig:senate_gamma_omega}}
\end{center}
\end{figure}

\begin{figure}[!htbp]
\begin{center}
\hspace*{-0.125\textwidth}\includegraphics[width = 1.3\textwidth]{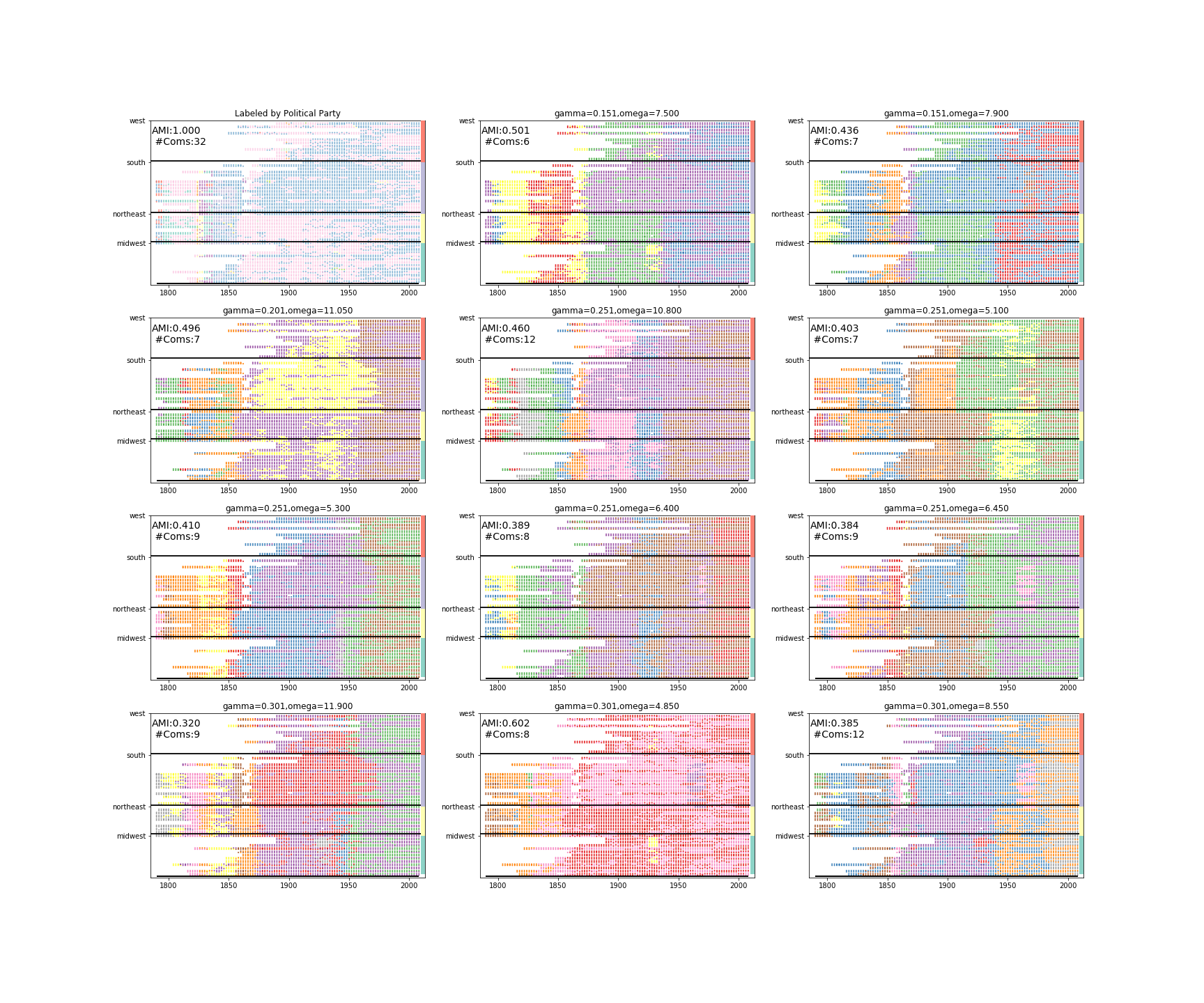}
\caption{ Top identified partitions based on minimization of the Bethe free energy on the US Senate voting similarity network.  In each, each row represents the Senator for a particular State, organized by region, while the x-axis denotes the year of each Congress.  Nodes are colored according to their identified partition, while the top left figure is colored by the political party affiliation of each senator.  \label{fig:senate_state}}
\end{center}
\end{figure}

\begin{figure}[!htbp]
\begin{center}
\hspace*{-0.1\textwidth}\includegraphics[width = 1.3\textwidth]{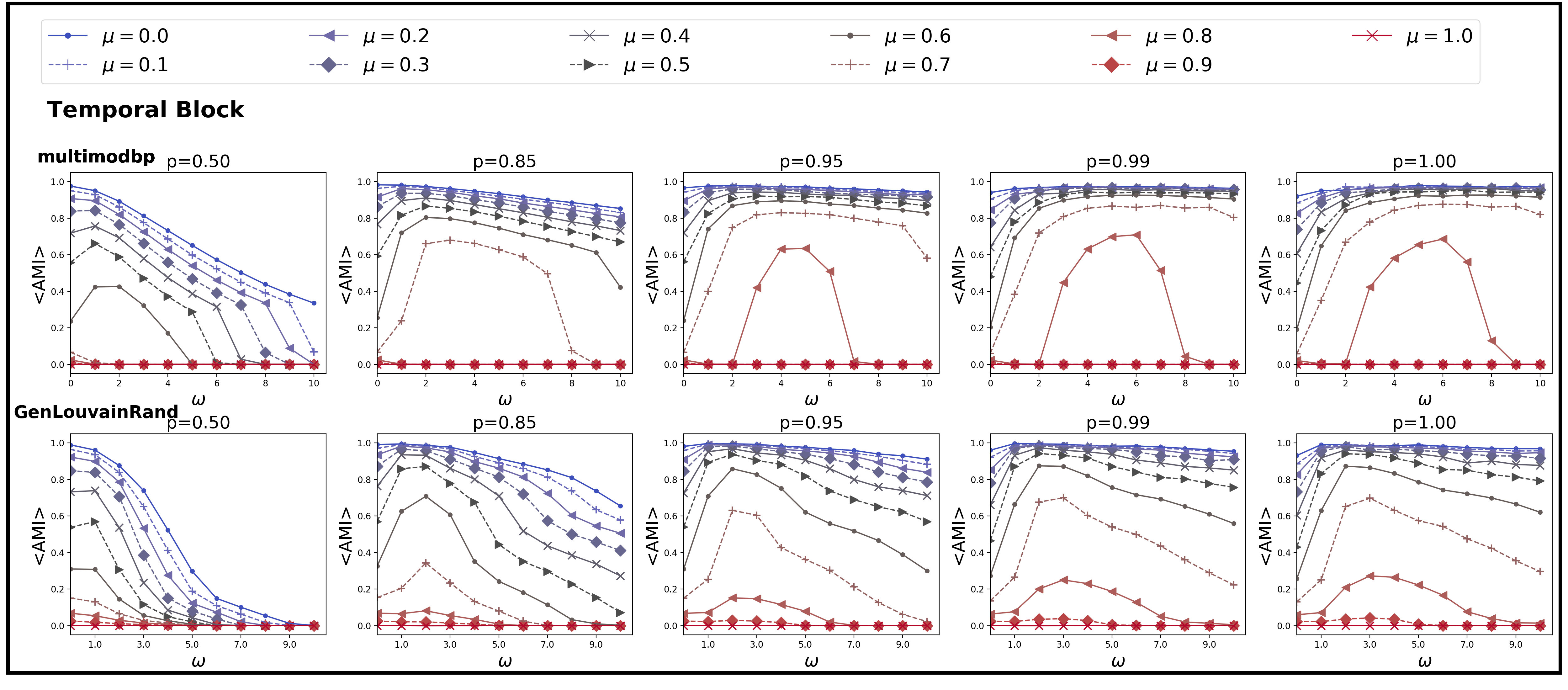}
\caption{ \label{fig:temp_block} Comparison of the performance of \textit{multimodbp} (top row) with \textit{GenLouvain} on the temporal block topology from Ref~\cite{Bazzi:vn}.  In this model discontinuities are introduces in the community structure such that the layers are divided into 4 equally sized, independent blocks, with a temporal topology within each (analogously to the multiplex block topology described in Section~\ref{sec:mulitlayer_benchmarking:multiplex}).  Each point represents the $\left<\text{AMI}\right>$ over 100 different realizations of the model.  Multilayer community partitions are drawn from Dirichlet distribution with $\theta=1$ and $q=5$.  Intralayer edges are samples from a DCSBM with $\eta_k=-2$, $k_{max}=30$, $k_{min}=3$.  Each network has 100 node-layers in each layer with 150 layers for a total of 15000 node-layers.}
\end{center}
\end{figure}

\begin{figure}[!htbp]
\begin{center}
\hspace*{-0.1\textwidth}\includegraphics[width = 1.3\textwidth]{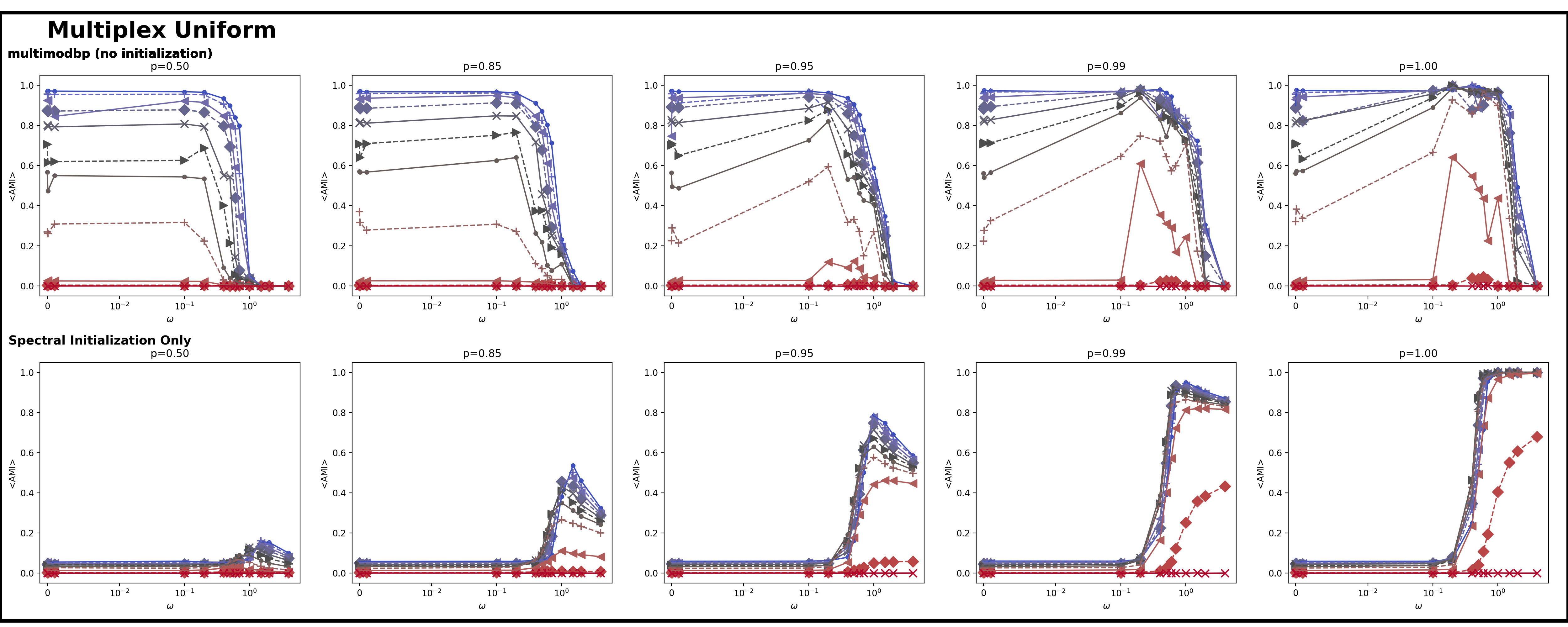}\\
\hspace*{-0.1\textwidth}\includegraphics[width = 1.3\textwidth]{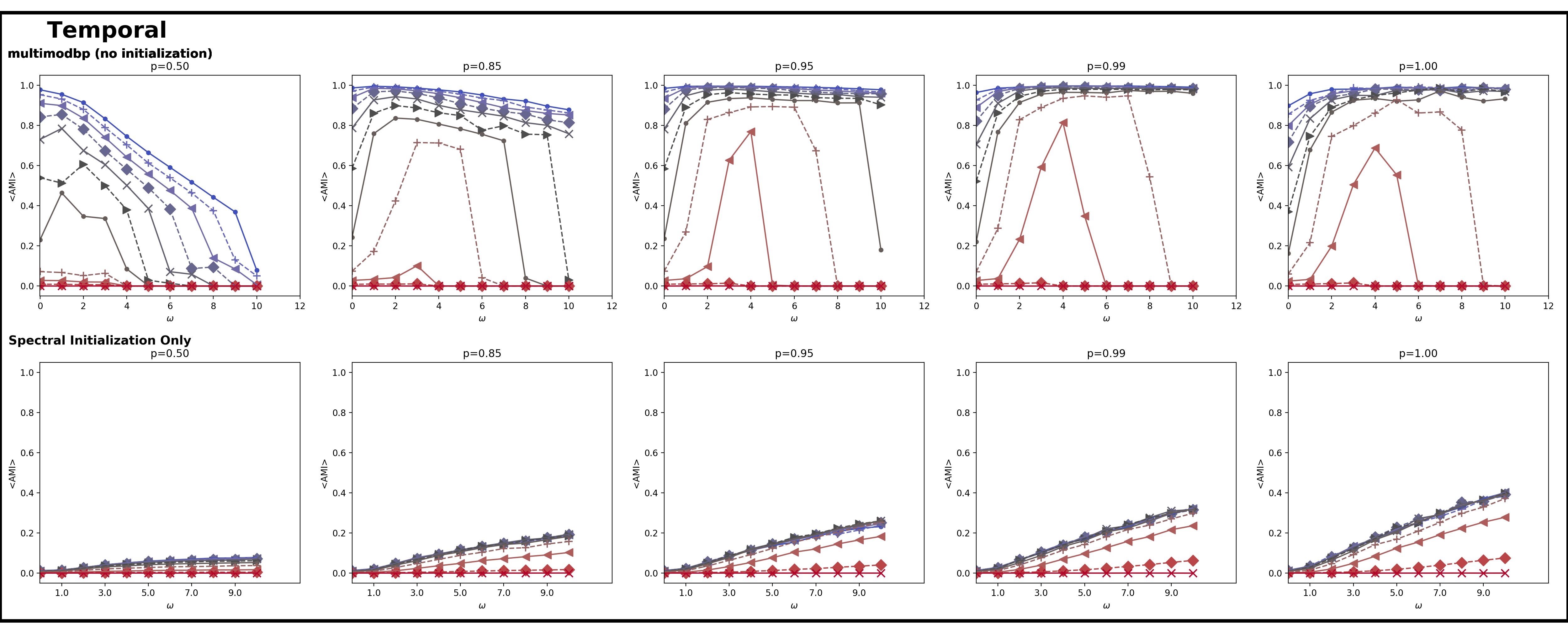}
\caption{ \label{fig:multiplex_noinit} Upper panel: The top row shows the performance of \textit{multimodbp} on the uniform multiplex network (as specified in Section~\ref{sec:mulitlayer_benchmarking:multiplex} of the main text) \emph{without} the spectral initialization detailed in the main text.  Performance of the algorithm at higher $\omega$ trails off abruptly.  For comparison with \textit{multimodbp} with spectral initialization, see Figure~\ref{fig:multilayer_benchmarks} in main text.  In the second row, we show performance of just the spectral initialization (without \textit{multimodbp}).  The spectral initialization's performance tends to be better at higher values of $\omega$, complementing the deficiencies in \textit{multimodbp}.  In the bottom two rows, we show the corresponding performance of \textit{multimodbp} without spectral initialization (row 3) and only using the spectral initialization (row 4).}
\end{center}
\end{figure}

\begin{figure}[!htbp]
\begin{center}
\includegraphics[width = 1\textwidth]{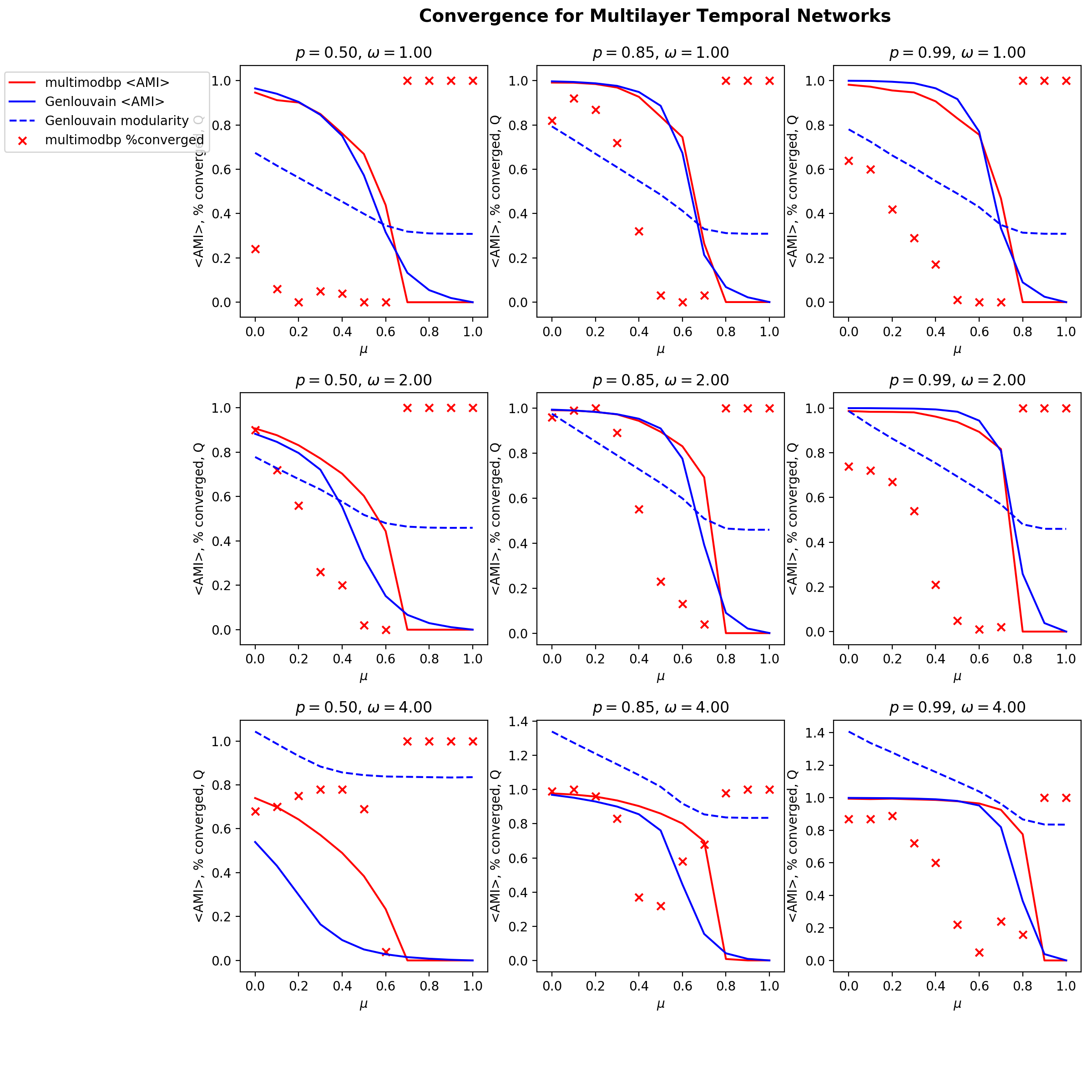}
\caption{ \label{fig:temp_across_params} Detectability of communities in the multilayer temporal benchmarking network as $\mu$ is varied.  Down the rows of the subfigures $\omega$ is increased, while across the columns, we increase $p$ the alignment of communities across the layers.   For each pair of $\omega$ and $p$, we plot the average $\left<\text{AMI}\right>$ of the detected communities for both \textit{multimodbp} (solid red line) and for \textit{GenLouvain} (solid blue line).  We also show the average modularity of the partitions identified by \textit{GenLouvain} (dashed blue line) as well as percentage of trials that converged to a non-trivial solution for \textit{multimodbp}. }
\end{center}
\end{figure}

\begin{figure}[!htbp]
\begin{center}
\includegraphics[width = 1\textwidth]{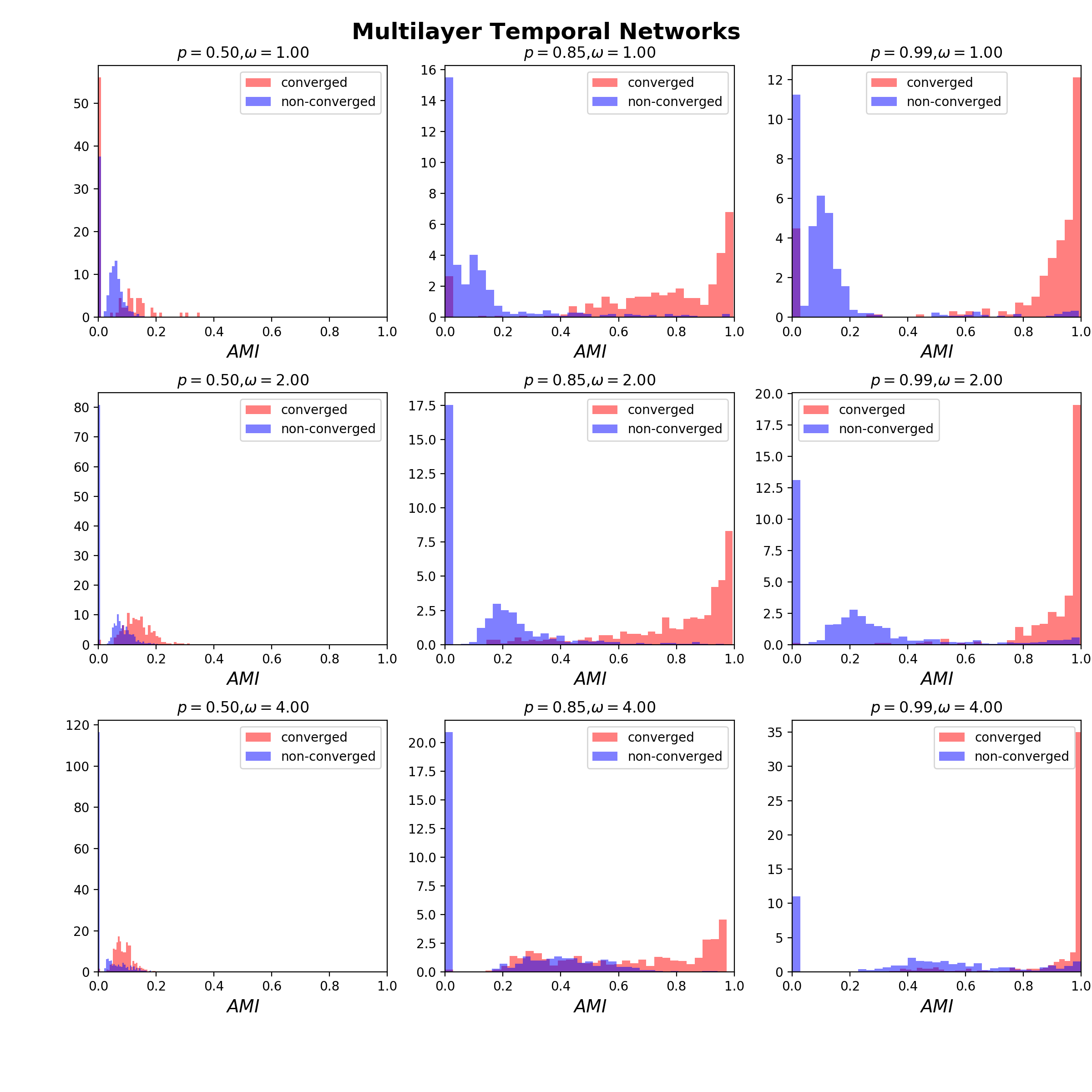}
\caption{ \label{fig:temp_hist_conv}  We show the distribution of  AMI for individual runs of \textit{multimodbp} split out by whether or not the run converged for the uniform temporal topology benchmark data.  Each panel represents a different combination of $(p,\omega)$ (runs are lumped together across different values of $\mu$).  The red histogram shows the distribution of AMI higher for the runs that converged vs those that did not (blue histogram).  There are a few runs that did not converge which nevertheless have high correlation with the known structure.}
\end{center}
\end{figure}

\begin{figure}[!htbp]
\begin{center}
\includegraphics[width = 1\textwidth]{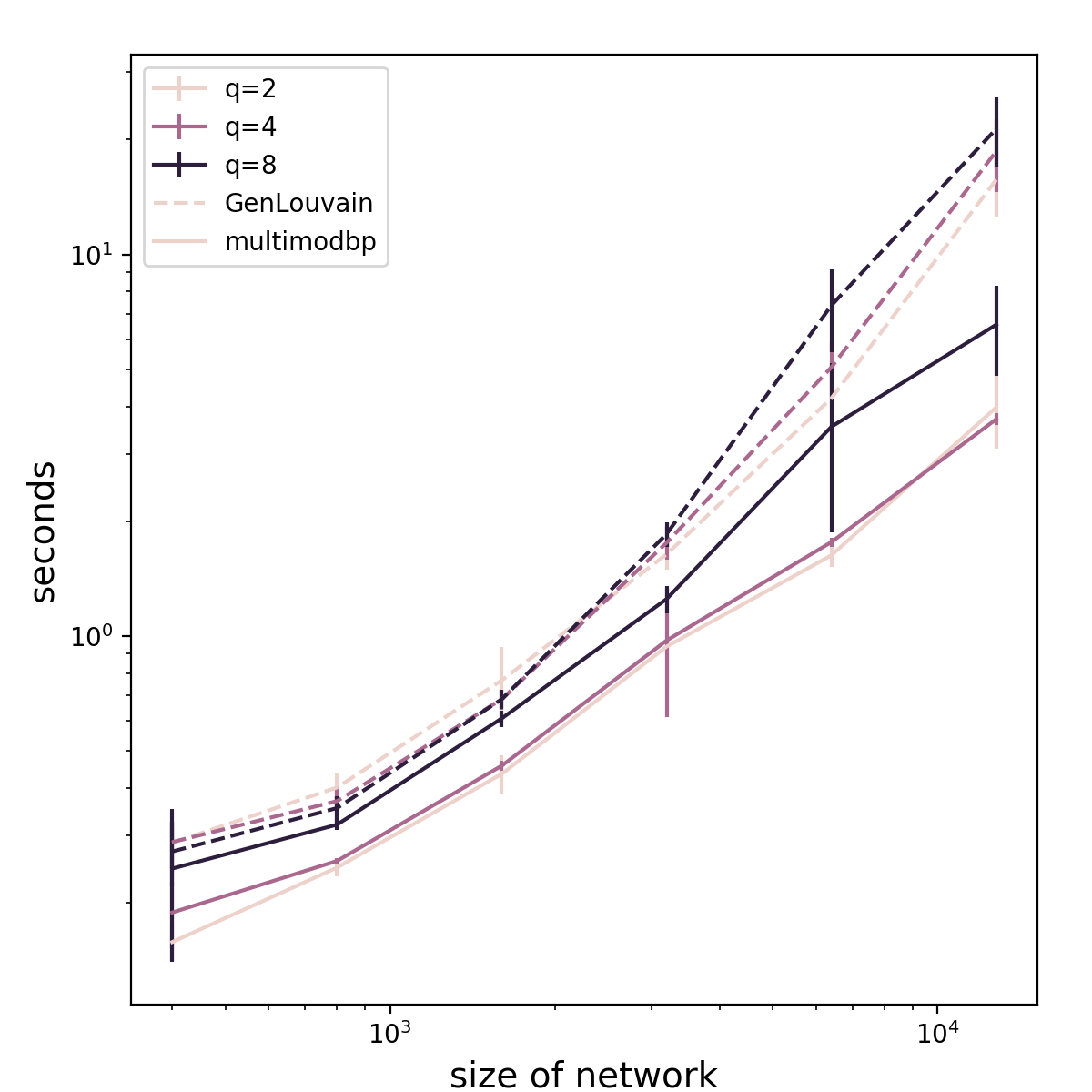}
\caption{ \label{fig:time_trial} We compare the run time for \textit{GenLouvain} (dashed lines) vs that of \textit{multimodb} (solid line) using the DSBM (see Section~\ref{sec:dsbm}).  We vary both the size of the network (x-axis), as well as the number of communities planted within the network, denoted by the color of each plot.  Each network has 5-layers, with mean degree $c=10$ and detectability, $\epsilon=.1$.  \textit{GenLouvain} was run using the iterated version using alignment method across the different layers.  \textit{multimodbp} was run using a single value of $\beta$, computed using the Equation~\ref{eq:bstar_cond}, without spectral initialization or any of the alignment post processing tool included in the package.  We note that for networks where the number of communities is not known \textit{a priori}, scanning a range of $\beta$ values might be required.  Although \textit{GenLouvain} is not necessarily the fasted approach for the specific null model we have used, we use it as a practical comparison because of it is widely used.}
\end{center}
\end{figure}

\FloatBarrier
\bibliographystyle{journalplain}
\bibliography{./supporting_files/biblio.bib}

\FloatBarrier

\end{document}